\documentclass[journal]{IEEEtran}
\usepackage{amsmath,amssymb,amsfonts,amsthm}
\usepackage[caption=false,font=footnotesize,labelfont=rm,textfont=rm]{subfig}
\usepackage{graphicx}
\usepackage[dvipsnames, svgnames, x11names]{xcolor}

\usepackage[colorlinks=true,linkcolor=red,citecolor=green]{hyperref}

\usepackage{algorithmic,algorithm}
\usepackage{setspace}

\usepackage{graphicx}
\usepackage{float}
\usepackage{subfig}
\usepackage{overpic}
\usepackage{epstopdf}

\usepackage{bm}

\begin{document}
\title{SCSC: A Novel Standards-Compatible Semantic Communication Framework for Image Transmission}
\author{	
\IEEEauthorblockN{Xue Han, Yongpeng Wu, \emph{Senior Member, IEEE}, Zhen Gao, \emph{Member, IEEE}, Biqian Feng, Yuxuan Shi, \\Deniz G{\"u}nd{\"u}z, \emph{Fellow, IEEE}, and Wenjun Zhang, \emph{Fellow, IEEE}}
\thanks{X. Han, Y. Wu, B. Feng, Y. Shi, and W. Zhang are with the Department of Electronic Engineering, Shanghai Jiao Tong University, Minhang 200240, China (e-mail: han.xue@sjtu.edu.cn; yongpeng.wu@sjtu.edu.cn;  fengbiqian@sjtu.edu.cn; ge49fuy@sjtu.edu.cn; zhangwenjun@sjtu.edu.cn) (Corresponding author: Yongpeng Wu).}
\thanks{Z. Gao is with the School of Information and Electronics, Beijing Institute of Technology, Beijing 100081, China (e-mail: gaozhen16@bit.edu.cn).}
\thanks{Deniz G{\"u}nd{\"u}z is with the Department of Electrical and Electronic Engineering, Imperial College London, SW7 2AZ London, U.K. (e-mail: d.gunduz@imperial.ac.uk).}
}
\maketitle

\begin{abstract}
Joint source-channel coding (JSCC) is a promising paradigm for next-generation communication systems, particularly in challenging transmission environments. In this paper, we propose a novel standard-compatible JSCC framework for the transmission of images over multiple-input multiple-output (MIMO) channels. Different from the existing end-to-end AI-based DeepJSCC schemes, our framework consists of learnable modules that enable communication using conventional separate source and channel codes (SSCC), which makes it amenable for easy deployment on legacy systems. Specifically, the learnable modules involve a preprocessing-empowered network (PPEN) for preserving essential semantic information, and a precoder \& combiner-enhanced network (PCEN) for efficient transmission over a resource-constrained MIMO channel. We treat existing compression and channel coding modules as non-trainable blocks.  Since the parameters of these modules are non-differentiable, we employ a proxy network that mimics their operations when training the learnable modules.
Numerical results demonstrate that our scheme can save more than 29\% of the channel bandwidth, and requires lower complexity compared to the constrained baselines. We also show its generalization capability to unseen datasets and tasks through extensive experiments.

\end{abstract}


\section{Introduction}
\IEEEPARstart{A}{s} the development towards the sixth-generation (6G) of mobile communication networks are at full speed, a widely accepted challenge is the explosive growth in multimedia transmission, which finds applications in many emerging verticals and services, e.g., augmented reality/virtual reality (AR/VR), autonomous driving, and intelligent transportation/factory \cite{Yang_2022}. To tame the increasing pressure on the bandwidth resources, a new communication paradigm has emerged, called \textit{semantic communication}, which aims at reducing the amount of transmitted information by only sending the relevant information to the receiver, and complements conventional approaches aimed at increasing the network capacity. Semantic communication benefits from the in-depth fusion of information and communication technology advances and artificial intelligence (AI) tools \cite{ping_sem}\cite{Gunduz_sem}, particularly to extract semantics from various complex input signals.  

While semantic encoding can be applied directly at the application layer as a source compression technique, it is increasingly becoming evident that to meet the performance and latency requirements of the aforementioned applications, we need to go beyond the conventional separation architecture, and consider semantic communication in an end-to-end fashion. This requires a joint source-channel coding (JSCC) approach, where the encoder maps the input signal directly to a channel input signal.
Recently, an increasing number of AI-based semantic communication frameworks have emerged relying on deep learning techniques. 
Bourtsoulatze et al. \cite{DJSCC} were the first to propose a deep learning-based JSCC scheme, called DeepJSCC, for wireless image transmission, and it was shown to outperform the separation based baselines that employ state-of-the-art compression and channel coding schemes. 
Jankowski et al. \cite{Jankowski} extended DeepJSCC to task-oriented applications.
Weng et al. \cite{Weng} investigated the speech recognition and speech synthesis as the tasks of the communication system and proposed a robust model to cope with different channel conditions. Dai et al. \cite{NTSCC} proposed nonlinear transform source-channel coding (NTSCC), and achieved content-aware variable-length JSCC via introducing an entropy model on the semantic latent representations.
The DeepJSCC approach has since been extended to feedback channels \cite{DJSCC-f}, to MIMO channels \cite{Wu_2024}, and to the relay channel \cite{11_DL}\cite{Bian_relay}.
Hu et al. \cite{15_DL} designed the masked vector quantized variational autoencoder (VQ-VAE) for combating the semantic noise.
However, it is worth mentioning that although these pioneering AI-based works show promising performance results, their adoption in practical systems is challenging as they require new standardization and hardware design.

This motivates JSCC schemes that are compatible with existing wireless communication modules.
Jiang et al. \cite{Jiang} proposed a DL-based JSCC framework for semantic communication with classic channel coding and HARQ, which can be easily implemented into existing HARQ systems.
Tung et al. \cite{Tung_Jun_2022} investigated the effects of constraining the transmission either to finite input alphabets, or to a predefined constellation, by introducing the so-called DeepJSCC-Q scheme. 
Huang et al. \cite{Huang_2023} considered data transmission simultaneously with semantic communication using low density parity check (LDPC) code and quantization in a separate source-channel coding system. 
Yao et al. \cite{Yao_2022}  proposed a class of novel semantic coded transmission (SCT) schemes over multiple-input multiple-output (MIMO) fading channels and designed a spatial multiplexing mechanism to realize adaptive coding rate allocation and stream mapping.

Semantic communication integrated with end-to-end JSCC designs for image transmission has demonstrated more satisfying performance; however, they have several limitations in front of their adoption in practical systems. First of all, they lose the modularity of SSCC approaches; that is, we need to design and employ a separate JSCC scheme for different source modalities. The second limitation is noise accumulation. Unlike digital schemes, in JSCC the decoded signal always has some residual noise, which accumulates when the signal is forwarded over multiple hops \cite{Bian_relay}. Last but not least, JSCC introduces a security risk since the transmitted signal is correlated with the underlying source signal, and limiting the information leakage results in performance loss \cite{Tung_ICC}. Another significant challenge for the adoption of JSCC in practical communication systems is the lack of coding standards for these learned compression and transmission methods, and hence, they can only be massively deployed after significant design and standardization efforts. 
		
Despite these limitations of end-to-end JSCC design, the benefits of jointly optimizing the parameters of separate source and channel codes, in the context of cross-layer design have been widely acknowledged \cite{Schaar_2005}\cite{Zhai_2007}. This motivates us to develop a data-driven approach to amalgamate the strengths of existing digital compression and communication codecs with the power of end-to-end design taking into account the semantic distortion measure. Since the proposed approach is built around existing codecs, it can be adopted in current systems by simply deploying the PPEN and PCEN modules in corresponding devices.

In this work, we go one step further and build a standards-compatible semantic communication framework upon conventional source-channel codecs, which benefits from low complexity and high efficiency \cite{edge_2022}. Different from the efforts primarily focused on image compression \cite{Luguo}\cite{Yangmingyi}, which is inherently a form of source coding without considering the communication process. Our work builds upon the principles of JSCC and complete digital source and channel coding. The proposed framework has the following advantages:  first, the semantic features related to downstream tasks are exploited to improve the transmission performance; second, it can adapt to stochastic fading in complex MIMO channels with finite alphabet signals \cite{MIMO}.  Specifically, for the former, we introduce an image preprocessing module before standard codecs while the latter is resolved through the implementation of a precoding-enhanced module. The proposed learnable modules are connected with the standard SSCC components, e.g., BPG and LDPC codes, which enhances the framework's applicability in real-world commercial systems. Different from DL-based networks, standard codecs that are not specifically designed for a particular source dataset or a specified downstream task, indicating a more general approach to data compression and transmission.

In this paper, we propose a generalized standards-compatible semantic communication (SCSC) framework based on conventional digital communication codecs, while leveraging the advantages of semantic communication for wireless image transmission. 
Specifically, we propose a learning-based preprocessing module before the standard codec such that the input images are processed before being fed to the compression codec.
We also employ a precoding enhanced network to further process the output of the conventional coding pipeline before transmission over the channel. The receiver, on the other hand, relies on a series of reverse operations to decode the image, or to carry out downstream vision tasks, like semantic segmentation \cite{Kang_2022}. These modules are trained jointly in an end-to-end fashion to improve the performance. Note that the training process involves non-differentiable components due to the discrete operation within the conventional SSCC components. This prevents the backpropagation of gradients. To overcome this challenge, we employ an end-to-end proxy network to simulate these components, which can then be used to train the learnable modules in an end-to-end fashion.
The main contributions of this paper can be summarized as follows.
\begin{itemize}
    \item {\textit{Semantic Preprocessing:} We propose a preprocessing-empowered network (PPEN) at the encoder that processes the input image before feeding it to the standard compression codec to reduce the semantic information redundancy while ensuring satisfactory performance for the desired downstream task. Specifically, a distortion-aware compensation (DAC) module combined with quantization adaptive (QA) layers is designed to output a preprocessed image, which can be effectively integrated into standard codecs with different combinations of compression ratios and channel coding rates.}

    \item {\textit{High-speed Wireless Transmission:} We consider practical MIMO fading communication channels with discrete constellations. A deep unfolding-based precoder \& combiner-enhanced network (PCEN) is employed for combating the channel variations and further increasing the end-to-end performance. PCEN consists of the precoder-enhancement network (PEN) at the transmitter (Tx) and the combiner-enhancement network (CEN) at the receiver (Rx). Specifically, by unfolding the training parameters into the proposed PCEN, it can greatly reduce the number of network parameters and improve the received signal-to-noise ratio (SNR).}

    \item {\textit{Proxy Network for End-to-End Training:} To enable end-to-end optimization of the entire network, we introduce a learnable proxy network to approximate non-differentiable standard codecs. This facilitates gradient backpropagation when training the parameters of the PPEN and PCEN.}

    \item {\textit{Performance Validation:} To demonstrate the superiority of our SCSC framework, we perform extensive experiments on the semantic segmentation task over Rayleigh fading channels. Experiments show that compared with the baselines, the proposed SCSC can save up to 29\% channel bandwidth when achieving the same end-to-end performance. Several ablation studies are conducted to validate the effectiveness and universal compatibility of PPEN and PCEN in practical systems.}
\end{itemize}

\begin{figure*}
	\centering
	\subfloat[]{
		\label{system(a)}
		\includegraphics[width=0.8\linewidth]{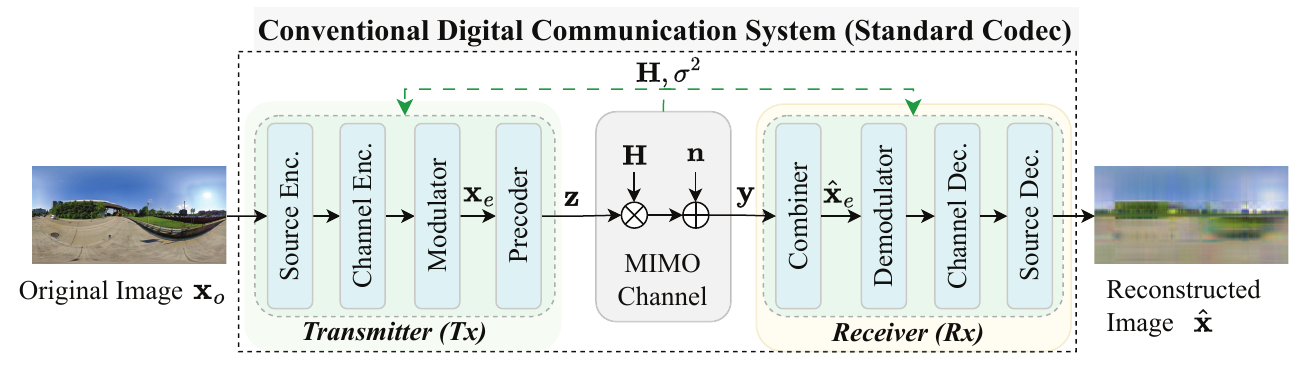}
	}\hfill
	\subfloat[]{
		\label{system(b)}
		\includegraphics[width=0.98\linewidth]{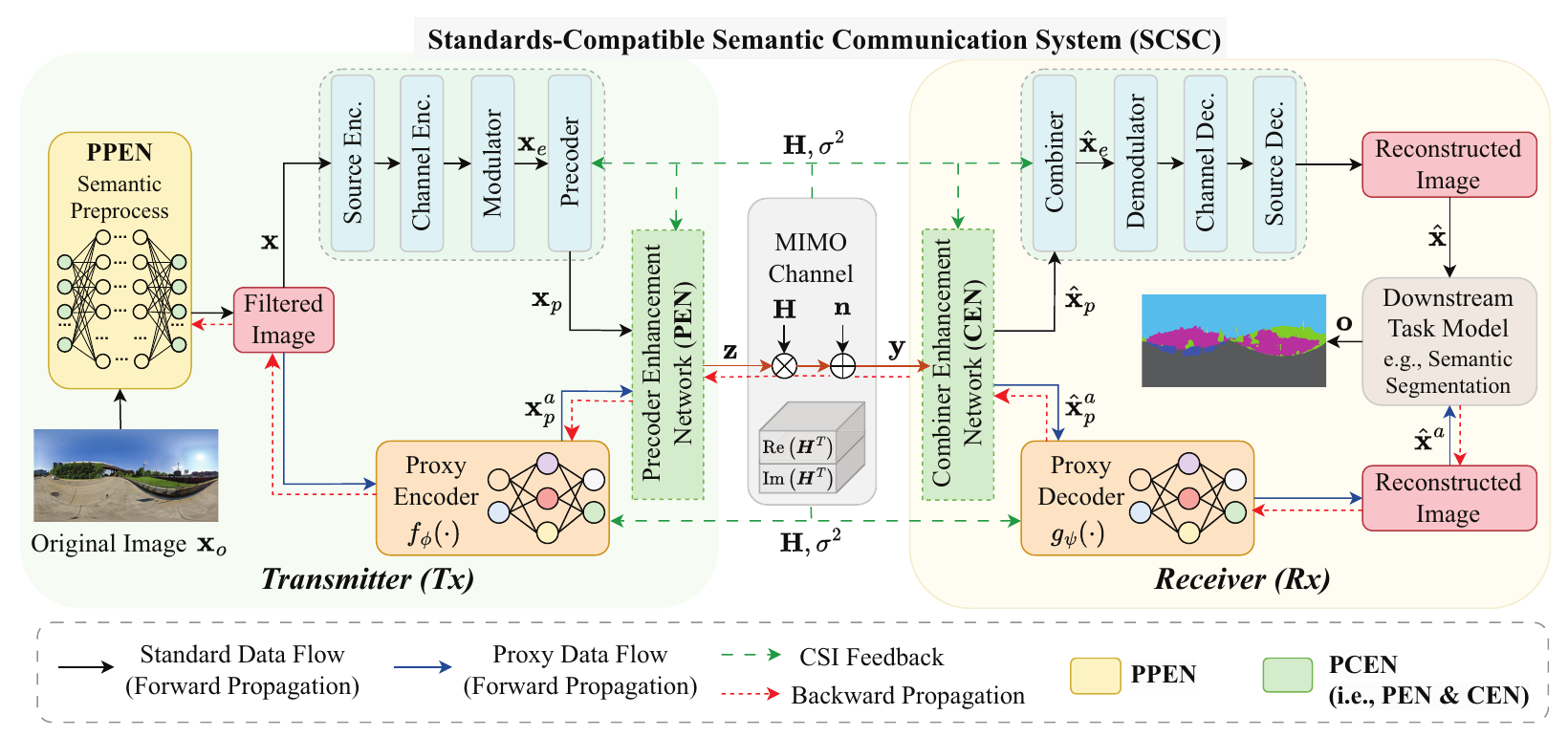}
	}
	\caption{(a) Conventional digital communication system. (b) Overview of the proposed SCSC framework for semantic communications.
	}
	\label{system}
\end{figure*}

The remainder of the paper is organized as follows. Section II introduces the system model and describes the overall structure of the proposed SCSC framework. Section III presents the detailed modules and key training methodologies. Section IV provides numerical comparisons with a number of baselines to quantify the performance gains from the proposed method. Finally, Section V concludes this paper.

\textit{Notations}: Vectors and matrices are denoted by boldface lower case, and boldface upper case letters respectively. The superscript, $\mathbf A^{-1} $ and $ \mathbf{A}^{H}$ are the inverse and Hermitian of matrix $\mathbf A$. Moreover, $ \mathbb{C} $ and $ \mathbb{R} $ represent the sets of all complex and real values, respectively. $ \mathbb{E}\{\cdot\} $  denotes the statistical expectation operation. $ \|\cdot\|_{\mathrm{F}} $ denotes the Frobenius norm and $ \|\cdot\|_{\infty}$  is the infinity norm. In addition, $\Pi_\mathcal{M}$ represents the projection onto the given finite alphabet set $\mathcal M$. 

\begin{figure*}[htpb]
	\vspace{-0.4cm}  
	\centering
	{\includegraphics[scale=0.63]{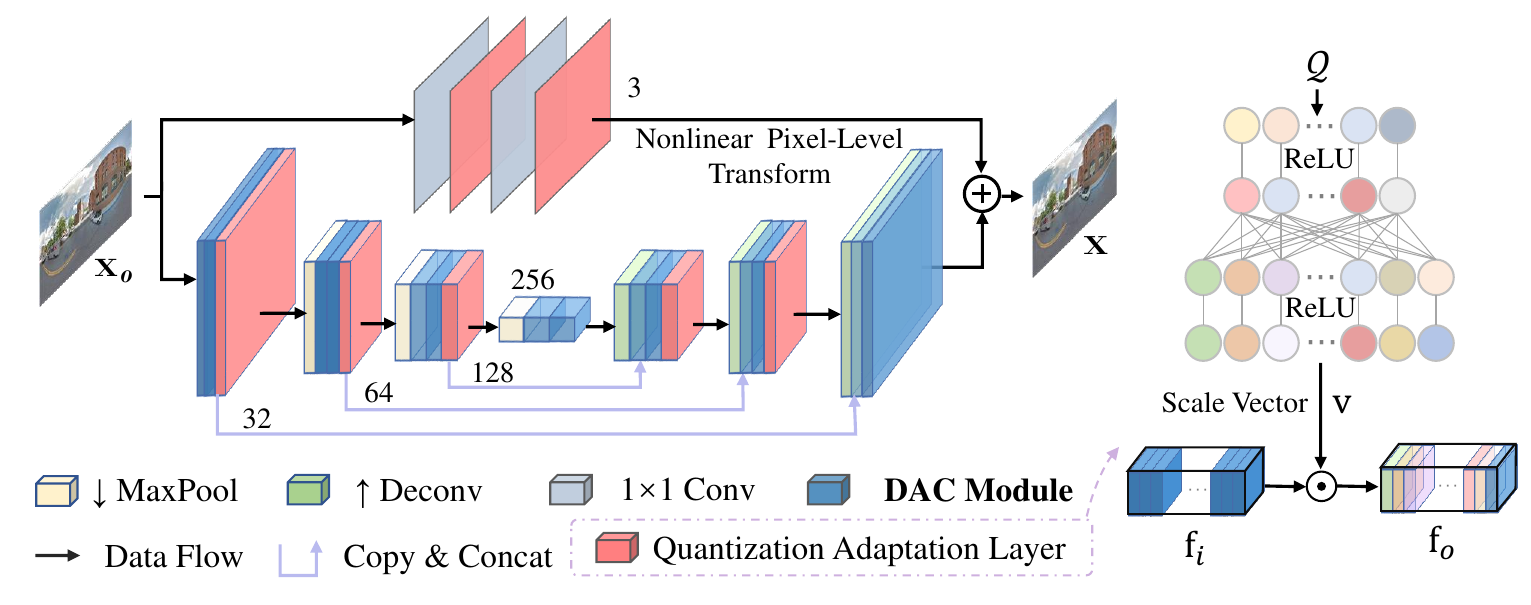}}
	\caption{The detailed framework of our PPEN module. The DAC module is used in the last stage of the distortion compensation layer.}
	\label{preprocessing}
\end{figure*}

\section{System Model}
We consider wireless image transmission over MIMO block fading channels for multiple downstream tasks. Fig. \ref{system}\subref{system(a)} illustrates a conventional communication system, in which the Tx maps the original image to a symbol sequence transmitted to the Rx over the channel. 
We seek to improve the performance of this conventional communication system by the inclusion of plug-and-play adaptation modules that can be deployed at the encoder and decoder for both the reconstruction of the input image and other downstream tasks. In particular, as illustrated in Fig. \ref{system}\subref{system(b)}, we propose a novel standards-compatible image semantic transmission framework that can be deployed over practical MIMO fading channels with finite-alphabet to optimize for various downstream tasks.

\subsection{Semantic-aware Transmitter}
Let $\mathbf{x}_o\in \mathbb{R}^{h\times w\times 3}$ denote the source image at the Tx, where $h$, $w$, and $3$ denote the height, width, and number of color channels in the RGB format, respectively. 
As shown in Fig. \ref{system}\subref{system(b)}, the encoding process at the Tx involves three parts: PPEN, standard transmit modules (which encompass source coding, channel coding, modulator, and precoder), and PEN. Specifically, the process begins by feeding the original image into PPEN, denoted by $E_{\bm \xi}(\cdot): \mathbb{R}^{h\times w\times 3} \mapsto \mathbb{R}^{h\times w\times 3}$ with parameter $\bm {\xi}$, to generate the preprocessed image $\mathbf{x}$, which effectively reduces redundant information, and potentially extracts and amplifies task-relevant features. For example, conventional image compression algorithms suppress high frequency components that are not perceived by human visual systems to reduce the communication rate without sacrificing the perceptual reconstruction quality much. However, this may result in poor performance in certain downstream tasks that would benefit from such information. PPEN, in this case, can potentially extract such useful information and embed it into low frequency components to make sure they will be conveyed through the conventional codec.

Next, let $L$, $N_t$, and $k$ denote the numbers of modulation symbols, antennas at the Tx, and the channel uses, respectively. The Tx encodes the filtered image $\mathbf{x}$ into a sequence of discrete constellation symbols $\mathbf x_e\in  \mathcal{M}^{L}$ via the function $S_{\bm e}(\cdot): \mathbb{R}^{h\times w\times 3} \mapsto \mathcal{M}^{L}$, which incorporates standard source coding (e.g., better portable graphics (BPG \cite{BPG}), joint photographic experts group (JPEG \cite{JPEG}) or JPEG2000 \cite{JPEG2000}), followed by channel coding (e.g., low-density parity-check (LDPC) codes \cite{LDPC} or polar codes \cite{Polar}), and a modulator. 
The modulation process maps encoded bits to symbols, which take values over a finite set  $\mathcal{M} = \{\mathcal M_0, \mathcal M_1, \ldots,  \mathcal M_{M-1}\}$ with $M = |\mathcal{M}|$, e.g., we have $M = 2$ for binary phase-shift keying (BPSK) and $M = 4$ for quadrature-amplitude modulation (QAM). 
To support the transmission of $N_s$ streams in a MIMO system, where $N_s$ is the number of data streams supported by the Rx, a precoder $S_{\bm p}( \cdot): \mathcal{M}^{L} \mapsto \mathbb{C}^{N_t\times k}$ is employed to transform the channel symbol vector $\mathbf x_e$ to codeword $\mathbf x_p \in \mathbb {C}^{N_t\times k}$ via linear or nonlinear precoding techniques. We assume that $N_{s} \leq N_{t}$ and the precoder maps the $N_{s}$ data streams to the $N_{t}$ antennas, with $L= N_{s} \times k$ symbols. Finally, to further improve the end-to-end performance, we propose a precoder-enhancement network (PEN), denoted by $P_{\bm \theta}(\cdot): \mathbb C^{N_t\times k} \mapsto \mathbb C^{N_t\times k}$, parameterized by  $\bm \theta$, to refine $\mathbf x_p$ to $\mathbf {z}\in\mathbb C^{N_t\times k}$, which should obey the power constraint
\begin{equation}
\frac{1}{N_t k}\|\mathbf{ \mathbf{z}}\|_{F}^{2} \leq P_z.
\end{equation}
We define the \textit{channel bandwidth ratio (CBR)}  $ R\triangleq k/n$ as the number of channel uses per pixel \cite{Kurka_CRB}, where $ n=3hw $ is the number of source symbols. We typically have $k<n$, but the framework is general enough to cover any $k, n$ values. The above process is summarized as follows 
\begin{equation}
\mathbf{x}_o\xrightarrow{E_{\bm \xi}(\cdot)}\mathbf{x}\xrightarrow{S_{\bm e}(\cdot)}\mathbf{x}_e\xrightarrow{S_{\bm p}(\cdot)} \mathbf{{x}}_p\xrightarrow{P_{\bm \theta}(\cdot)} \mathbf{{z}},
\end{equation}
where the transmit signal $\mathbf z$ can be expressed as
\begin{equation}
\mathbf z = P_{\bm \theta}(S_{\bm p}( S_{\bm e}(E_{\bm \xi}(\mathbf{x}_o)))).
\vspace{-0.4cm}
\end{equation}

\subsection{Channel Model}
Let $\mathbf H\in \mathbb{C}^{N_r \times N_t}$ be the coefficients of the linear channel between the Tx and Rx, where $N_r$ denotes the number of antennas at the Rx. We assume a block-fading channel, where $\mathbf H$ remains constant for a block of $k$ consecutive symbols and changes to an independent realization in the next block \cite{Marzetta_block_fading}. The received signal $\mathbf y\in\mathbb C^{N_r\times k}$ at the Rx is given by
\begin{equation}
	\mathbf{y}=\mathbf{Hz+n},
	\label{MIMO}
\end{equation}
where $\mathbf{n}\in\mathbb C^{N_r\times k}$ denotes the zero-mean additive white Gaussian noise with variance $\sigma^2$.  The CSI is assumed to be known perfectly at both the Tx and Rx.

\begin{figure}[htpb]
	\centering
	\includegraphics[scale=0.42]{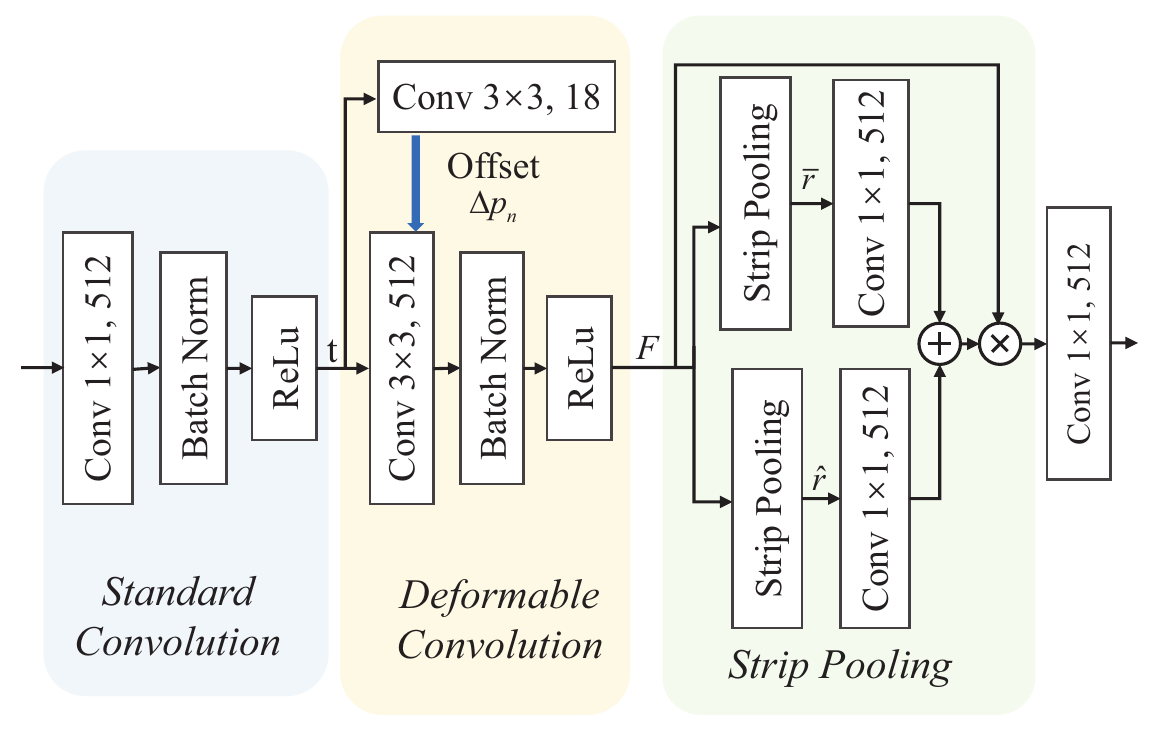}
	\includegraphics[scale=0.5]{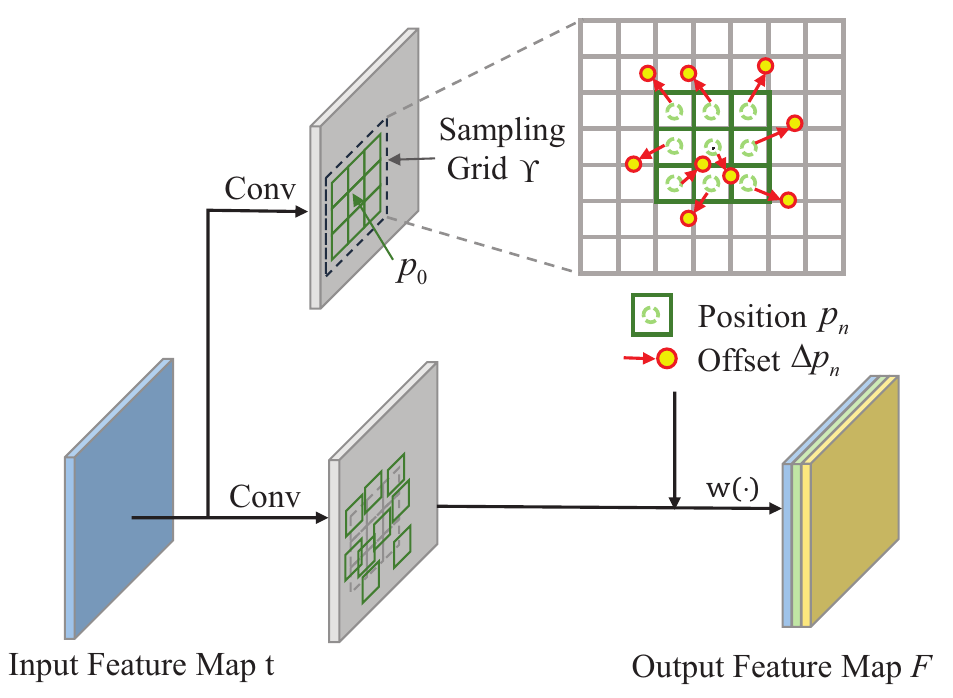}
	\caption{Illustration of the DAC semantic feature extraction block and the detailed process of deformable convolution.}
	\label{DAC}
\end{figure}

\subsection{Semantic-aware Receiver}
Rx aims both to reconstruct the input image and to facilitate downstream tasks based on the received signal $\mathbf y$. The inverse operations at the Rx also consist of three parts: CEN followed by standard receiver modules (which encompass MIMO combiner \cite{Albreem_MIMO_detection}, demodulator, channel decoder, and source decoder), and the downstream task module (e.g., image segmentation). Specifically, the received signal $\mathbf y$ is first fed to CEN, $Q_{\bm \eta}(\cdot): \mathbb C^{N_r\times k} \mapsto \mathbb C^{N_r\times k}$ with parameters $\bm \eta$, such that the processed signal $\hat{\mathbf x}_p$ is more compatible with existing standard decoding modules. Using a standard MIMO detection method (e.g., maximum likelihood (ML), matched filter (MF), zero-forcing (ZF), minimum mean square error (MMSE) etc. \cite{Albreem_MIMO_detection}), the Rx estimates the transmitted discrete signal vector, i.e., $\hat{\mathbf x}_e = S_{\bm p}^{-1}(\hat{\mathbf x}_p)$,  where $\hat{\mathbf x}_e \in \mathbb C^{L}$. Then the reconstructed image $\hat{\mathbf x}\in\mathbb R^{h\times w\times 3}$ is recovered via the function $ S_{\bm e}^{-1}(\cdot): \mathbb C^{L} \mapsto \mathbb{R}^{h\times w\times3}$, comprising demodulator, channel decoding, and source decoding modules. Finally, the reconstructed image $\mathbf{\hat x}$ can be fed into the downstream module $D_{\bm \varsigma}(\cdot): \mathbb{R}^{h\times w\times3} \mapsto \mathcal{O}$ to facilitate desired downstream tasks, where $\mathbf o = D_{\bm \varsigma}(\hat{\mathbf x})$. Here $D_{\bm \varsigma}(\cdot)$ can represent multiple tasks, e.g., segmentation, classification, etc.
The above process is summarized as follows
\begin{equation}
\mathbf{y}\xrightarrow{Q_{\bm \eta}(\cdot)}\hat{\mathbf{x}}_p\xrightarrow{S_{\bm p}^{-1}(\cdot)}\hat{\mathbf x}_e\xrightarrow{ S_{\bm e}^{-1}(\cdot) }\hat{\mathbf x}\xrightarrow{D_{\bm \varsigma}(\cdot)} \mathbf{o}.
\end{equation}
Overall, the output signal for the downstream task, $\mathbf o$, can be expressed as
\begin{equation}
\mathbf o = D_{\bm \varsigma}(S_{\bm e}^{-1}( S_{\bm p}^{-1}(Q_{\bm \eta}(\mathbf{y})))).
\end{equation}

\subsection{Proxy Network}
Since the parameters of the traditional codecs, modem, precoder, and MIMO detection are not differentiable, it is not possible to conduct backpropagation of weights through the conventional codecs during training. To overcome this, we introduce a learnable image processing and transmission network as a proxy for the traditional modules during the training stage. This allows the gradients of the proxy network to be propagated to PPEN and PCEN, facilitating the joint optimization of these modules. Specifically, at the Tx, we employ a learnable network denoted by $f_{\bm \phi}(\cdot): \mathbb{R}^{h\times w\times3} \mapsto \mathbb{C}^L$, parameterized by ${\bm \phi}$, which consists of source coding, channel coding, modulation, and precoding to efficiently approximate the true signal $\mathbf x_p$ as $\mathbf x_p^a = f_{\bm \phi}(\mathbf x)$. Likewise, at the Rx, another learnable network denoted as $g_{\bm \psi}(\cdot): \mathbb{C}^L \mapsto \mathbb{R}^{h\times w\times3}$, parameterized by ${\bm \psi}$, which represents the MIMO detection, demodulation, channel decoding, and source decoding operations to efficiently approximate the true reconstructed image $\hat{\mathbf x}$ as $\hat{\mathbf x}^a = g_{\bm \psi}(\hat{\mathbf x}_p)$. By introducing these learned networks as proxies, we replace the non-differentiable coding and modulation operations in standard codecs with differentiable counterparts and enable efficient end-to-end optimization.

\section{Detailed Transceiver Design}
In this section, we discuss the details and key methodologies of the proposed SCSC framework, which mainly include PPEN, PCEN, proxy network, and the procedure of training and deployment.

\subsection{Preprocessing-Empowered Network (PPEN)}
PPEN is proposed as a distortion-aware module, which is realized via deformable convolutions and non-linear transformations in our framework, to preserve critical semantic information for scene parsing and generation of the downstream task-friendly source image $\mathbf x$.
The proposed PPEN architecture is demonstrated in Fig. \ref{preprocessing}, where the original image $\mathbf{x}_o$ is input to two parallel branches. The upper branch uses 1$\times$1 convolutional layers and quantization adaptive layers to realize nonlinear pixel-level transformation. The lower branch is a distortion-aware compensation (DAC) module for calibrated semantic feature extraction \cite{Fangchang_encoder_decoder}. The outputs of the two branches are added together to form the filtered image $\mathbf{x}$, which preserves the useful texture and semantic information by conducting both shallow and deep transforms.
In the following, we will introduce the details of the two key modules, the DAC module, and the quantization adaptive layer, in detail.

\textbf{(1) DAC module.} The DAC module aims to enhance the modeling ability of the network for object deformation. As illustrated in Fig. \ref{DAC}, the input image initially passes through a standard convolutional layer to capture the overall information. Then, a deformable convolution module \cite{Jifeng_deformable} is introduced to enable flexible sampling with the consideration of object deformation and distortion information.  
After that, the strip pooling module \cite{Qibin_SPM} is adopted to effectively capture useful semantic context information from irregular regions for machine analysis tasks.  

\begin{figure*}[htpb]
	\centering
	\includegraphics[scale=0.8]{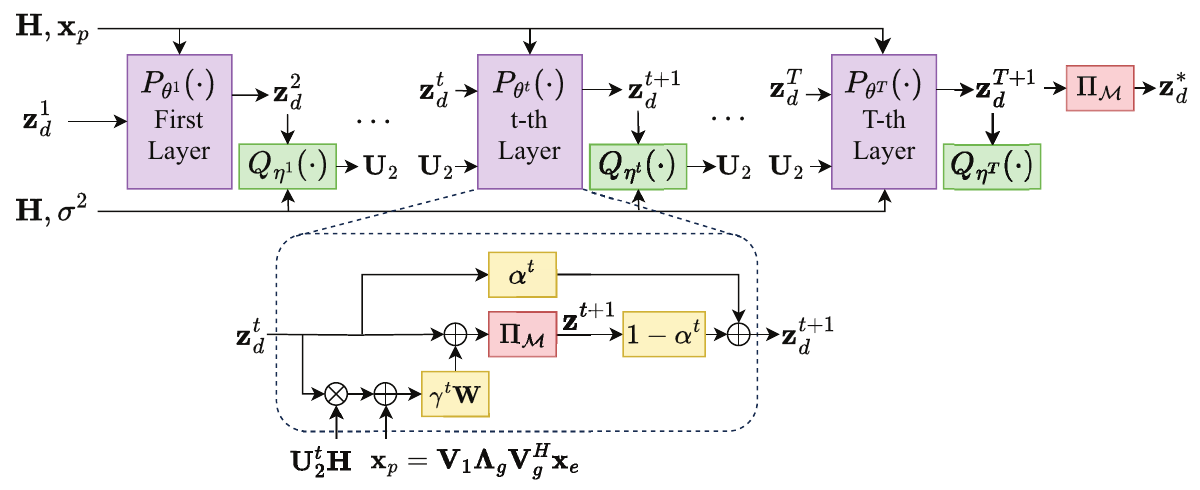}
	\caption{The diagram of the PEN and CEN models with parameters $\bm \theta$ and $\bm \eta$, respectively. The PEN consists of total $T$ iteration rounds and each layer has the same structure, which contains the linear estimator $\mathbf{W}$, nonlinear estimator $\Pi_{\mathcal{M}}$, trainable variables $\gamma^t$, and $\alpha^t$. $\mathbf z_d^*$  is the final output of PEN. The CEN is primarily implemented by the fully connected layer with trainable parameter $\bm \eta$. }
	\label{unfolding}
\end{figure*}

\textit{Deformable convolution} is adopted in our module to learn the distortion pattern and achieve better semantic preservation and protection. It allows for better adaptation to irregular target shapes, capturing localized features, and enhancement of the sensory field. Given a set of sampling locations on a regular grid $\Upsilon$, the input feature map $\mathbf t$, and the weights  $\mathbf w$ for the kernel, a convolutional layer is applied to produce the output feature map $F$ in convolution neural networks (CNNs). Deformable convolution can be defined as
\begin{equation} F\left(\boldsymbol{p}_{\mathbf{0}}\right)=\sum_{\boldsymbol{p}_{\boldsymbol{n}} \in \Upsilon} 
\mathbf w \left(\boldsymbol{p}_{\boldsymbol{n}}\right) \cdot \mathbf t\left(\boldsymbol{p}_{\mathbf{0}}+\boldsymbol{p}_{\boldsymbol{n}}+\Delta \boldsymbol{p}_{\boldsymbol{n}}\right),
\label{deformable}
\end{equation}
where $ \boldsymbol{p}_{\mathbf{0}}$ is a location on the output feature map $F$, $ \boldsymbol{p}_{\boldsymbol{n}} $ is a position on the regular sampling grid $\Upsilon$, and $\Delta \boldsymbol{p}_{\boldsymbol{n}}$ represents the offset corresponding to the position of $ \boldsymbol{p}_{\boldsymbol{n}}$.
In this way, we can obtain distortion-compensated features after standard convolution. 
Note that, deformable convolution can only improve the local feature representation capability. Meanwhile, global context is also important for semantic information extraction. 

\textit{Strip pooling} selectively extracts contextual information from feature maps along horizontal and vertical dimensions: 
\begin{equation}
\left\{\begin{aligned}
\bar {\mathbf{r}}_{i,c}&=\max _{0 \leq j<w'} \mathbf{t}_{i, j, c}, \\
\hat {\mathbf{r}}_{j,c}&=\max _{0 \leq i<h'} \mathbf{t}_{i, j, c},
\end{aligned}\right.
\end{equation}
where $\mathbf{t}_{i, j, c}$ is a location of the sampling grid on the input feature map, $h'$ and $w'$ represent the kernel size of strip pooling along horizontal and vertical dimensions, respectively. $ \bar {r}_{i,c}$ and $\hat {r}_{j, c}$ represent the $ i $-th and $ j $-th grids on the feature map of channel $ c $ for horizontal and vertical strip pooling, respectively. The two feature maps are then added and processed by element-wise multiplications to focus on texture-rich areas. 

Consequently, our DAC network can learn both local and global information about input images by using several DA modules and coarse-to-fine semantic information can be obtained through different stages.

\begin{algorithm}[t]
	\caption{PCEN.}
	\label{Alg: training}
	{\bf Input:} $\mathbf{x}_p=\mathbf{V}_1 \mathbf{\Lambda}_g \mathbf{V}_{\mathrm{g}}^{H} \mathbf x_e$, $\mathbf{H}$, noise variance $\sigma^2$, $\mathbf{z}_{\mathrm{d}}^{1}=0$, $\alpha=0.95$.\\
	{\bf Output:} \{$\bm \theta, \bm \eta$\}.
	\begin{algorithmic}[1]
		\STATE $\mathbf{W} =[\text{diag}(({\mathbf{U}_2 \mathbf H})^{H}{\mathbf{U}_2 \mathbf H})]^{-1}({\mathbf{U}_2 \mathbf H})^{H}$
		\FOR {$ t=0,1,2,...,T $}
		\STATE $\mathbf{r}^{t}=\mathbf{z}_\mathrm{d}^t+\gamma^t\mathbf{W}(\mathbf{x}_p-\mathbf{U}_2 \mathbf H\mathbf{z}_\mathrm{d}^t)$
		\STATE $\mathbf{z}^{t+1} =\Pi_\mathcal{M}(\mathbf{r}^t)$
		\STATE $\mathbf{z}_{\mathrm{d}}^{t+1}=\alpha^{t}\mathbf{z}_{\mathrm{d}}^{t}+(1-\alpha^{t})\mathbf{z}^{t+1}$
		\ENDFOR
		\STATE Update model parameters \{$\bm \theta, \bm \eta$\}.
	\end{algorithmic}
	\label{PEN}
\end{algorithm}  

\textbf{(2) Quantization Adaptive Layer}. As illustrated in Fig. \ref{preprocessing}, the incorporation of the quantization adaptive layer into the semantic preprocessing module facilitates the adaptive filtering process by scaling the intermediate features accordingly. Considering that the standard source compression, such as BPG, has different compression ratios, i.e., quantization parameters $\mathcal{Q} = \{ Q_0, Q_1, \ldots, Q_{q-1}\}$ with $q = |\mathcal{Q}|$, we introduce the quantization adaptive (QA) layer in the preprocessing module to generate the filtered image $\mathbf{x}$ for each specific compression ratio in practical applications. Specifically, a two-layer perceptron is employed to generate the scale vector, represented by $\mathbf{v}$. Then, given the input feature vector $\mathbf f_{i}$, we can obtain the output feature vector $\mathbf f_{o}= \mathbf f_{i} \odot \mathbf{v} $, which means the whole preprocessing module is able to adjust the intermediate features and ultimately produce a satisfactory filtered image $\mathbf{x}$ according to the given $Q$ values in the standard image codec.

\subsection{Precoder \& Combiner-Enhanced Network (PCEN)}
The goal of the proposed PCEN module (consisting of PEN and CEN) is to make the system more compatible with the discrete signal than the standard precoder, where we formulate the finite-alphabet precoding as an integer programming problem by unfolding the PCEN and adding several trainable parameters, so as to further increase the throughput in MIMO systems. One of the standard precoders $\mathbf{G} \in \mathbb{C}^{N_t \times N_s}$ admits the  SVD $\mathbf{G}=\mathbf{U}_g \mathbf{\Lambda}_g \mathbf{V}_{{g}}^{H}$, where the power allocation matrix $\mathbf{\Lambda}_g \in \mathbb{C}^{ N_t \times N_s}$ is diagonal and the diagonal entries represent the power allocation weights for each antenna signal stream. The matrix $\mathbf{\Lambda}_g$ can be obtained by applying the water-filling algorithm \cite{water-filling}. $\mathbf{U}_g \in \mathbb{C}^{ N_t \times N_t}$ and $\mathbf{V}_g \in \mathbb{C}^{ N_s \times N_s}$ are unitary. The standard precoder design obeys the following rule: i) if the channel $\mathbf{H}=\mathbf{U}_1 \mathbf{\Lambda}_1 \mathbf{V}_{1}^{H}$ is known by the transmitter, then from \cite[Prop. 2]{chengshan_precoder}, the optimal design satisfies $\mathbf{U}_g = \mathbf{V}_{1}$; ii) $\mathbf V_g$ is selected from the codebook specified in 3GPP  \cite{3GPP} to maximize the transmission performance. Then, the transmitted signal after the standard precoder is given by
\begin{equation}
\mathbf x_p = S_{\bm p}(\mathbf x_e)=\mathbf{V}_1 \mathbf{\Lambda}_g \mathbf{V}_{{g}}^{H} \mathbf x_e,
\end{equation}
where each entry of $\mathbf{x}_e$ is assumed to be drawn independently from the finite-alphabet $\mathcal{M}$.

In the proposed approach, we further process the output of the conventional precoder through the PEN model at the Tx, and the channel output is passed through the CEN module at the Rx before being fed to the channel combiner. To guarantee the output of PEN satisfies the input power constraint, we employ a power normalization layer.
We formulate the mean square error (MSE) between $\mathbf x_p$ and $\hat{\mathbf x}_p$ in Eq. \eqref{MSE} as the loss function when training PCEN. Different from DL-based approaches trained by using a huge volume of data, the proposed method unfolds a set of trainable parameters into a layer-wise structure, resulting in reduced model complexity and a smaller parameter count \cite{unfolding}. The loss function is specified as:
\begin{equation}
\begin{aligned}
\mathcal L_{pcen} 
&\triangleq \mathbb{E}_{\mathbf x_p,\mathbf n}\left\{\|\mathbf x_p-\hat{\mathbf x}_p\|^2\right\}\\
&=\mathbb{E}_{\mathbf x_p,\mathbf n}\left\{\|\mathbf x_p-Q_{\bm \eta}(\mathbf{H} P_{\bm \theta}(\mathbf x_p)+\mathbf n )\|^2\right\}.
\end{aligned}
\label{MSE}
\end{equation}
Note that if we take $P_{\bm \theta}(\mathbf x_p ) = \mathbf x_p$ and $Q_{\bm \eta}(\mathbf y) = \mathbf y$, we recover the standard precoder. To reduce the complexity at the Rx, we simplify the design of CEN by modelling it as a single fully connected layer, i.e., $Q_{\bm \eta}(\mathbf y) = \mathbf U_2 \mathbf y$. Then, Eq. \eqref{MSE} reduces to
\begin{equation}
\begin{aligned}
&\mathbb{E}_{\mathbf x_p,\mathbf n}\left\{\|\mathbf x_p-\hat{\mathbf x}_p\|_2^2\right\}\\
&=\mathbb{E}_{\mathbf x_p,\mathbf n}\left\{\|\mathbf x_p-\mathbf{U}_2\mathbf{(Hz+n)}\|^2\right\}\\
&=\mathbb{E}_{\mathbf x_p,\mathbf n}\left\{\|\mathbf x_p-\mathbf{U}_2\mathbf{Hz}\|^2\right\}+\|\mathbf{U}_2\|^2\sigma^2.
\label{MSE1}
\end{aligned}
\end{equation}
We employ alternating optimization of $\bm \theta$ and $\bm \eta$ to minimize the loss function, where at each iteration, we optimize one of the blocks while the other is fixed.

First, we consider the optimization of $\mathbf{U}_2$ for a given $\mathbf{z}$  in an iteration step:
\begin{equation}
\mathbf{U}_2^* = (\mathbb{E}_{\mathbf{x}_p}\{\|\mathbf{H}\mathbf{z}\mathbf{z}^H\mathbf{H}^H\|\}+\sigma^2\mathbf I)^{-1} \mathbb{E}_{\mathbf{x}_p}\{\mathbf{H}\mathbf{z}\mathbf{x}_p^{H}\},
\label{prop1}
\end{equation}
which is obtained from the first-order optimality condition of the unconstrained convex problem in \eqref{MSE1}.

\emph{Remark 1:} Although the optimal $\mathbf U_2$ can be obtained using \eqref{prop1}, the distribution of $\mathbf x_p$ is unknown in a practical system. Hence, we model it as a fully connected layer $\bm \eta$ and use stochastic gradient descent to approximate the optimal value.

Next, we turn to the optimization of $\mathbf{z}$ for a given $\mathbf{U}_2$, which can be formulated as follows:
\begin{equation}
\begin{array}{ll}
\min\limits_{\mathbf {z}}
&\mathbb{E}_{\mathbf {x}_p}\{\|\mathbf{x}_p - \mathbf{U}_2 \mathbf H \mathbf {z}\|_2^2\}.
\end{array}
\label{MSE2}
\end{equation}
The common alternating direction method of multipliers (ADMM) framework enables us to rewrite problem \eqref{MSE2} in a consensus form as follows:
\begin{equation}
\begin{array}{ll}
\min\limits_{\mathbf z, \mathbf z_1}
&\|\mathbf{x}_p - \mathbf{U}_2 \mathbf H \mathbf {z}_1\|_2^2+ I_\mathcal{M}(\mathbf{z}) \\\mathrm{s.t.}
&\mathbf{z}_1 -\mathbf{z}=0,
\end{array}
\label{ADMM1}
\end{equation}
where $I_{\mathcal{M}}(\mathbf{z})$ is the indicator function of $\mathcal{M}$, i.e.,
\begin{equation}
I_{\mathcal{M}}(\mathbf{z})=\left\{
\begin{array}{ll}
0,&\text{if }\mathbf{z}\in\mathcal{M}^{N_t\times k},\\
\infty,&\text{otherwise.}
\end{array}\right.
\end{equation}
The augmented Lagrangian of \eqref{ADMM1} is expressed as 
\begin{equation}
\begin{aligned}
&L_r\left(\mathbf{z}_1,\mathbf{z},\mathbf{u}\right)=\|\mathbf{x}_p - \mathbf{U}_2 \mathbf H \mathbf {z}_1\|_2^2+I_{\mathcal{M}}(\mathbf{z})\\
&\quad\quad\quad+\mathbf{u}^H(\mathbf{z}_1-\mathbf{z})+\gamma\left\|\mathbf{z}_1-\mathbf{z}\right\|_2^2,
\end{aligned}  
\end{equation}
where $\mathbf{u}$ is the dual variable corresponding to the constraint $\mathbf{z}_1-\mathbf{z}=0$, and $\gamma$ is the non-negative penalty parameter. The problem can be efficiently solved by algorithm 2 in  \cite{damping}, and we can obtain the following iterative process:

\begin{equation}
\begin{aligned}
\mathbf{W}& =[\text{diag}(({\mathbf{U}_2 \mathbf H})^{H}{\mathbf{U}_2 \mathbf H})]^{-1}({\mathbf{U}_2 \mathbf H})^{H},  \\
\mathbf{r}^{t}& =\mathbf{z}_\mathrm{d}^t+\gamma^t\mathbf{W}(\mathbf{x}_p-\mathbf{U}_2 \mathbf H\mathbf{z}_\mathrm{d}^t),  \\
\mathbf{z}^{t+1}& =\Pi_\mathcal{M}(\mathbf{r}^t),  \\
\mathbf{z}_{\mathrm{d}}^{t+1}& =\alpha^{t}\mathbf{z}_{\mathrm{d}}^{t}+(1-\alpha^{t})\mathbf{z}^{t+1}, 
\end{aligned}
\end{equation}
where $\bm \theta = \{\gamma^t, \alpha^{t}\}$ are the learnable parameters in PEN. The aforementioned equations are executed iteratively until convergence. 

The above procedure of PCEN in our SCSC framework is formally summarized in Fig. \ref {unfolding} and Alg. \ref{Alg: training}. Thanks to its low complexity and flexibility, PCEN holds a promising potential to be deployed in practical applications  \cite{finite}. Unlike conventional designs, which often rely on Gaussian distribution assumptions, thanks to its data-driven nature, PCEN can adapt to arbitrary distributions.

\begin{figure}[t]
	\centering
	\includegraphics[scale=0.72]{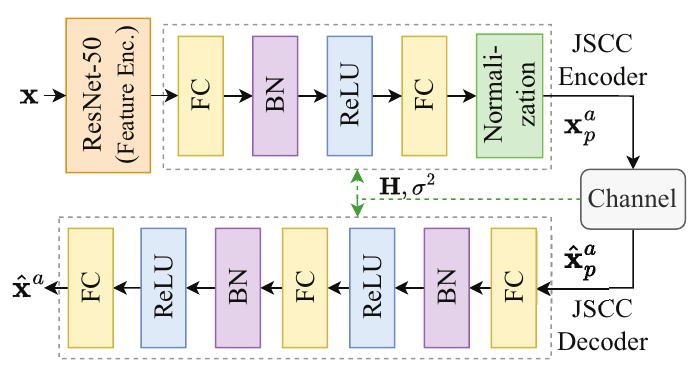}
	\caption{The structure of the proxy network.}
	\label{proxy_network}
\end{figure}

\subsection{Proxy Network}
In this subsection, we introduce a neural network as a proxy to mimic the non-differentiable operations of the standard source codec, channel codec, modulation, and precoding, which will allow SCSC to be trained in an end-to-end fashion. Here, we use the JSCC AE approach in \cite{Jankowski} as our proxy network.
To make sure that the proxy network can well approximate the standard codec, the reconstruction quality of the digital scheme $\hat{\mathbf{x}}$ and the proxy approach $\hat{\mathbf{x}}^a$ should be similar. If the discrepancy is too significant, the proxy network cannot substitute for the conventional codecs, indicating a failure in training. The proxy network knows the channel realization $\mathbf{H}$.
As illustrated in Fig. \ref{proxy_network}, the proxy network consists of two symmetrical structures: i) the encoder transforms the input image data $\mathbf{x}$ into the preliminary processed transmit signal, and a JSCC encoder is utilized for anti-noise and efficient wireless transmission; ii) the decoder executes an approximate inverse function to map the preliminary processed received signal back to the pixel-level image.

 \begin{algorithm}[tpb]
	\caption{End-to-End SCSC Training Procedure}
	\label{Alg: overall}
	{\bf Input:} Training dataset, trade-off parameters $\lambda_1$, $\lambda_2 $, $\lambda_3$, learning rate $\alpha$.\\
	{\bf Output:} The trained SCSC model.\\
	\textbf{Training of Proxy Network:} Based on the fine-tuning procedure $\mathcal{L}_{pro}= d(\mathbf{\hat{x}}, \mathbf{\hat{x}}^a)$, we can obtain several proxy networks that mimic the standard codec with different settings.\\
	\textbf{End-to-End Training}
	\begin{algorithmic}[1]
		\FOR {each epoch $k=1,2, \cdot \cdot \cdot K$} 
		\FOR {all $\mathbf{x}_o$}
		\STATE PPEN: $E_{\bm \xi}(\mathbf x_o)\rightarrow \mathbf{x}$;
		\STATE Standard codec: $\hat{\mathbf x}$, $R_t$;
		\STATE Downstream task: $ D_{\bm \varsigma}(\hat{\mathbf x}) \rightarrow \mathbf{o} $; 
		\STATE Compute the loss $\mathcal{L}$ by (\ref{loss}): \\$\mathcal{L}= \lambda_1 \mathcal{L}_{pcen} +\lambda_2 \mathcal{L}_{seg}+\lambda_3 \mathcal{L}_{pre}$;
		\STATE Proxy codec: $\hat{\mathbf x}^a$, $R_p$;
		\STATE Gradients computation:\\
		\setstretch{1.6}
		$\displaystyle \frac{\partial{\mathcal{L}}}{\partial{\mathbf{o}}} \cdot \frac{\partial{\mathbf{o}}}{\partial{\hat{\mathbf x}^a}} \rightarrow g_{task} $ ;\\
		\setstretch{2}
		$\displaystyle  g_{task} \cdot \frac{\partial{\hat{\mathbf{x}}^a}}{\partial{\mathbf{x}}}  + \frac{\partial{\mathcal{L}}}{\partial{R_p}} \cdot \frac{\partial{R_p}}{\partial{\mathbf{x}}} \rightarrow g_{pro} $;\\
		$\displaystyle  g_{pro} \cdot \frac{\partial{\hat{\mathbf x}}}{\partial{\mathbf x}} + \frac{\partial{\lambda_2 \mathcal{L}_{seg}}}{\partial \hat{\mathbf x}_p} \cdot \frac{\partial{\hat{\mathbf x}_p}}{\partial{{\mathbf x}_p}} \rightarrow g_{pcen} $;\\
		$\displaystyle g_{pcen} \cdot \frac{\partial{\mathbf{x}}}{\mathbf{x}_o} + g_{pro} \cdot \frac{\partial{\mathbf{x}}}{\mathbf{x}_o} + \frac{\partial{\lambda_2 \mathcal{L}_{seg}}}{\mathbf{x}} \cdot \frac{\partial{\mathbf{x}}}{\mathbf{x}_o}  \rightarrow g_{ppen} $;\\
		\setstretch{1.6}
		$\displaystyle \bm \xi- \alpha \cdot g_{ppen} \rightarrow \bm \xi $.
		\ENDFOR
		\setstretch{1.4}
		\ENDFOR
	\end{algorithmic}
\end{algorithm}

\subsection{Training and Deployment}
Our goal is to obtain suitable PPEN and PCEN parameters to enhance task-oriented communication under the standard transmission protocols. To realize it, we first choose a pre-trained JSCC model with suitable parameters to approximate the digital scheme, then finetune the proxy network by using the MSE loss 
\begin{equation}
	\begin{aligned}
		\mathcal{L}_{pro} = d(\hat{\mathbf{x}}, \hat{\mathbf{x}}^a)  = \frac{1}{h\times w\times 3 } \| \mathbf{\hat{x}} - \mathbf{\hat{x}}^a \|_{2}^{2},
	\end{aligned}
	\label{proxy_loss}
\end{equation}
where $ d(\hat{\mathbf{x}}, \hat{\mathbf{x}}^a) $ represents the distortion between the reconstructed image $\mathbf{\hat{x}}$ and the output $ \mathbf{\hat{x}}^a$ from the proxy network.
We train the proxy network for different SNRs and CBRs to minimize the distortion, so that we can obtain the proxy network to mimic the traditional codec. We assume that CSI is known perfectly to both the encoder and decoder. Once the proxy network is trained, it replaces the standard codec during the backward propagation stage of SCSC.

Next, we provide more details on the end-to-end training of SCSC in Algorithm \ref{Alg: overall}. After executing the downstream task, we can calculate the total loss function $\mathcal{L}$ as
\begin{equation}
\mathcal{L}=   \lambda_1 \mathcal{L}_{pcen}  + \lambda_2 \mathcal{L}_{seg}+ \lambda_3 \mathcal{L}_{pre},
\label{loss}
\end{equation}
where $\lambda_1$, $\lambda_2$ and $\lambda_3$ are the trade-off parameters among different loss terms. $\mathcal{L}_{pcen}$ is defined in Eq. \eqref{MSE}, and $\mathcal{L}_{seg}$ denotes the loss from the specific downstream task. In addition, to stabilize the training process and evaluate the reconstruction performance, we consider the MSE loss between the original image $\mathbf{x}_o$ and the reconstructed image $\hat{\mathbf{x}}$, denoted by $\mathcal{L}_{pre}$.

At the same time, we obtain the corresponding reconstructed image $\hat{\mathbf{x}}^a$ and CBR $R_p$ from the proxy network. Based on this operation, we can use the CBR and reconstructed image from the standard codec in forward propagation and calculate the value of loss function while using the gradients of the proxy network in backward propagation to optimize the weights of PPEN and PCEN. The backpropagation procedure involves sequentially computing the gradients of the downstream task model, PCEN, proxy network, and PPEN, denoted by $ g_ {task} $, $ g_{pro} $, $ g_{pcen} $, and $ g_{ppen} $, respectively. The weights of PPEN and PCEN are optimized while the weights of the other modules remain fixed.

For the practical deployment of SCSC, we directly implement the trained PPEN and PCEN at the base station, where the remaining standard codecs and wireless transmission are specified by the current protocol.

\section{Numerical Results}
In this section, numerical results are provided to validate the effectiveness of the proposed SCSC framework. We first describe the simulation setup, and then demonstrate the superior performance in the performance of the downstream machine vision tasks by comparing the proposed framework with several recent state-of-the-art schemes. Finally, several ablation studies are carried out to motivate our architectural choices, and the computational complexity of the proposed approach is investigated.

\subsection{Experimental Setups}
\label{setups}
\subsubsection{Datasets}
In our simulations, a $2\times 2$ MIMO system is considered for the physical layer transmission. For the transmitted images, we consider the RGB image datasets  Cityscapes \cite{Cityscapes} and CVRG-Pano \cite{CVRG}. Cityscapes is a dataset of street scene images captured under similar weather and lighting conditions in 50 different locations. It consists of 2975 training, 500 validation, and 1525 test images with fine panoptic annotations. The high-resolution CVRG-Pano dataset comprises 600 pixel-level annotated panoramic images with size $1664 \times 832$ pixels. It covers a wide range of subjects and contains 20 semantic classes grouped into 7 categories. Data augmentation is adopted to increase sample diversity, and improve model generalization ability and robustness, which encompass random horizontal flips, vertical flips, and random cropping to dimensions of $256 \times 256$.

\subsubsection{Downstream Task Module}
We consider semantic segmentation as the task to be completed at the receiver and leverage ERF-PSPNet \cite{ERF-PSPNet} to implement it. The semantic loss for semantic segmentation is evaluated by the mean intersection over union (mIoU) defined as
\begin{equation}
	\mathcal{L}_{seg} \triangleq 1-\frac{1}{B} \sum_{b=1}^{B} \frac{\mathbf{o}_{g}^{b} \cap \mathbf{o}_{d}^{b}}{\mathbf{o}_{g}^{b} \cup \mathbf{o}_{d}^{b}},
\end{equation} 
where $\mathbf{o}_{d}^{b}$ and $\mathbf {o}_{g}^{b}$ denote the semantic segmentation results of the reconstructed image $\hat{\mathbf{x}}$ and original image $\mathbf{x}_o$, respectively. 
$B$ denotes the total number of classes in the original dataset and $b$ represents a specific class.

\begin{figure*}[tpb]%
	\centering
	\subfloat[PSNR versus SNR]{
		\label{PSNR_SNR}
		\includegraphics[width=0.46\linewidth]{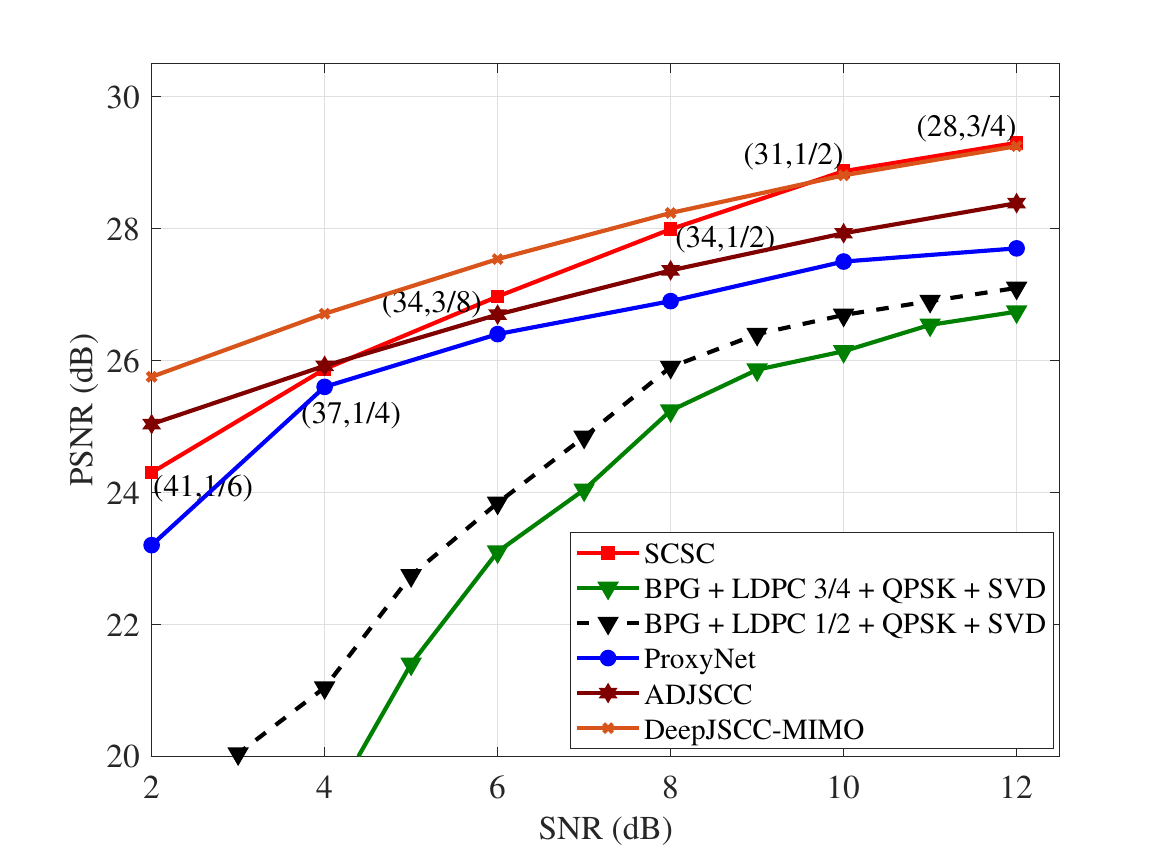}
	}\hfill
	\subfloat[PSNR versus CBR]{
		\label{PSNR_CBR}
		\includegraphics[width=0.46\linewidth]{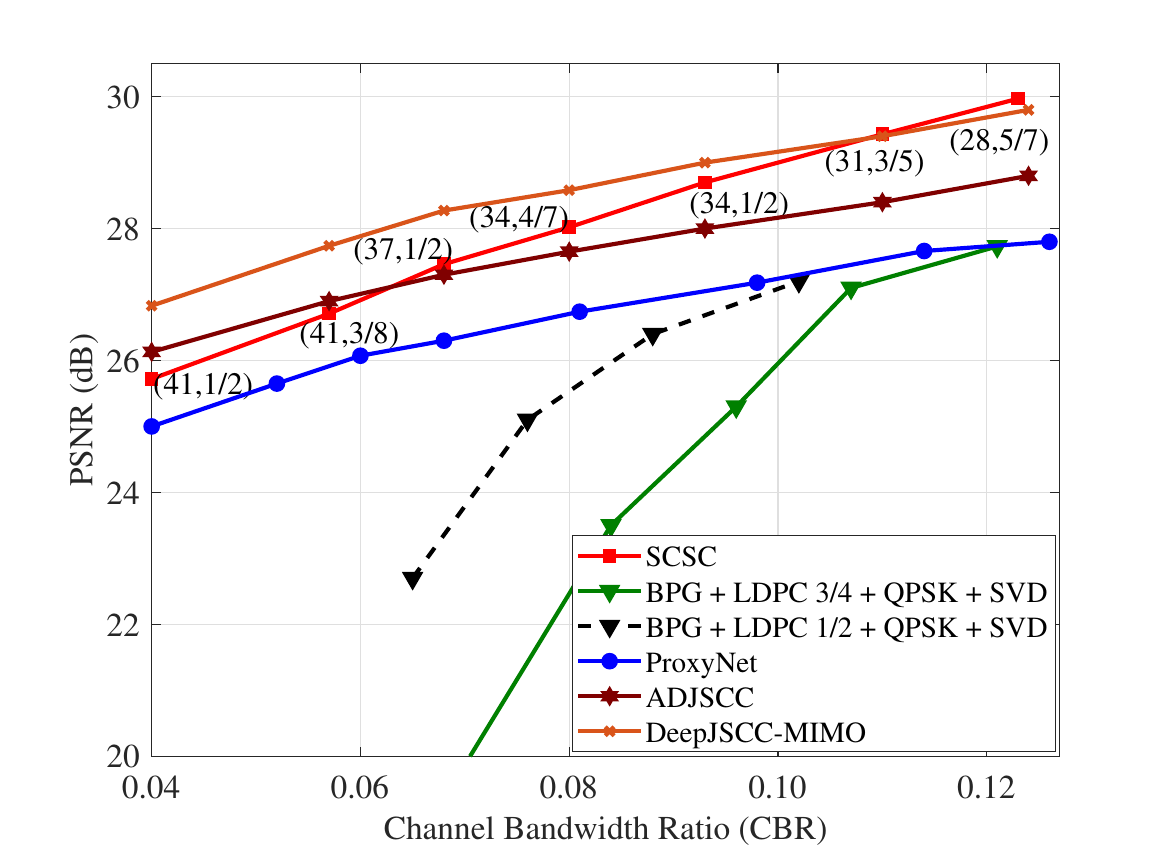}
	}
	\caption{Comparison of the PSNR metric for our proposed SCSC framework and the other baselines over a $2 \times 2$ MIMO fading channel with CBR $R = 1/12$ and SNR = 6 dB, respectively.}
	\label{SNR}
\end{figure*}

\subsubsection{Metrics}
We quantify the image transmission performance with the following three metrics: pixel-wise metric PSNR, perceptual metric multi-scale structural similarity index (MS-SSIM), and semantic segmentation metric mIoU. Overall, higher PSNR/MS-SSIM/mIoU indicates better performance.

\subsubsection{Model Deployment Details}
We train and evaluate the whole SCSC model on eight RTX 4090 GPU cards. The optimizer is AdamW \cite{Adam} and the learning rate at epoch $n$, based on the polynomial function, is denoted as ${\beta^{n}}=\beta^0\left(1-\frac{n}{K}\right)^{p}$, where the initial learning rate is $\beta^0=5\times 10^{-4}$, the power of the polynomial is $p=0.9$, and $K$ denotes the number of epochs that the scheduler decays the learning rate. 
To compress the source image according to different requirements, we consider BPG at different compression ratios, i.e., $\mathcal Q = \{28, 31, 34, 37, 41\}$, where lower $\mathcal Q$ value means less compression and better image quality. The weights of the loss function $\mathcal L$ in \eqref{loss} are set as $\lambda_1=\lambda_3=0.1$, $\lambda_2 = 0.5$. For the SCSC framework, we sample the channel SNR uniformly from the range of [2, 12] dB, combined with different selected $Q$ values in each training iteration. The average input power $P_z$ is w.l.o.g. restricted to 1. The weights of the downstream network ERF-PSPNet are fixed during both training and testing. For the channel model, unless specifically indicated, we consider a Rayleigh fading channel. To simulate practical 5G MIMO transmission scenarios, we also consider a general non-stationary 5G wireless channel model with MIMO CSI matrices generated according to \cite{MIMOCSI}. We utilize 1000 MIMO channel matrix samples for training during image transmission and 200 samples for testing to guarantee robust performance under realistic channel conditions.

\subsubsection{Comparison Schemes}

We compare the proposed scheme with the following baselines.
\begin{itemize}
	
	\item {SCSC.} For the proposed scheme, we consider all possible combinations of image compression parameters in $\mathcal{Q}$ with LDPC, QPSK, and SVD, and the best-performing digital transmission scheme is selected. Then, the chosen combination is regarded as the training baseline for SCSC. The annotations in the following figures denote the selected combination of the image compression parameter and LDPC rate, e.g., (41,1/6). The smaller $Q$, the better image quality. Moreover, for the test, we utilize the optimal parameters for the standard transmission. 

	\item {Digital transmission scheme.} The baseline employs BPG for source coding, 1/2 and 3/4 rate LDPC for channel coding, QPSK for modulation, and SVD for precoding. 
   
	\item {ProxyNet scheme.} The CNN-based ProxyNet scheme represents the proposed PPEN and PCEN modules only integrated with the proxy network.

    \item {ADJSCC.} We adopt the DL-based JSCC (ADJSCC) scheme proposed in \cite{ADJSCC} as a baseline, where the MIMO CSI and SNR are provided to the joint source-channel encoder and decoder. The ADJSCC model outperforms the basic JSCC model across diverse SNR and bandwidth ratios, highlighting its adaptability to various channel conditions. 
    
    \item {DeepJSCC-MIMO.} The vision transformer-based Deep JSCC scheme \cite{Wu_2024} for wireless image transmission over MIMO channels applies SVD to the channel matrix and uses the gains of the obtained parallel channels instead of the channel matrix.
	
\end{itemize}

\begin{figure*}[htpb]
	\centering
	\subfloat[MS-SSIM versus SNR]{
		\label{SSIM_SNR}
		\includegraphics[width=0.46\linewidth]{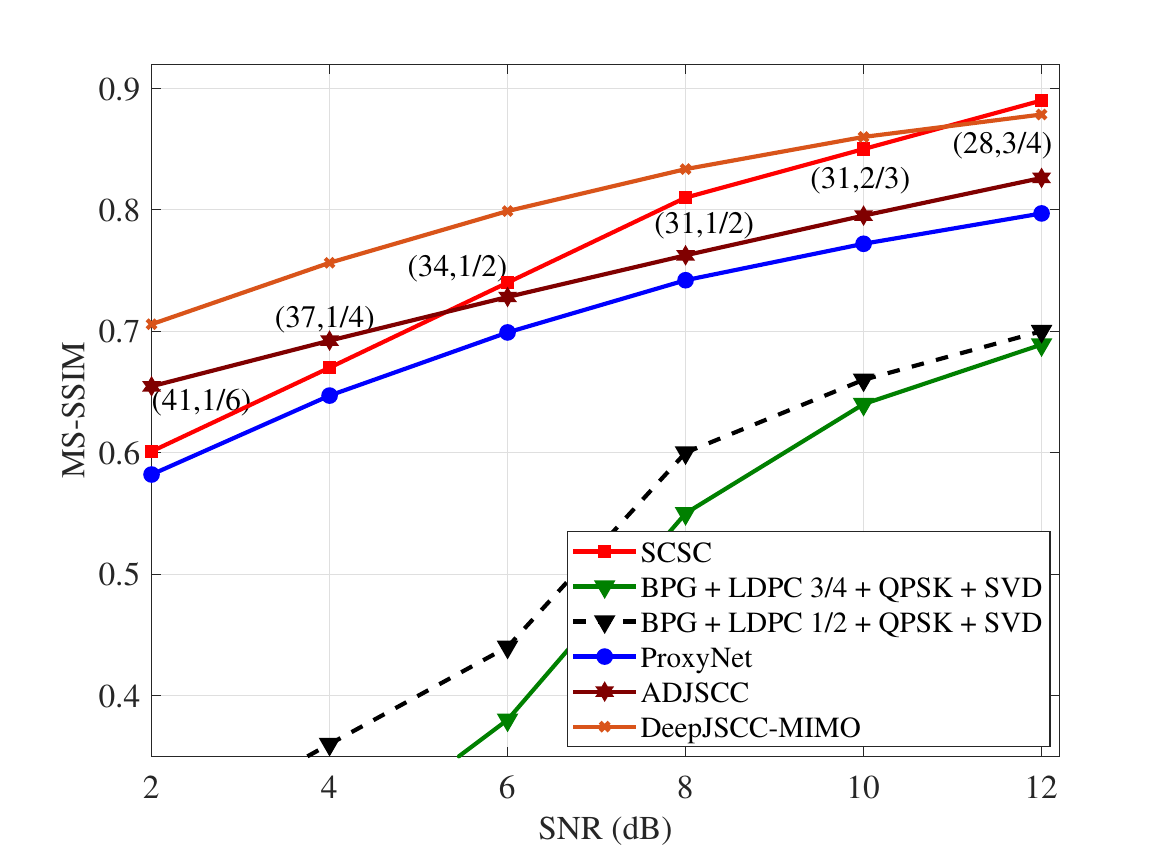}
	}\hfill
	\subfloat[MS-SSIM versus CBR]{
		\label{SSIM_CBR}
		\includegraphics[width=0.46\linewidth]{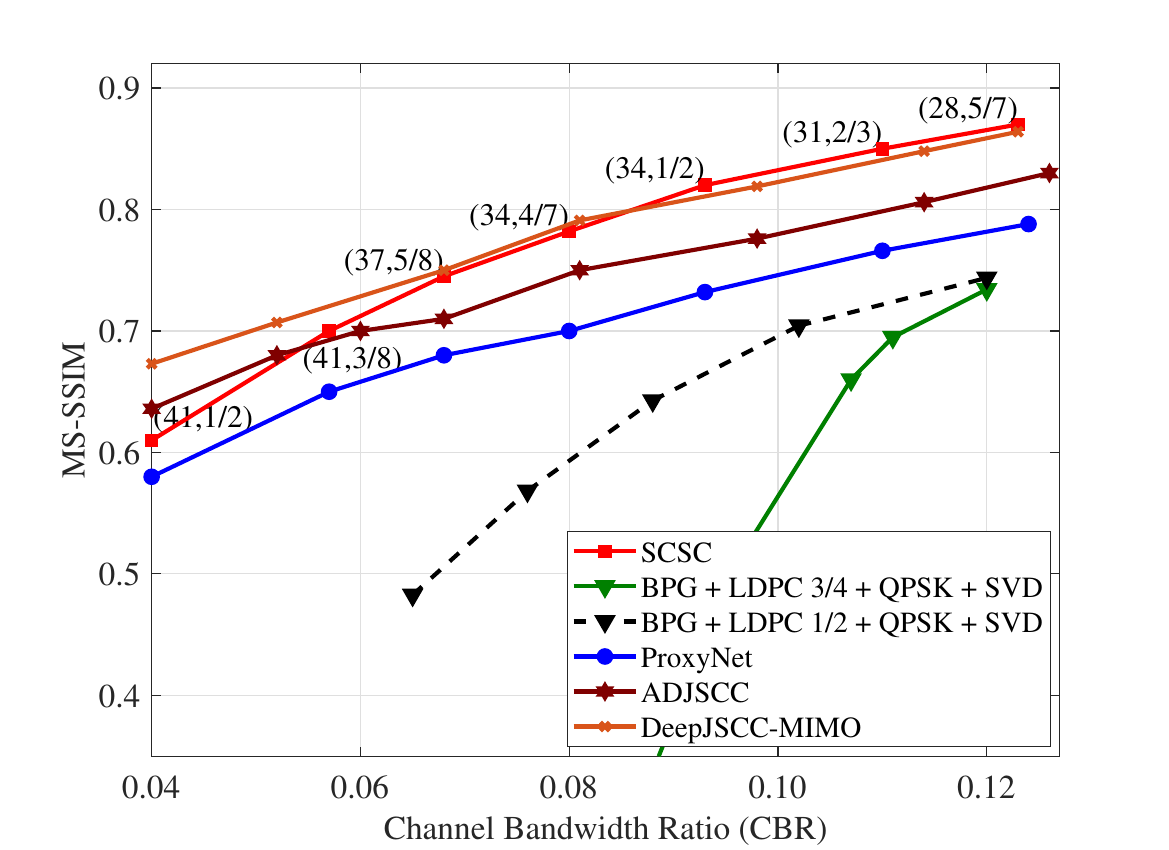}
	}
	\caption{Comparison of MS-SSIM performance in a $2 \times 2$ MIMO fading channel; (a) shows the MS-SSIM performance versus channel SNR, where the average CBR is set to $R = 1/12$. (b) shows the MS-SSIM performance versus CBR over the fading channel at SNR = 6 dB.}
	\label{SSIM}
\end{figure*}

\begin{figure*}[htpb]%
	\centering
	\subfloat[mIoU versus SNR]{
		\label{mIOU_SNR}
		\includegraphics[width=0.45\linewidth]{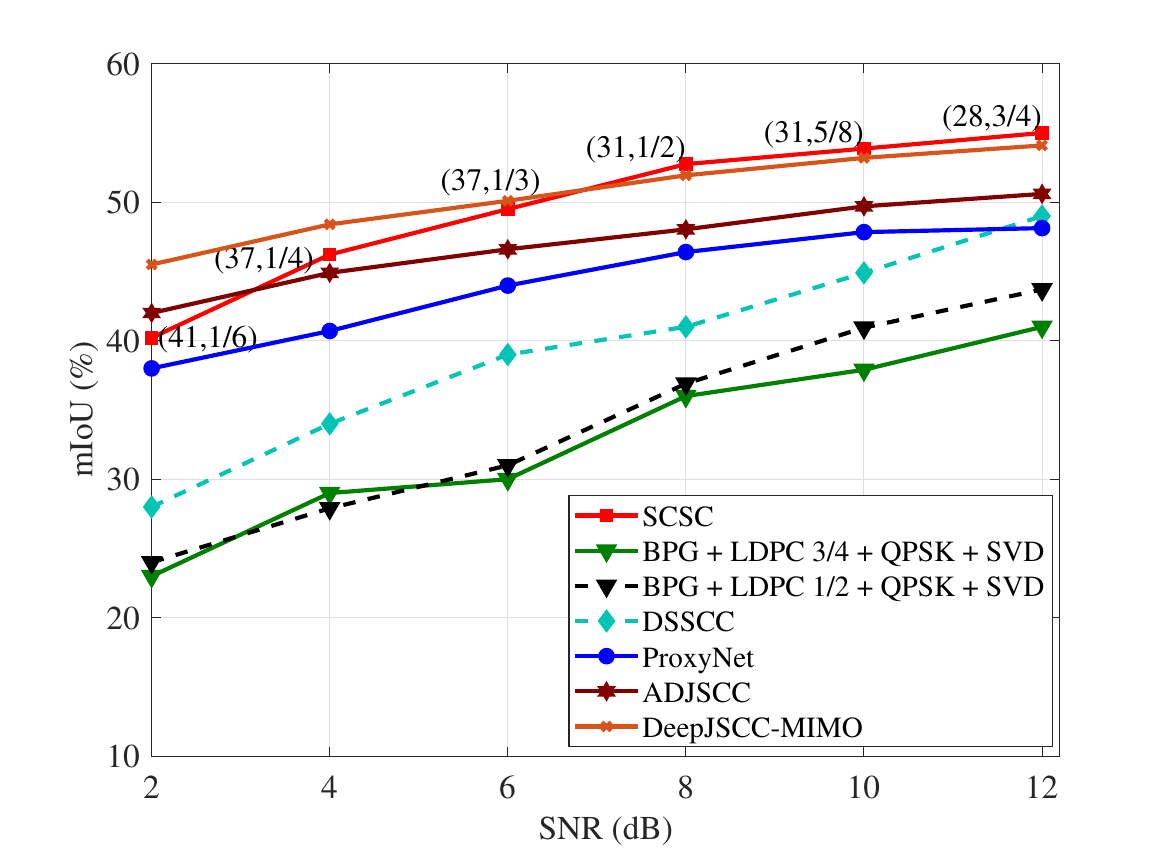}
	}\hfill
	\subfloat[mIoU versus CBR]{
		\label{mIOU_CBR}
		\includegraphics[width=0.45\linewidth]{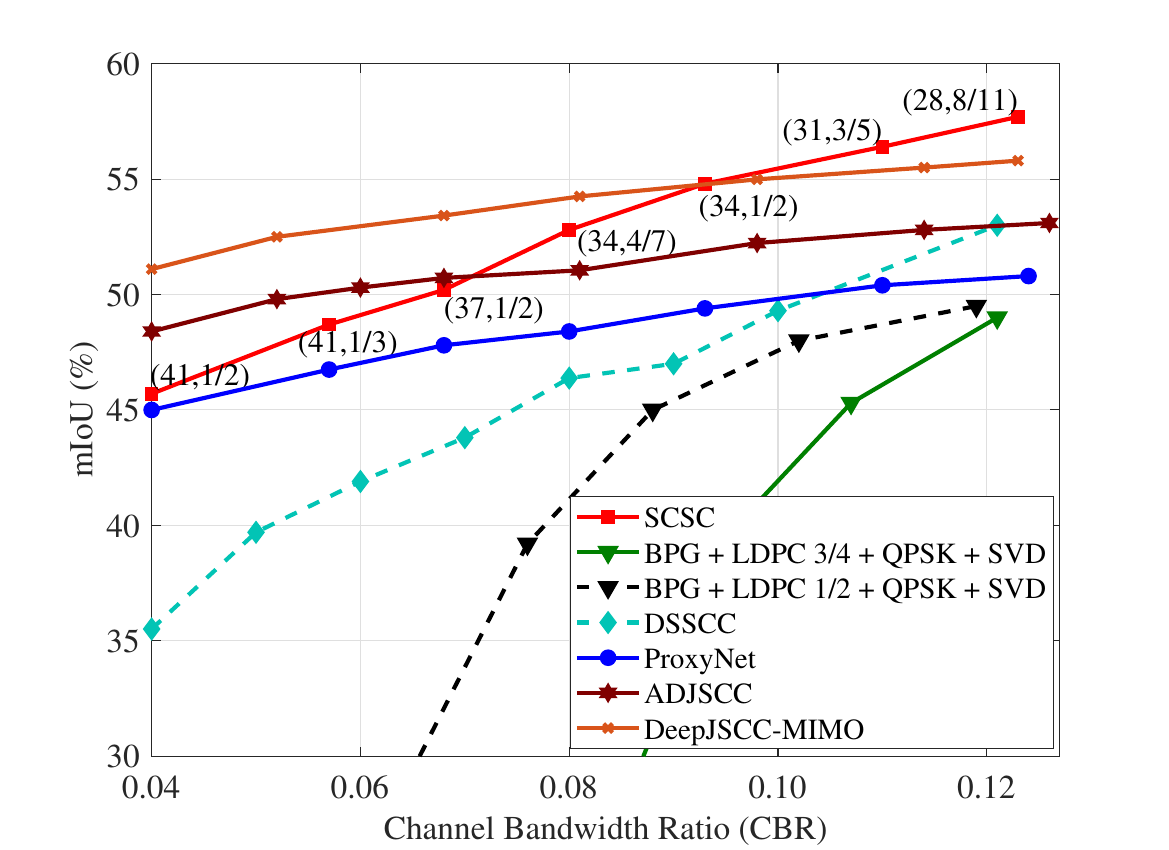}
	}
	\caption{Performance of semantic segmentation on CVRG-Pano dataset over a $2 \times 2$ MIMO fading channel; (a) mIoU performance against SNR with $R$ = 1/12; (b) mIoU performance against CBR with SNR = 6 dB.}
	\label{mIOU}
\end{figure*}

\subsection{Experimental Results}
\subsubsection{PSNR Performance}

In Fig. \ref{SNR}\subref {PSNR_SNR}, we investigate the PSNR performance of the proposed scheme and other baselines as a function of the test channel SNR over MIMO fading channels while maintaining a fixed CBR of $R = 1/12$.
Higher SNR values imply better channel conditions and more reliable wireless transmission. We can observe that our SCSC can generally outperform the traditional digital schemes in all SNR values and exhibit a significant performance improvement of around 1.6 dB as SNR increases when compared to ADJSCC. This gap arises from the unique strengths of the digital-based SCSC, which combines the benefits of semantic communication and digital communication.
In Fig. \ref{SNR}\subref {PSNR_CBR}, we plot the PSNR versus CBR result for the compared schemes at SNR = 6 dB. It is observed that our proposed scheme outperforms ADJSCC, ProxyNet, as well as the separation-based baseline over a large range of CBR values, and as the SNR increases, it gradually approaches and slightly surpasses the DeepJSCC-MIMO scheme. Taking the case $R = 0.04$ as an example, when 96\% transmission overhead is saved, the SCSC still has the ability to reconstruct relatively high-quality images. Overall, our framework exhibits strong adaptability to varying channel conditions with different SNRs and CBRs, resulting in comparable or superior performance to the other baselines.

\subsubsection{MS-SSIM Performance}

Fig. \ref{SSIM}\subref{SSIM_SNR} indicates that in terms of the MS-SSIM performance, SCSC achieves better performance for all SNR regions compared with digital baselines under the same conditions. Conventional schemes are inferior to the DL-based counterparts in the low SNR regime, because BPG compression is designed mainly for PSNR, and does not consider perceptual quality. We can note that, compared with ADJSCC, at the same level of MS-SSIM performance, our SCSC is able to save the CBR by approximately 33\%. In Fig. \ref{SSIM}\subref{SSIM_CBR}, SCSC scheme achieves relatively satisfying performances as DeepJSCC-MIMO, ADJSCC, and ProxyNet schemes, compared to the digital scheme in low CBR region ($R \leq 0.08$). This outcome is attributed to the fact that noise exerts a more substantial impact on communication performance at low SNR, and DL-based schemes demonstrate greater resilience to noise than traditional SSCC.

\begin{figure*}[htpb]
	\centering
	\captionsetup[subfloat]{labelsep=none,format=plain,labelformat=empty}
	
	\subfloat[ \small{$R$ / PSNR (dB)}]{
		\begin{minipage}{0.24\linewidth}
			\includegraphics[width=\textwidth]{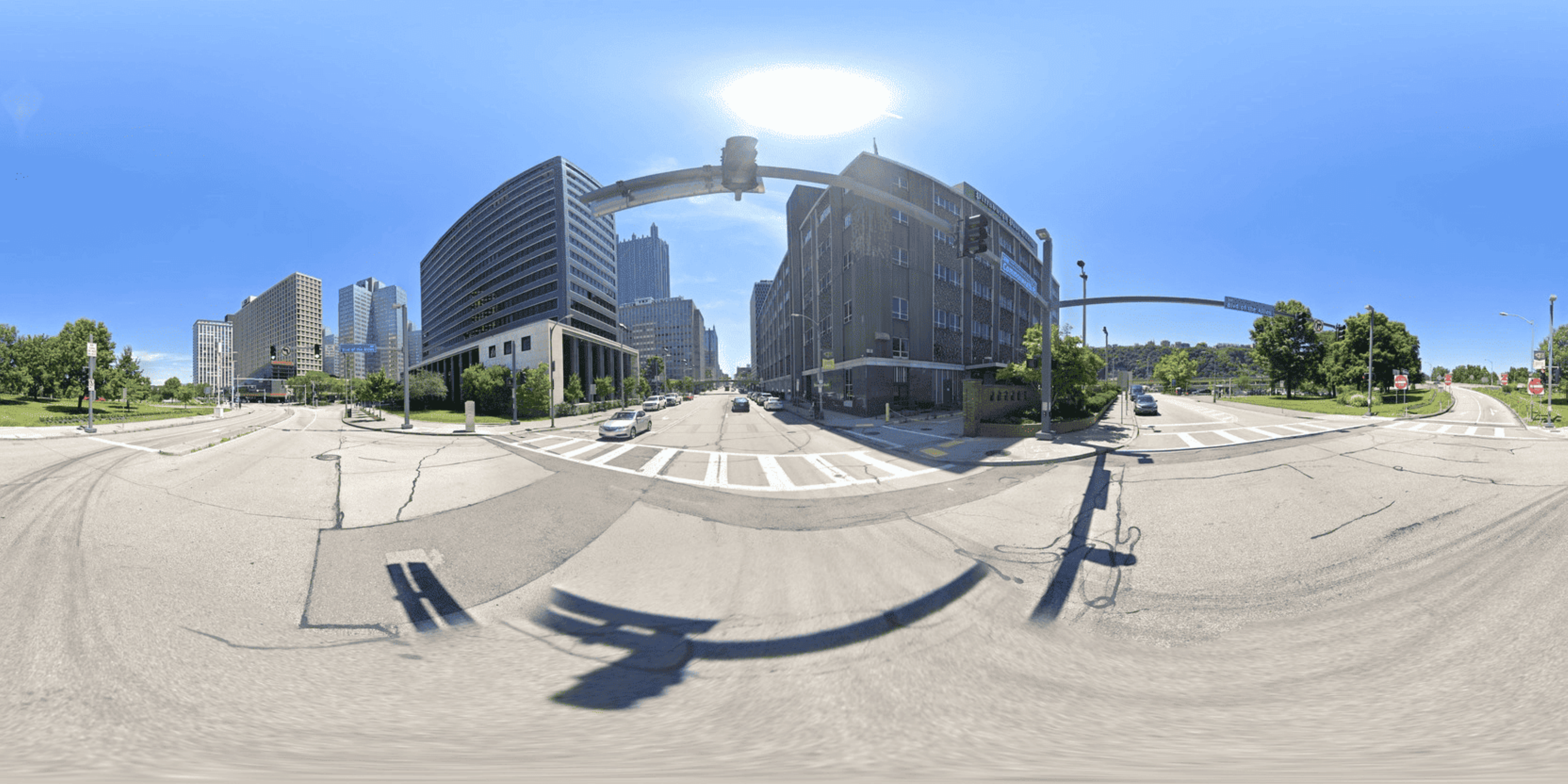}
			\vspace*{-21ex} 
			\begin{center}
				\text{Ground Truth}
			\end{center}
			\vspace*{13ex}
		\end{minipage}%
	}\hfill
	\subfloat[\small{0.0253 / 18.72}]{
		\begin{minipage}{0.24\linewidth}
			\includegraphics[width=\textwidth]{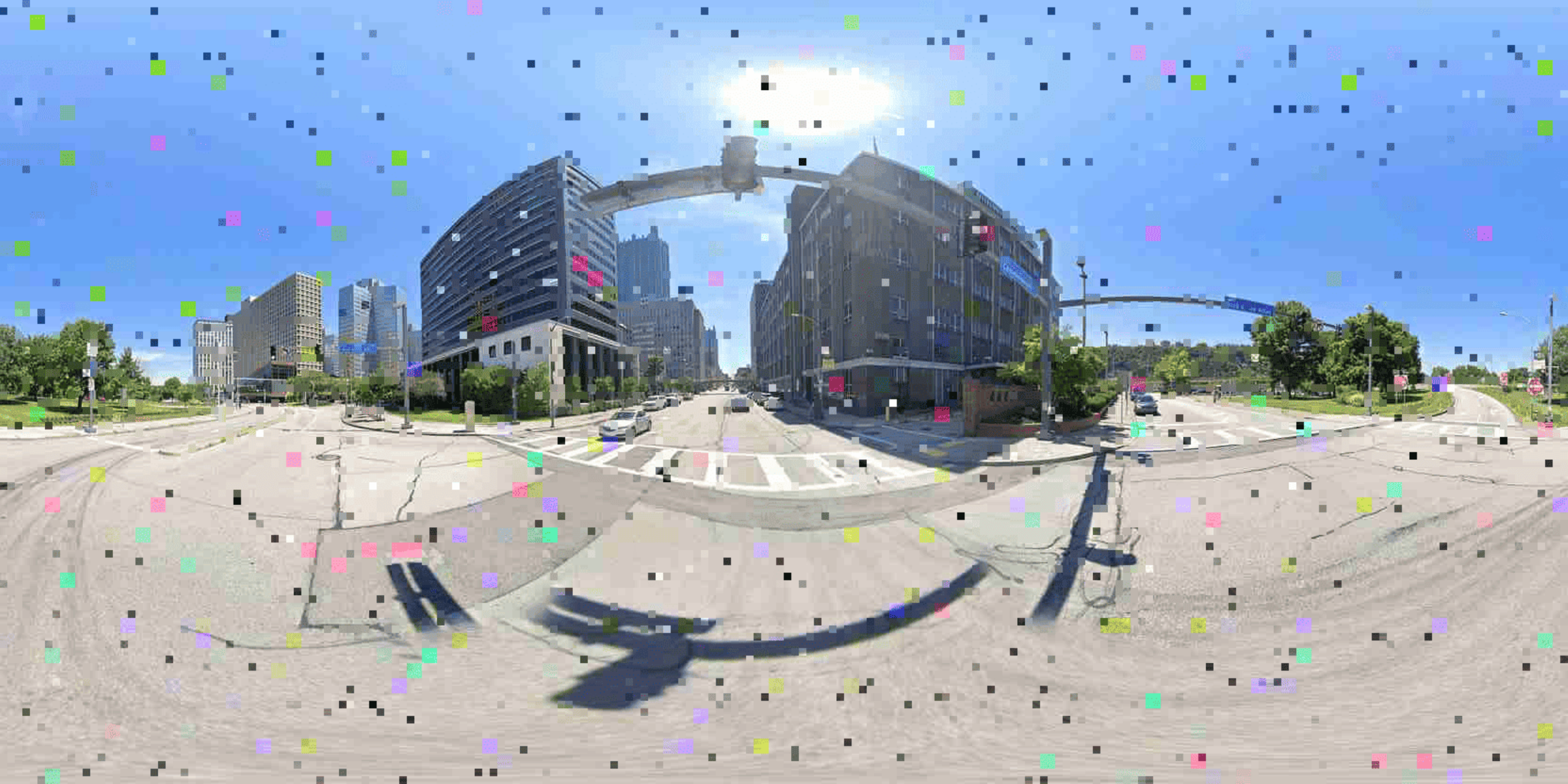}  
			\vspace*{-21ex} 
			\begin{center}
				\text{Standard Scheme}
			\end{center}
			\vspace*{13ex}
		\end{minipage}%
	}\hfill
	\subfloat[\small{0.0216 / \textbf{25.32}}]{
		\begin{minipage}{0.24\linewidth}
			\includegraphics[width=\textwidth]{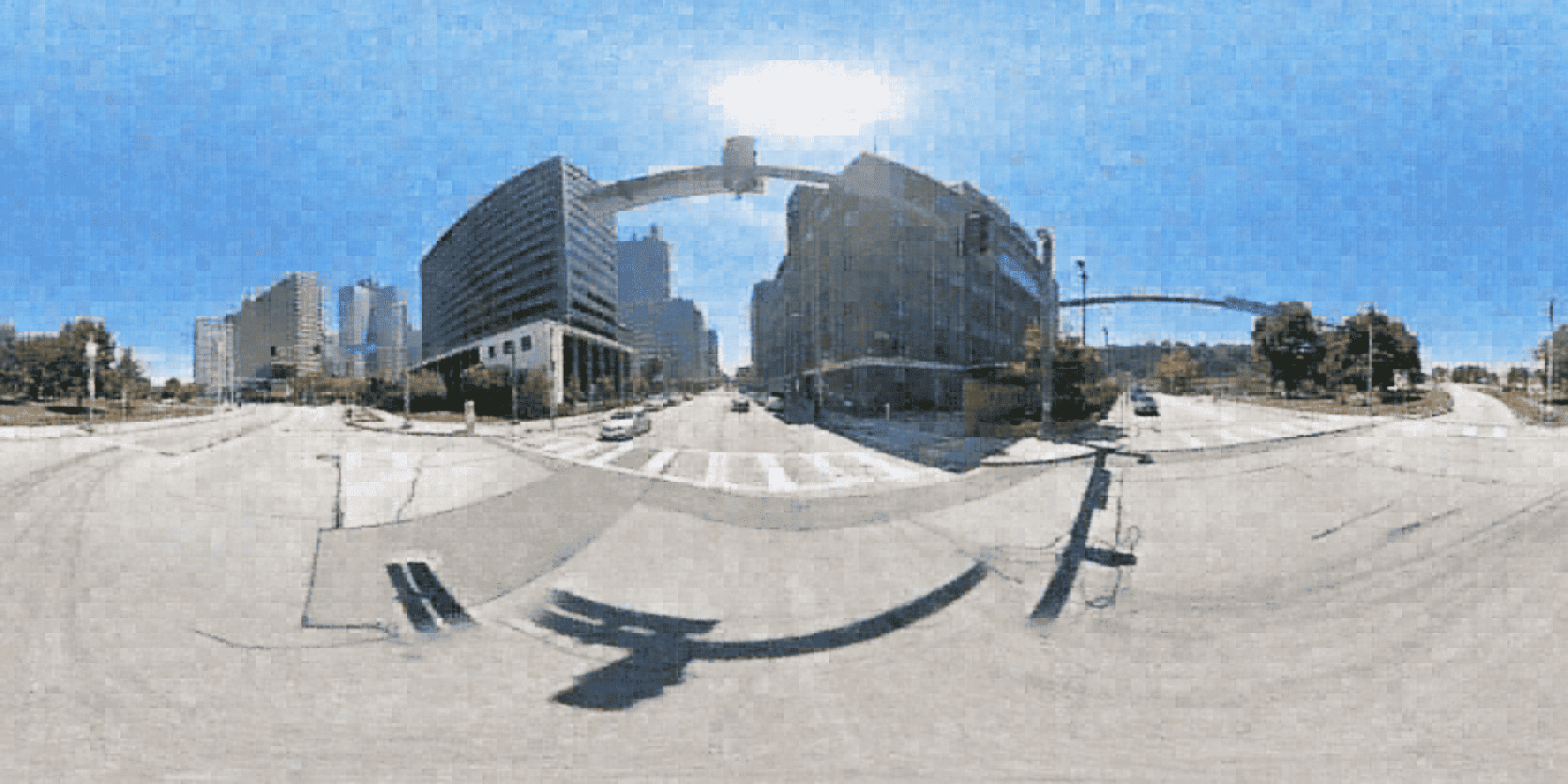}
			\vspace*{-21ex} 
			\begin{center}
				\text{ADJSCC}
			\end{center}
			\vspace*{13ex}
		\end{minipage}%
	}\hfill
	\subfloat[\small{0.0216 / 25.08}]{
		\begin{minipage}{0.24\linewidth}
			\includegraphics[width=\textwidth]{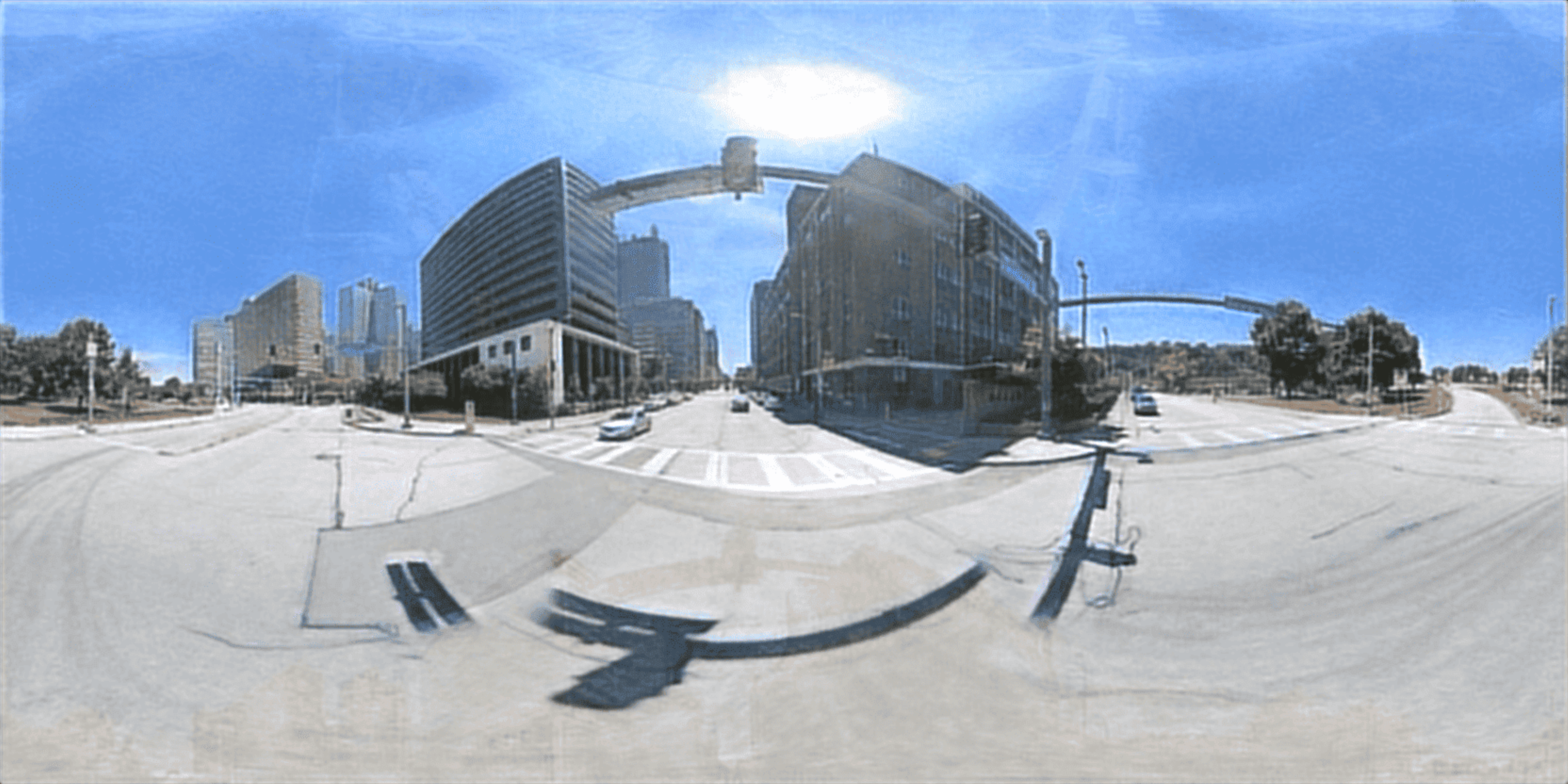}
			\vspace*{-21ex} 
			\begin{center}
				\text{SCSC}
			\end{center}
			\vspace*{13ex}
		\end{minipage}%
	}
	
	\subfloat{
		\begin{minipage}{0.24\linewidth}
			\vspace*{-2ex} 
			\begin{center}
				\text{\small{(a) SNR = 2 dB}}
			\end{center}
			\vspace*{-3ex}
		\end{minipage}%
	}\hfill

	\subfloat[\small{$R$ / PSNR (dB)}]{
		\begin{minipage}{0.24\linewidth}
			\includegraphics[width=\textwidth]{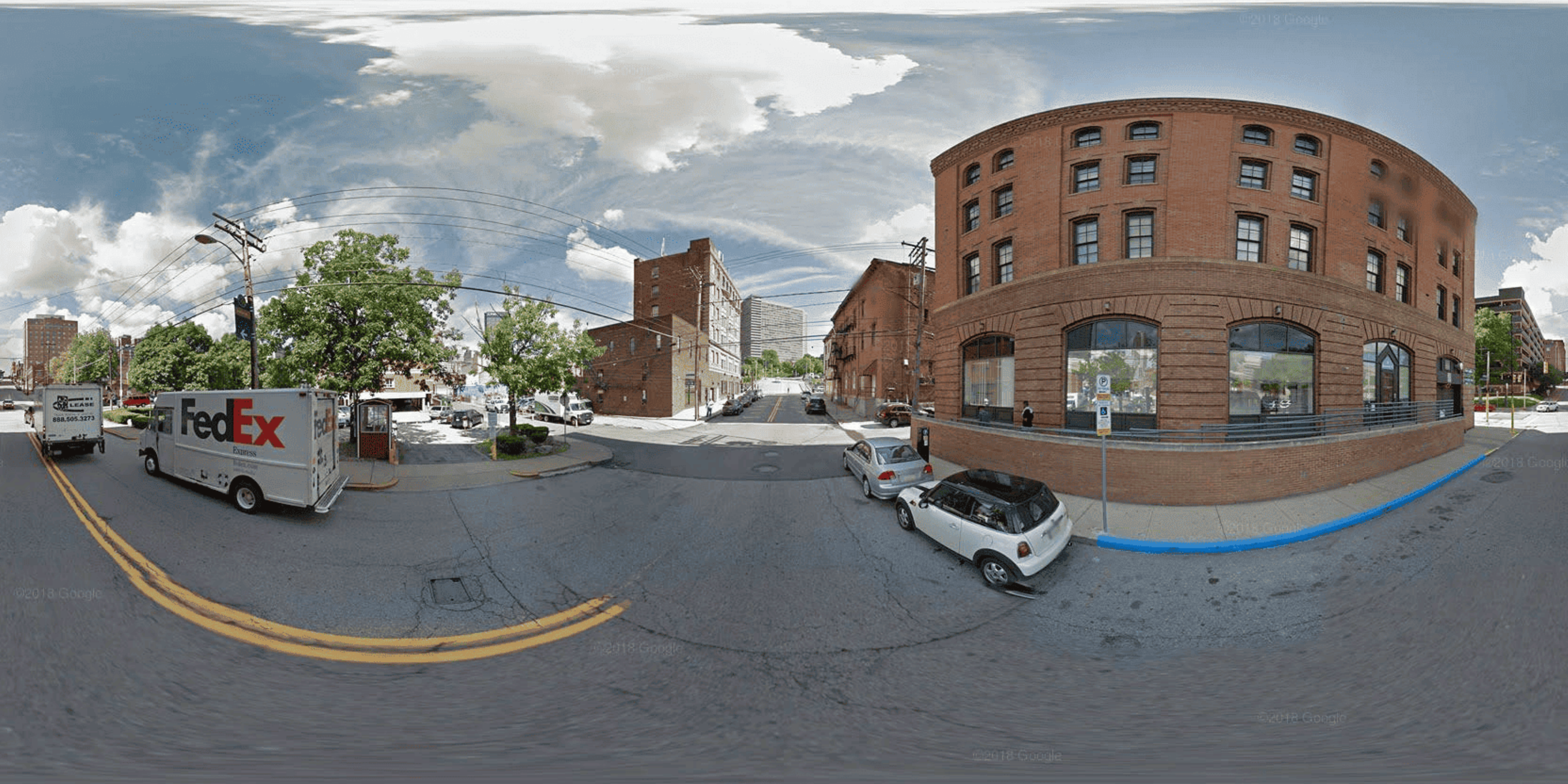}
		\end{minipage}%
	}\hfill
	\subfloat[\small{0.0236 / 26.31}]{
		\begin{minipage}{0.24\linewidth}
			\includegraphics[width=\textwidth]{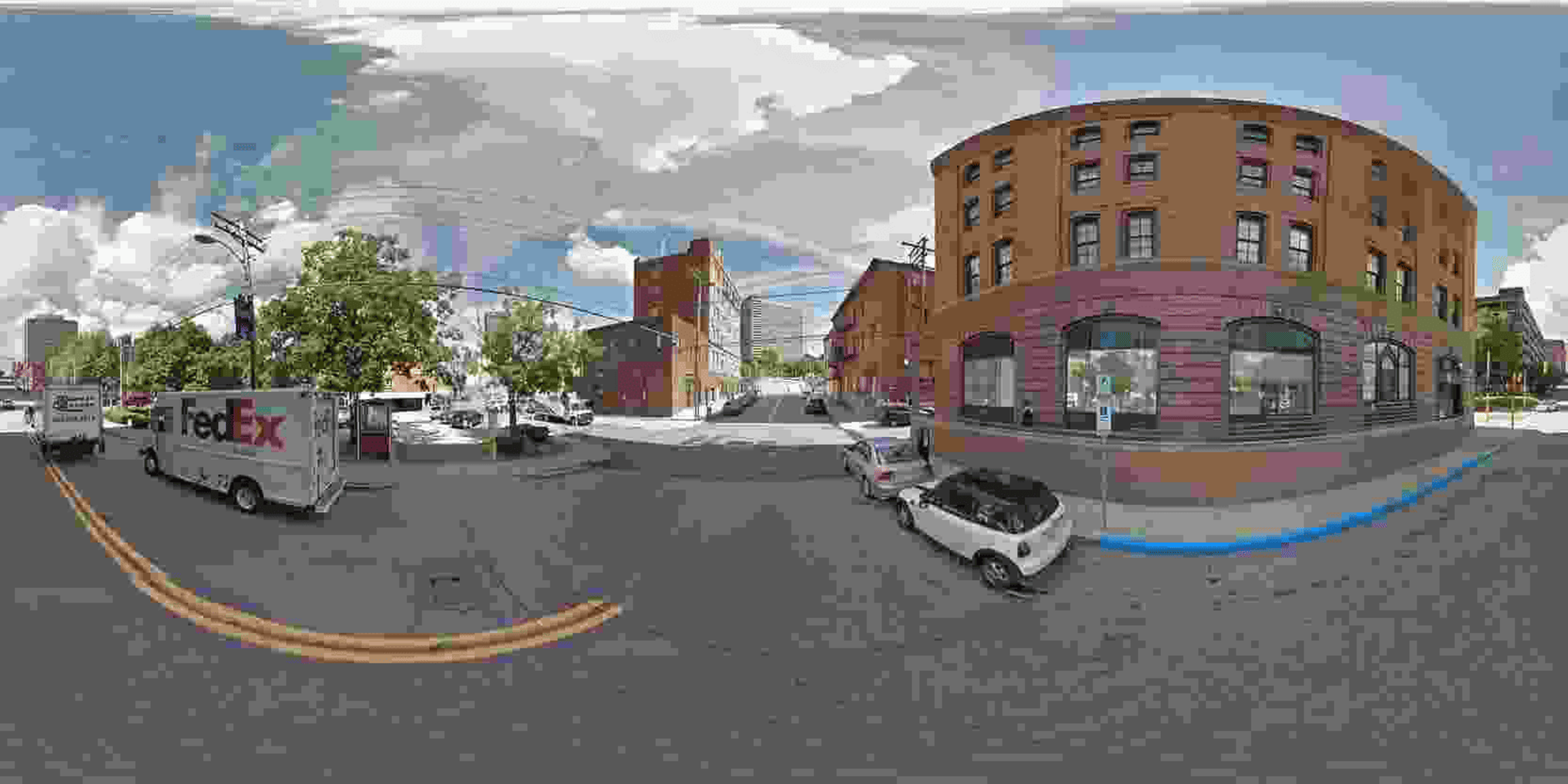}
		\end{minipage}%
	}\hfill
	\subfloat[\small{0.0223 / 26.53}]{
		\begin{minipage}{0.24\linewidth}
			\includegraphics[width=\textwidth]{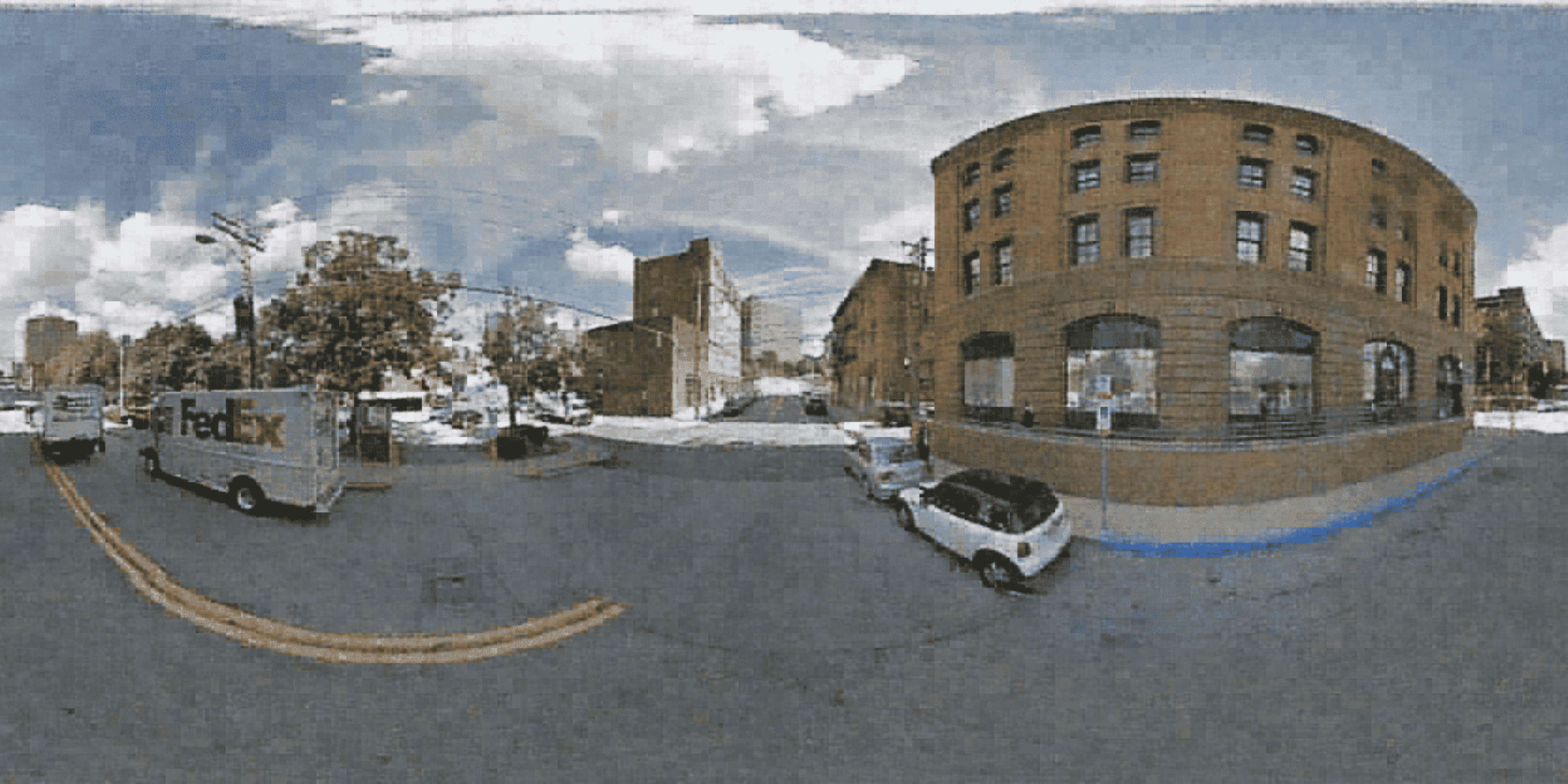}
		\end{minipage}%
	}\hfill
	\subfloat[\small{0.0216  / \textbf{27.62}}]{
		\begin{minipage}{0.24\linewidth}
			\includegraphics[width=\textwidth]{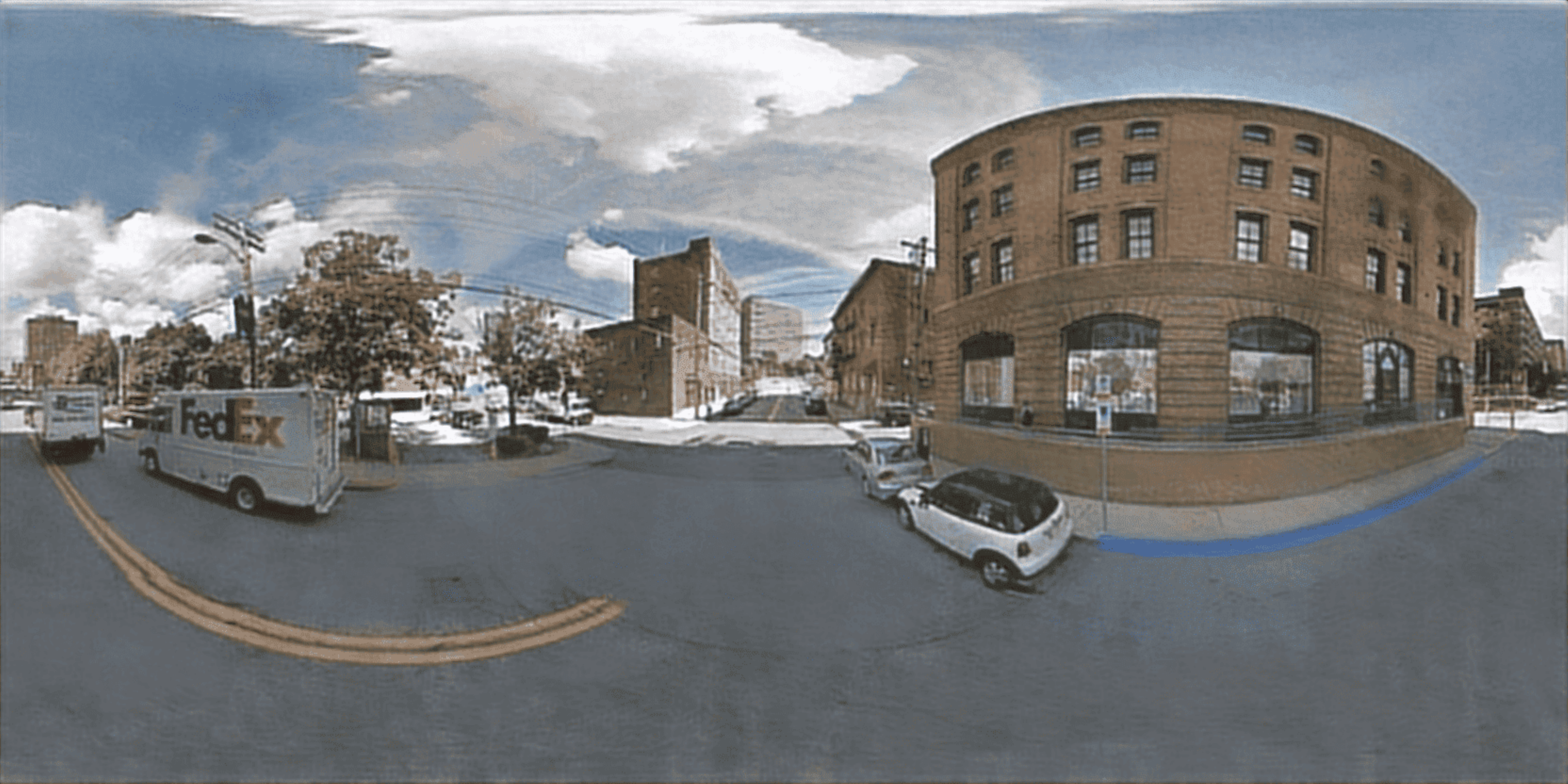}
		\end{minipage}%
	}
	
	\subfloat{
		\begin{minipage}{0.24\linewidth}
			\vspace*{-2ex} 
			\begin{center}
				\text{\small{(b) SNR = 6 dB}}
			\end{center}
			\vspace*{-3ex}
		\end{minipage}%
	}\hfill

	\subfloat[\small{$R$ / PSNR (dB)}]{
		\begin{minipage}{0.24\linewidth}
			\includegraphics[width=\textwidth]{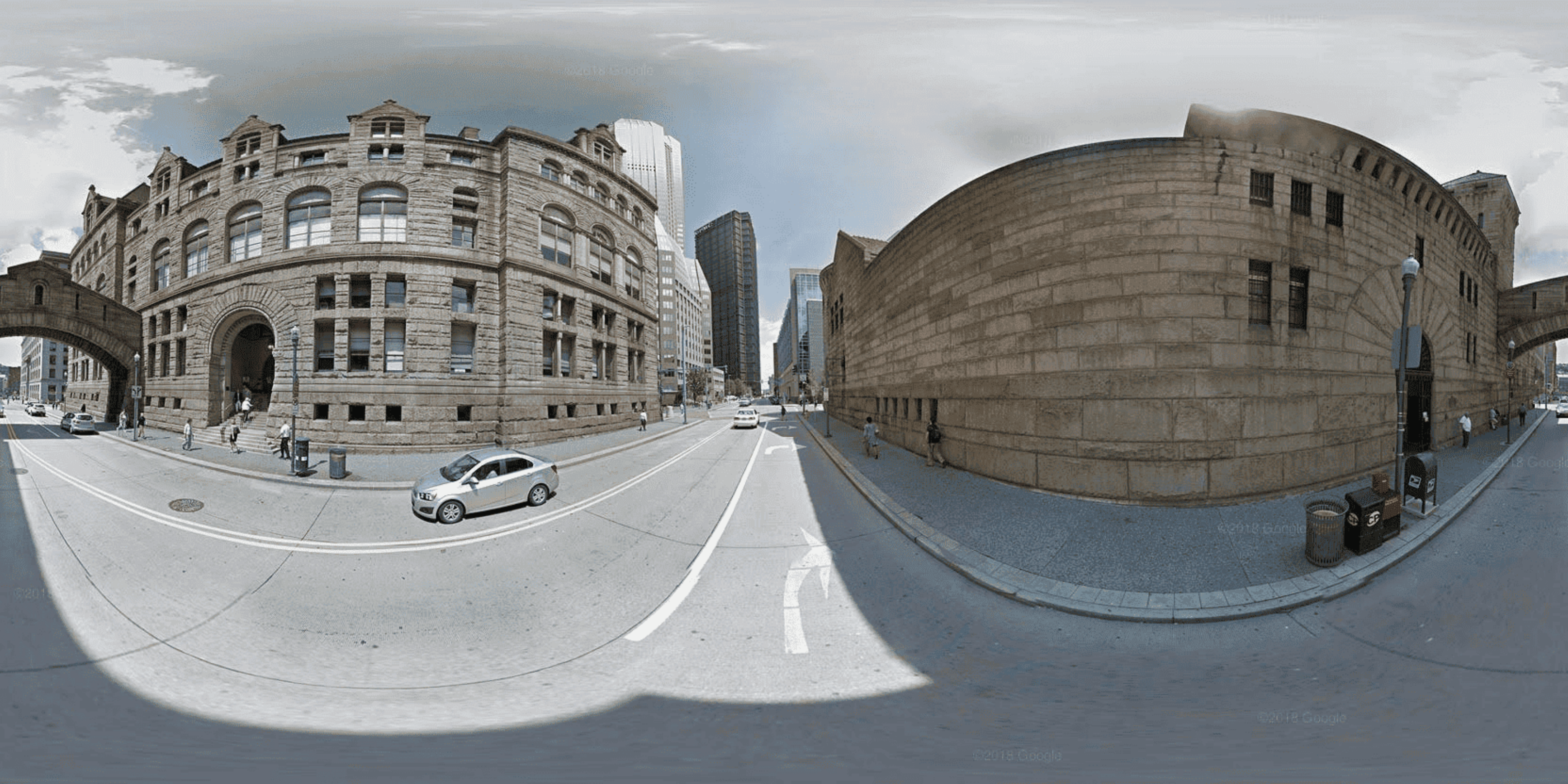}
		\end{minipage}%
	}\hfill
	\subfloat[\small{0.026 / 29.47}]{
		\begin{minipage}{0.24\linewidth}
			\includegraphics[width=\textwidth]{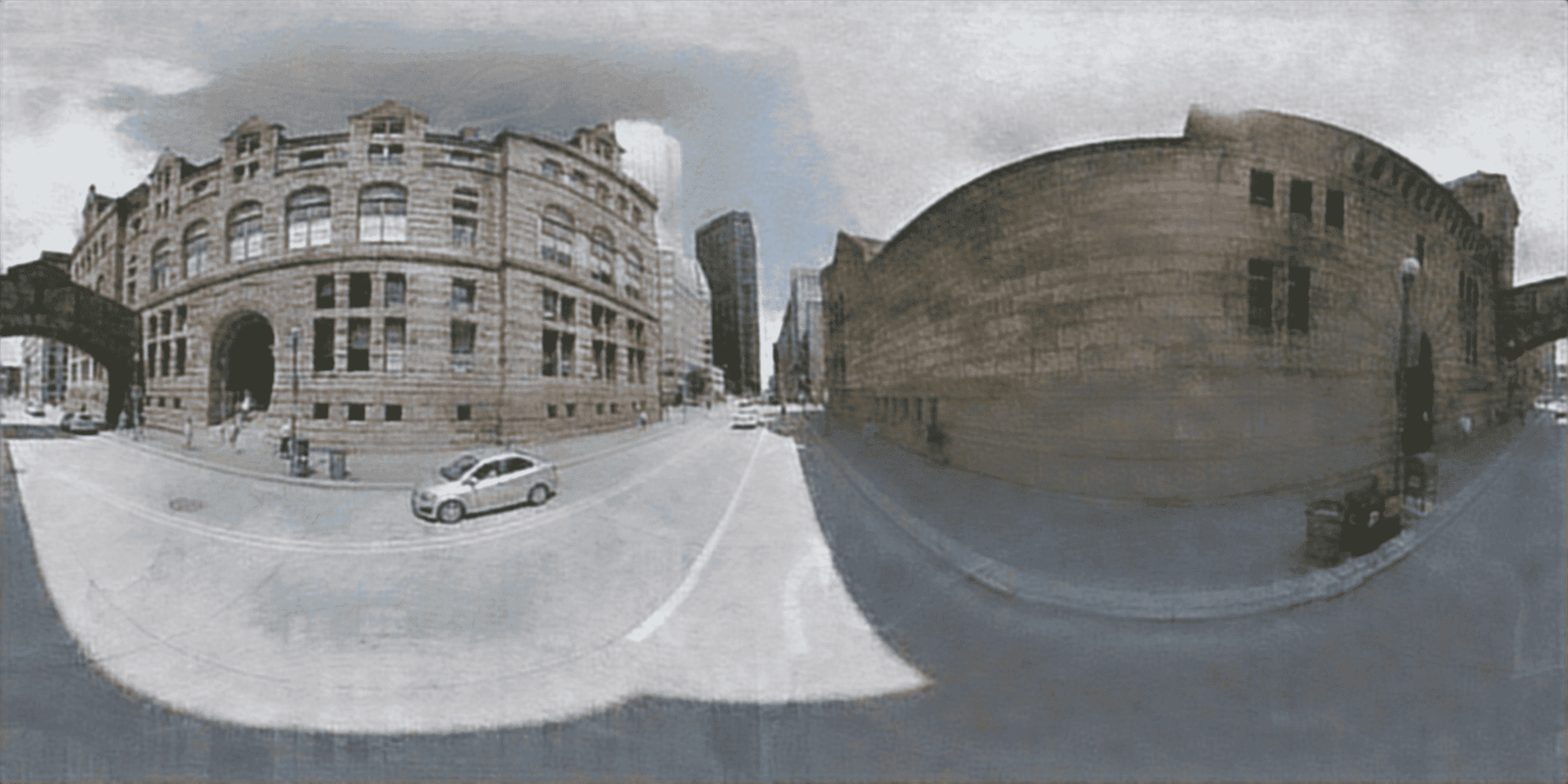}
		\end{minipage}%
	}\hfill
	\subfloat[\small{0.0249 / 31.26}]{
		\begin{minipage}{0.24\linewidth}
			\includegraphics[width=\textwidth]{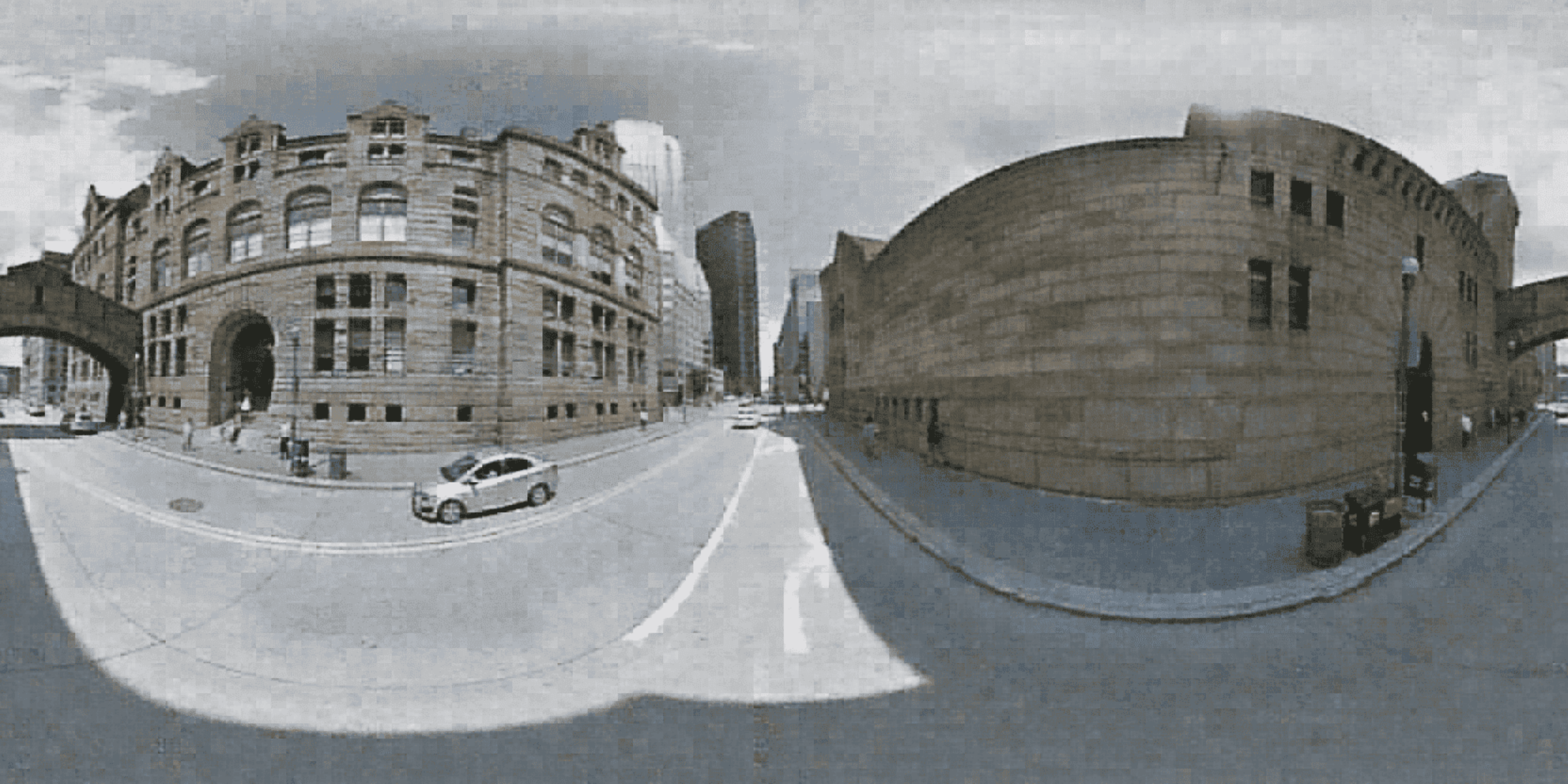}
		\end{minipage}%
	}\hfill
	\subfloat[\small{0.0216 / \textbf{32.04}}]{
		\begin{minipage}{0.24\linewidth}
			\includegraphics[width=\textwidth]{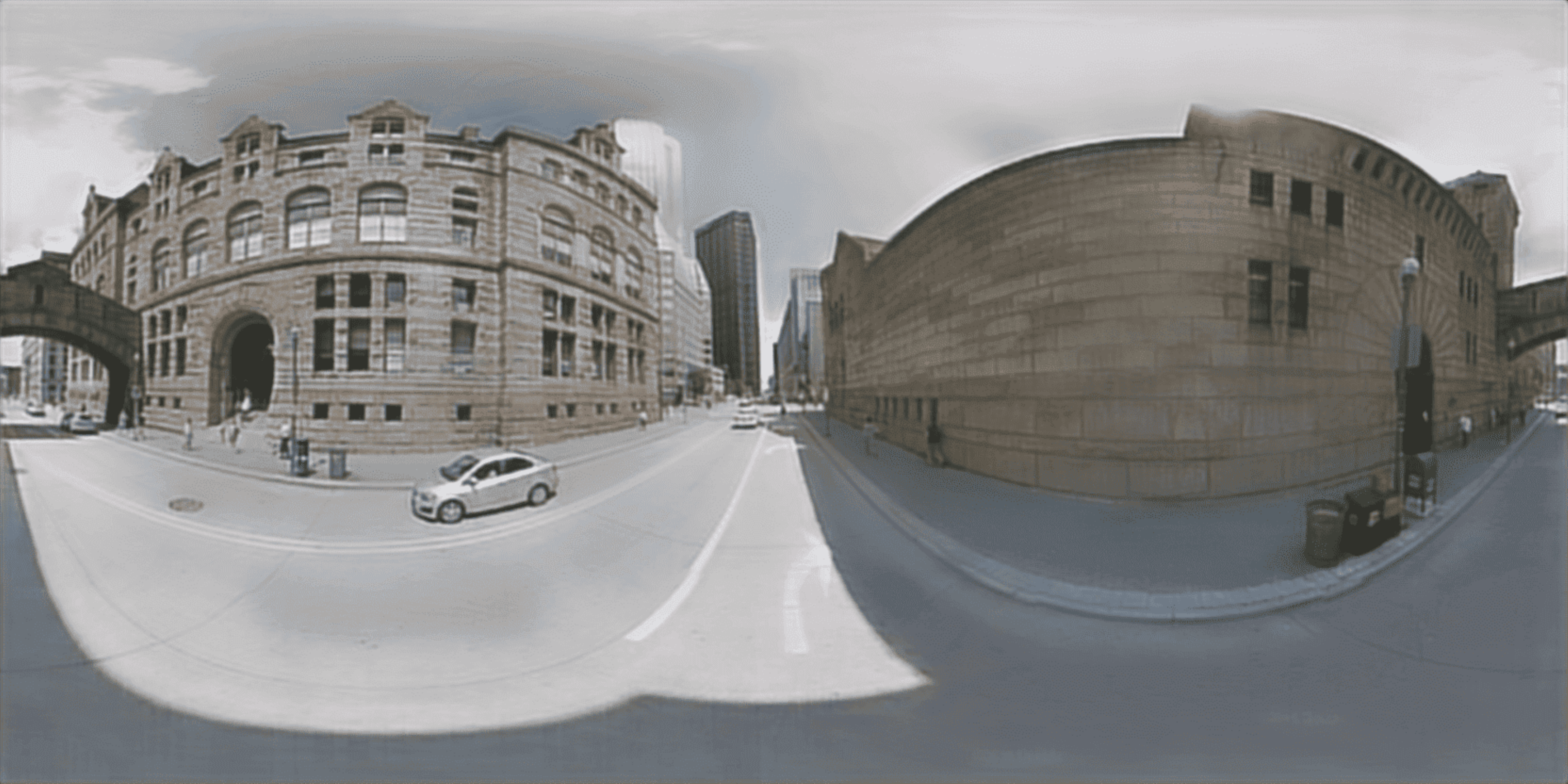}
		\end{minipage}%
	}
	
	\subfloat{
		\begin{minipage}{0.24\linewidth}
			\vspace*{-2ex} 
			\begin{center}
				\text{\small{(c) SNR = 10 dB}}
			\end{center}
			\vspace*{-3ex}
		\end{minipage}%
	}\hfill

	\caption{Visual comparison of reconstructed images. The first, second, and third rows correspond to SNR = 2, 6, and 10 dB, respectively. The first column shows the ground truth images, while the second to fourth columns show the reconstructions of the standard coding scheme, ADJSCC, and SCSC, respectively.}
	\label{visual1}
\end{figure*}

\begin{figure*}[htpb]
	\centering
	\captionsetup[subfloat]{labelsep=none,format=plain,labelformat=empty}
	\subfloat{
		\rotatebox{90}{\small{Gound Truth}}
		\begin{minipage}[t]{0.235\linewidth}
			\includegraphics[width=\textwidth]{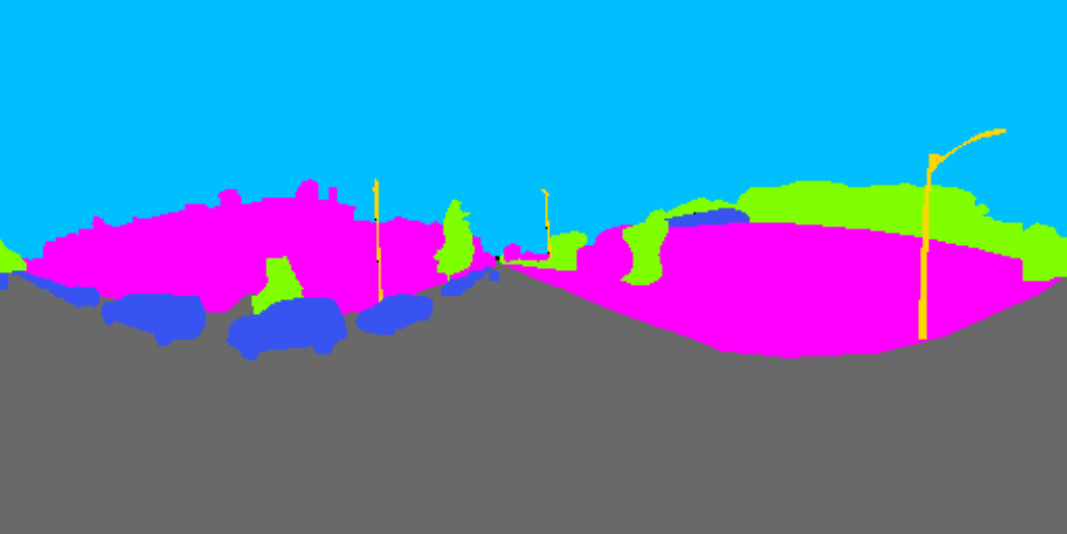}
		\end{minipage}%
	}\hfill
	\subfloat{
		\begin{minipage}[t]{0.235\linewidth}
			\includegraphics[width=\textwidth]{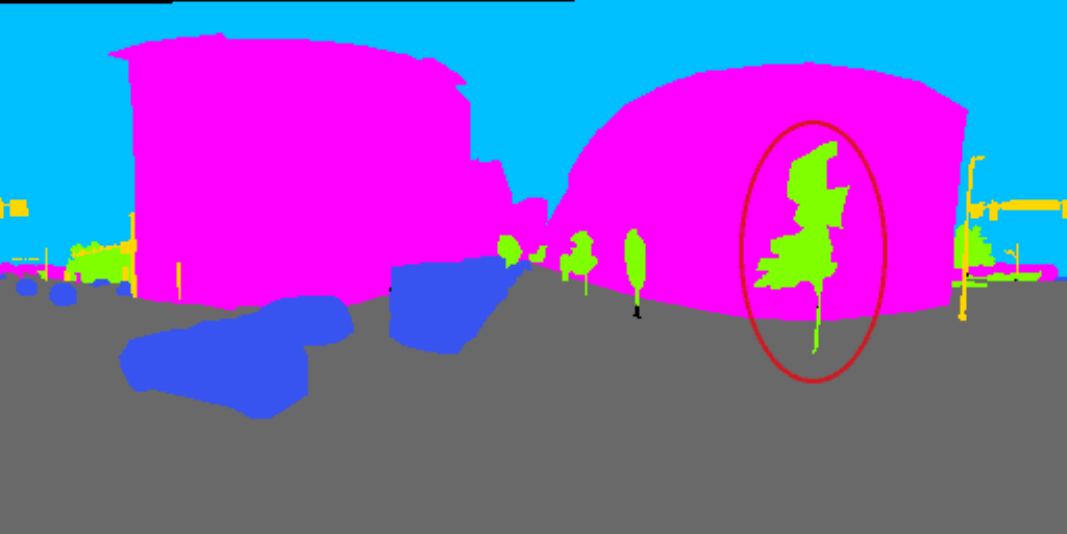}
		\end{minipage}%
	}\hfill
	\subfloat{
		\begin{minipage}[t]{0.235\linewidth}
			\includegraphics[width=\textwidth]{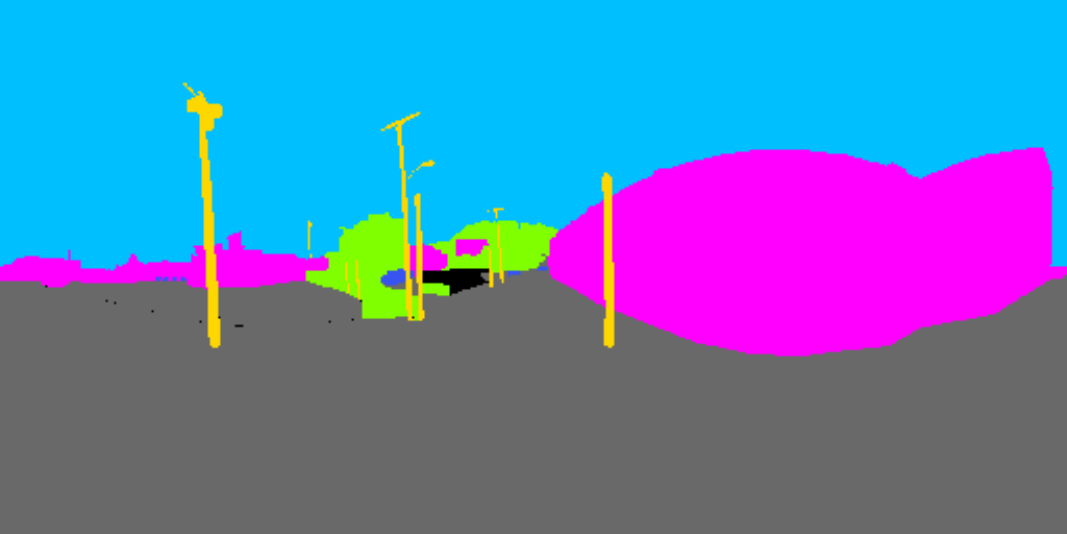}
		\end{minipage}%
	}\hfill
	\subfloat{
		\begin{minipage}[t]{0.235\linewidth}
			\includegraphics[width=\textwidth]{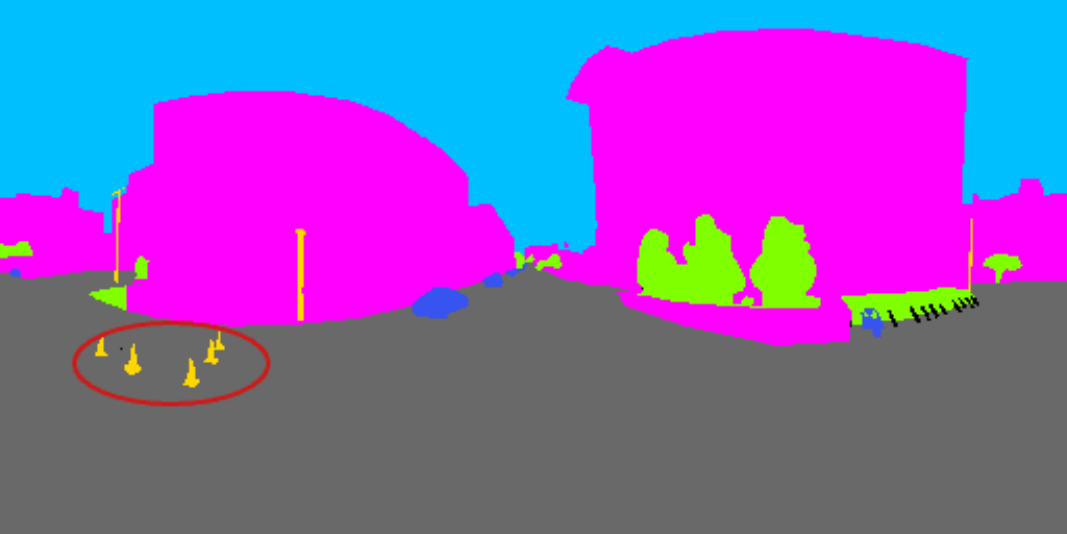}
		\end{minipage}%
		
	}
	
	\subfloat{
		\rotatebox{90}{\footnotesize{Standard Scheme}}
		\begin{minipage}[t]{0.235\linewidth}
			\includegraphics[width=\textwidth]{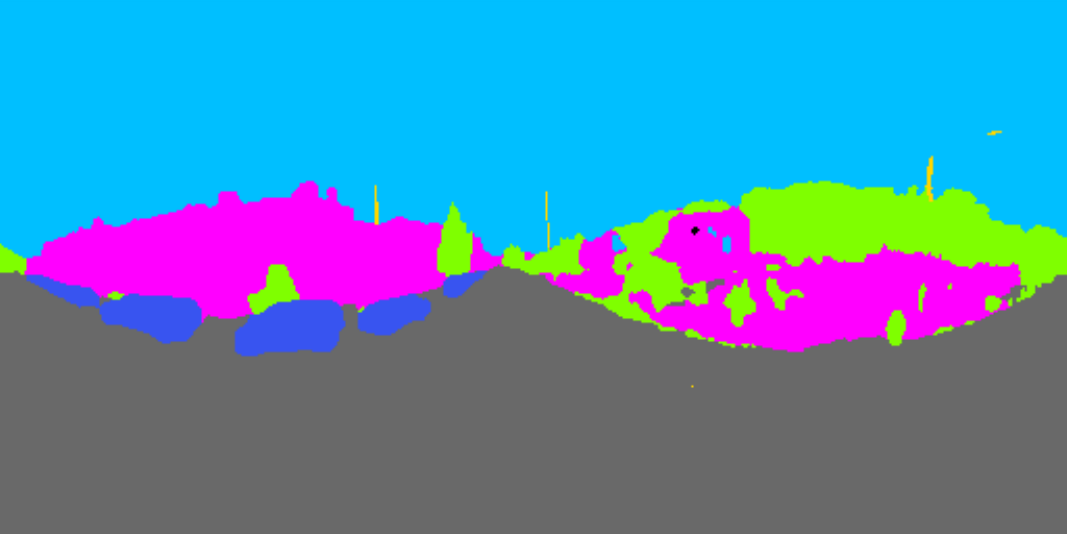}
		\end{minipage}%
	}\hfill
	\subfloat{
		\begin{minipage}[t]{0.235\linewidth}
			\includegraphics[width=\textwidth]{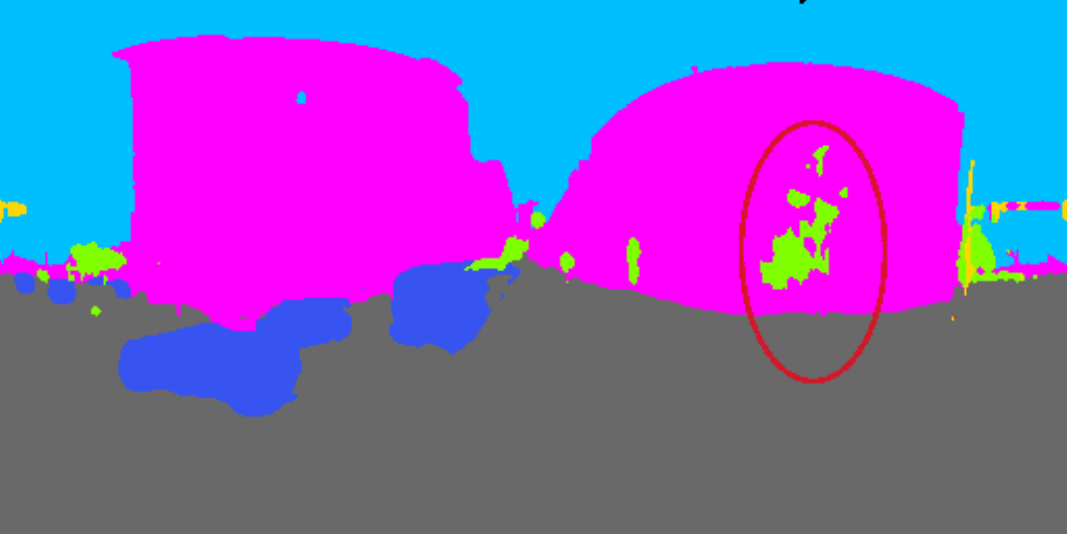}
		\end{minipage}%
	}\hfill
	\subfloat{
		\begin{minipage}[t]{0.235\linewidth}
			\includegraphics[width=\textwidth]{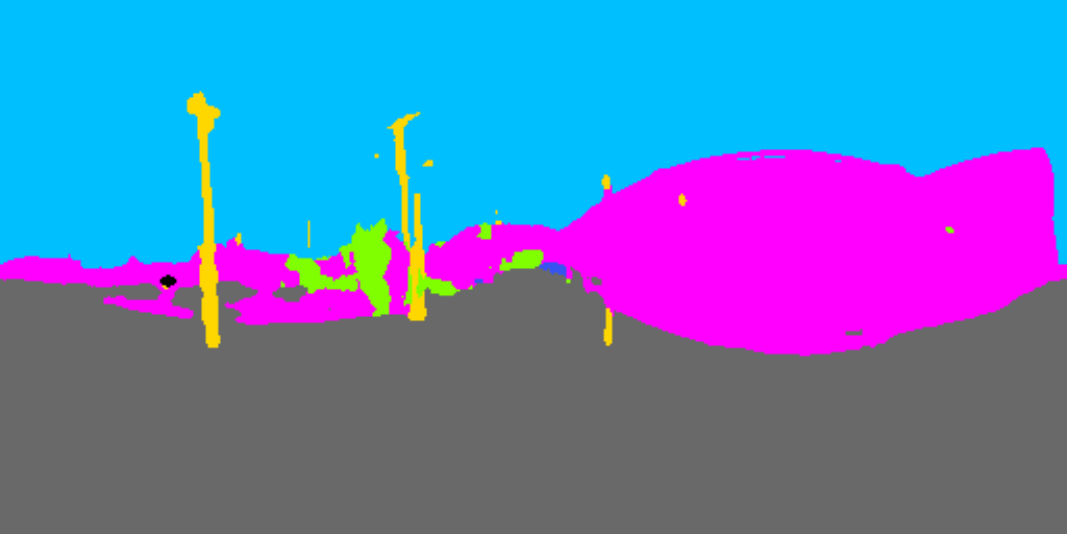}
		\end{minipage}%
	}\hfill
	\subfloat{
		\begin{minipage}[t]{0.235\linewidth}
			\includegraphics[width=\textwidth]{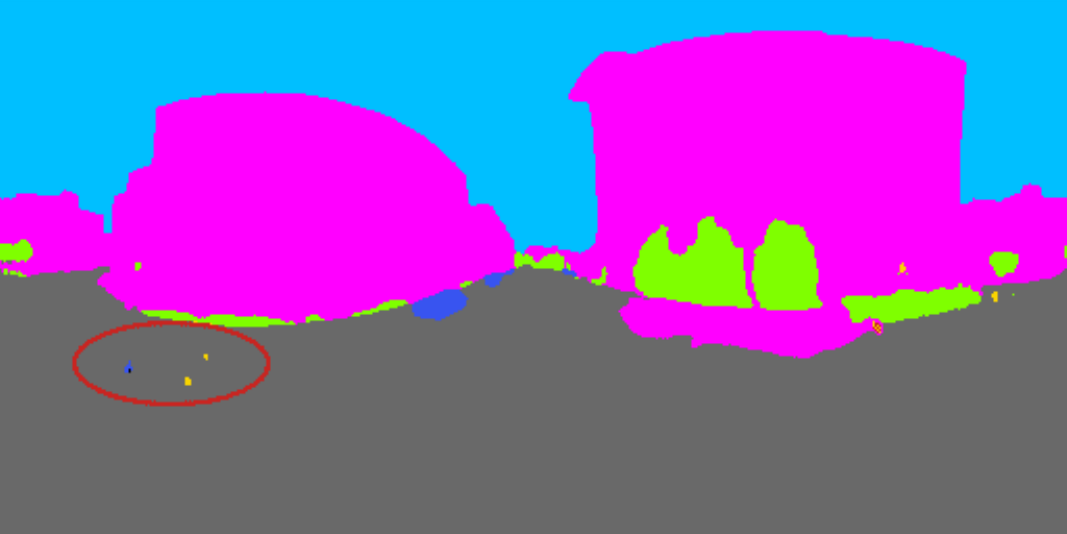}
		\end{minipage}%
	}
	
	\subfloat{
		\rotatebox{90}{\footnotesize{~~ADJSCC}}
		\begin{minipage}[t]{0.234\linewidth}
			\includegraphics[width=\textwidth]{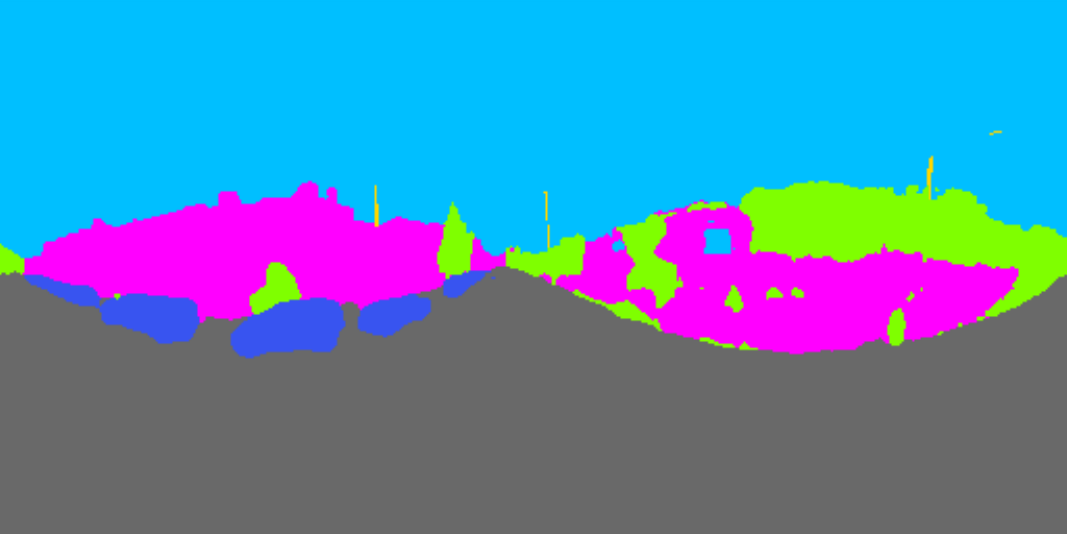}
		\end{minipage}%
	}\hfill
	\subfloat{
		\begin{minipage}[t]{0.234\linewidth}
			\includegraphics[width=\textwidth]{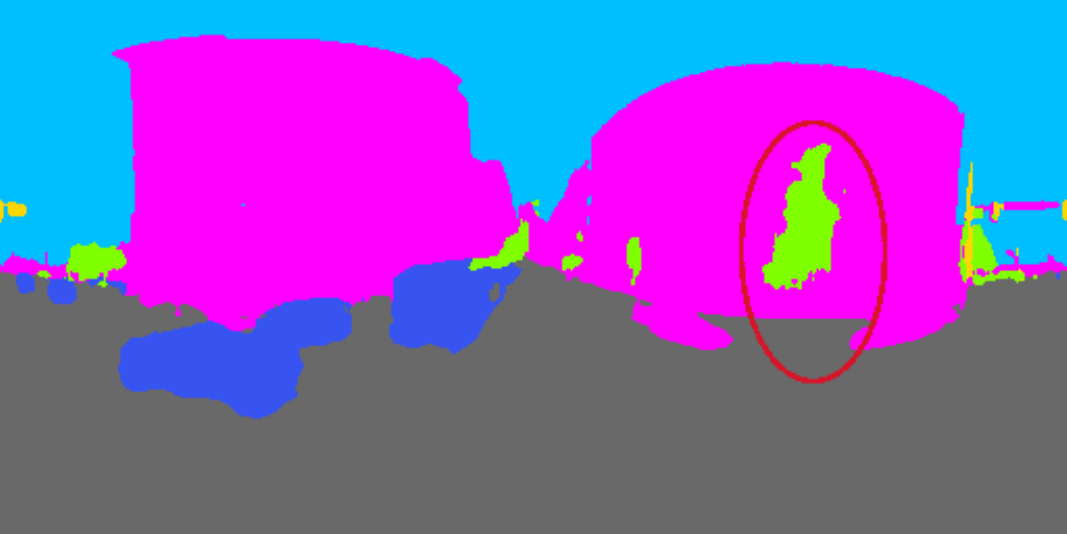}
		\end{minipage}%
	}\hfill
	\subfloat{
		\begin{minipage}[t]{0.234\linewidth}
			\includegraphics[width=\textwidth]{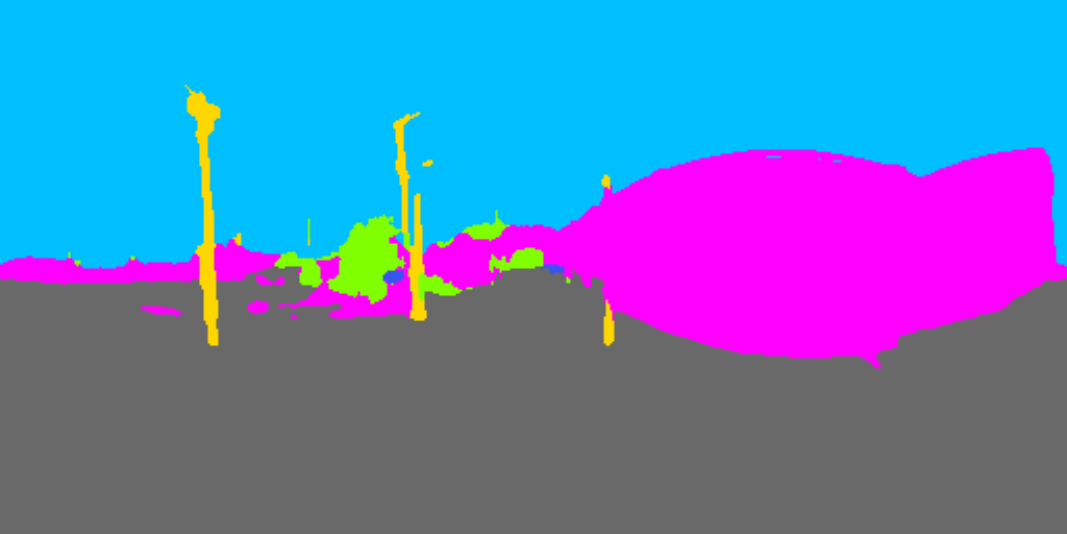}
		\end{minipage}%
	}\hfill
	\subfloat{
		\begin{minipage}[t]{0.234\linewidth}
			\includegraphics[width=\textwidth]{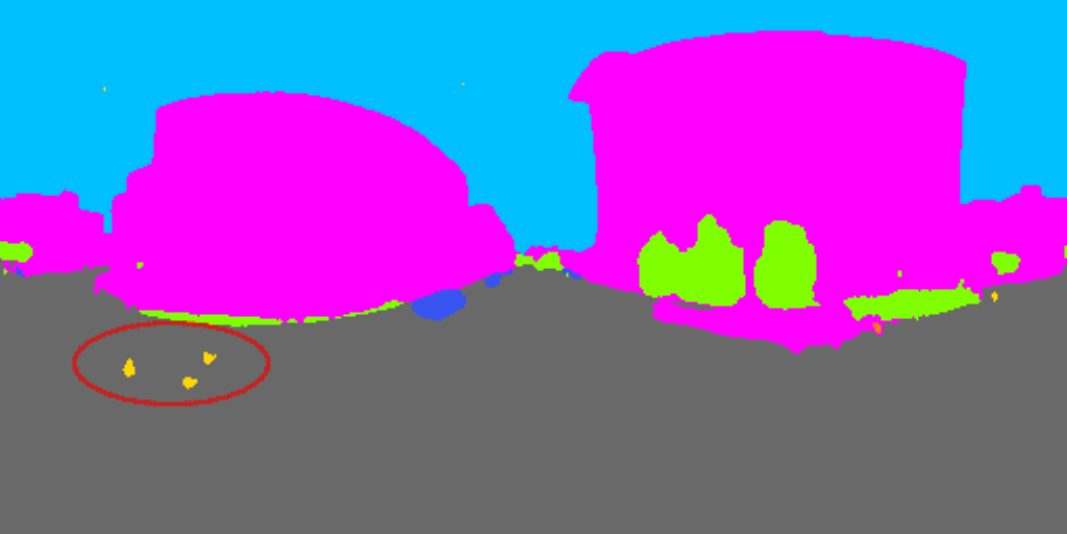}
		\end{minipage}%
	}
	
	\subfloat{
		\rotatebox{90}{\footnotesize{~~~~~SCSC}}
		\begin{minipage}[t]{0.233\linewidth}
			\includegraphics[width=\textwidth]{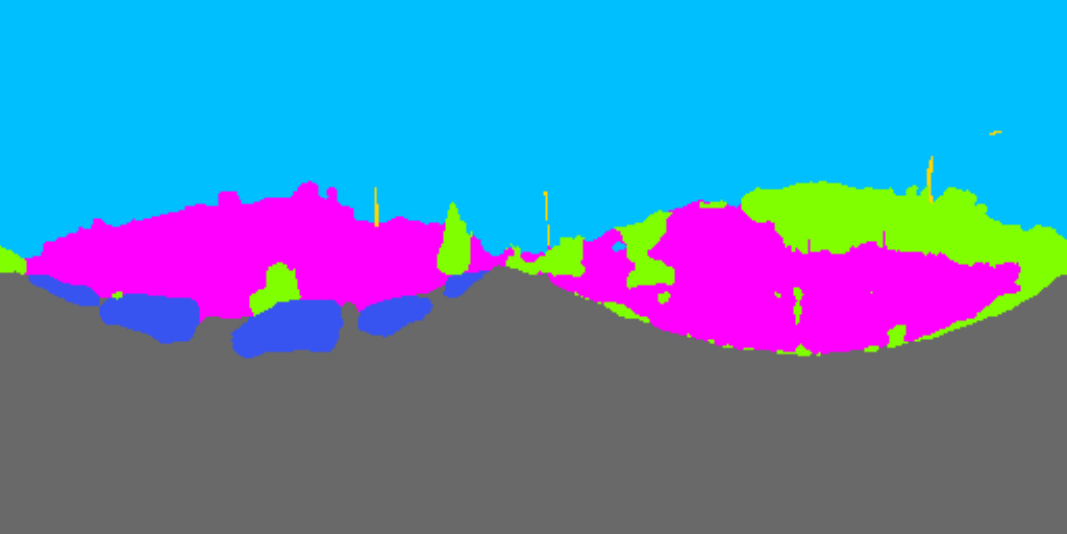}
		\end{minipage}%
	}\hfill
	\subfloat{
		\begin{minipage}[t]{0.235\linewidth}
			\includegraphics[width=\textwidth]{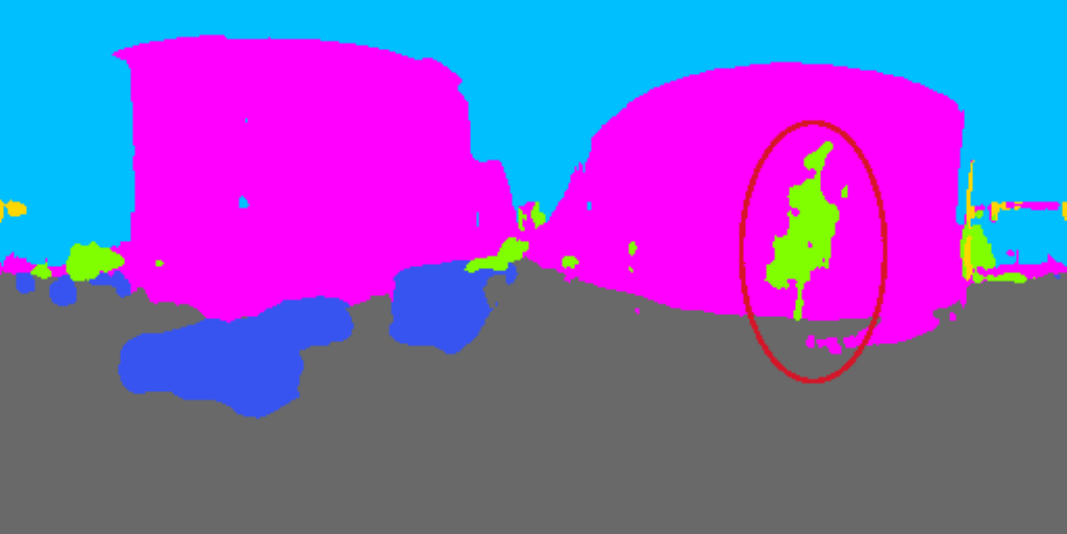}
		\end{minipage}%
	}\hfill
	\subfloat{
		\begin{minipage}[t]{0.235\linewidth}
			\includegraphics[width=\textwidth]{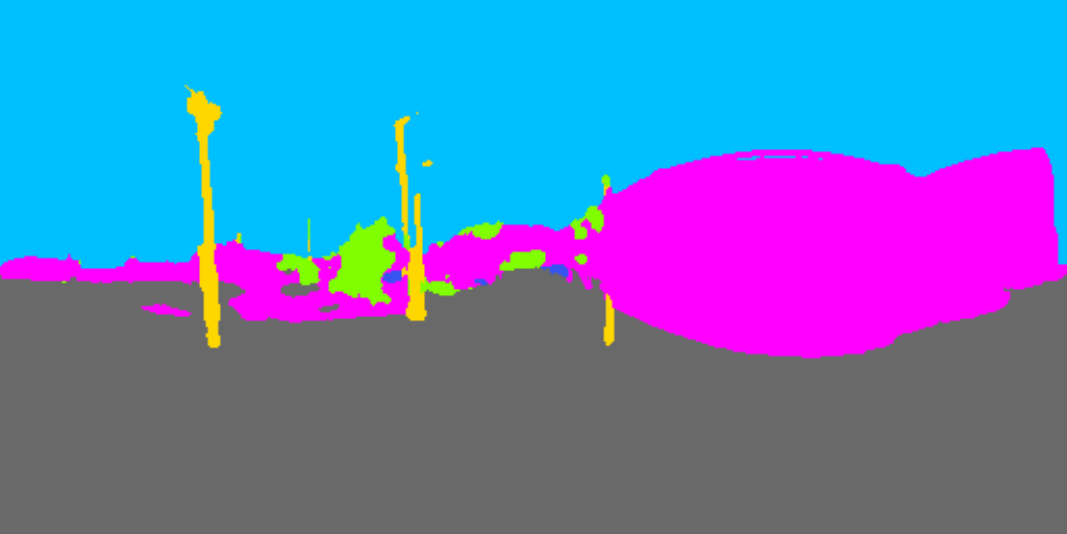}
		\end{minipage}%
	}\hfill
	\subfloat{
		\begin{minipage}[t]{0.235\linewidth}
			\includegraphics[width=\textwidth]{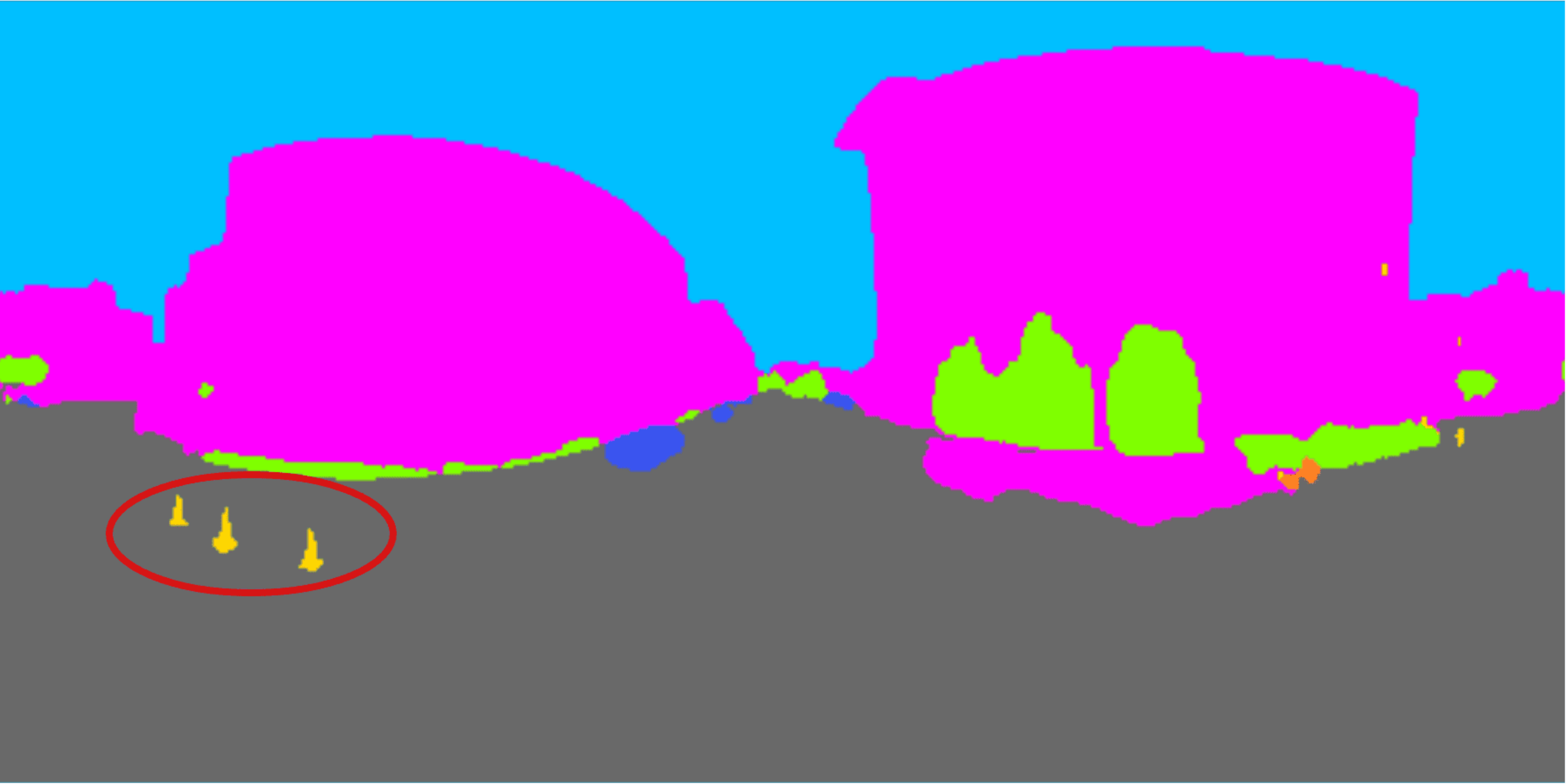}
		\end{minipage}%
	}
	
	\subfloat{
		{\footnotesize{~~unlabeled}}
		\begin{minipage}{0.03\linewidth}
			\includegraphics[width=\textwidth]{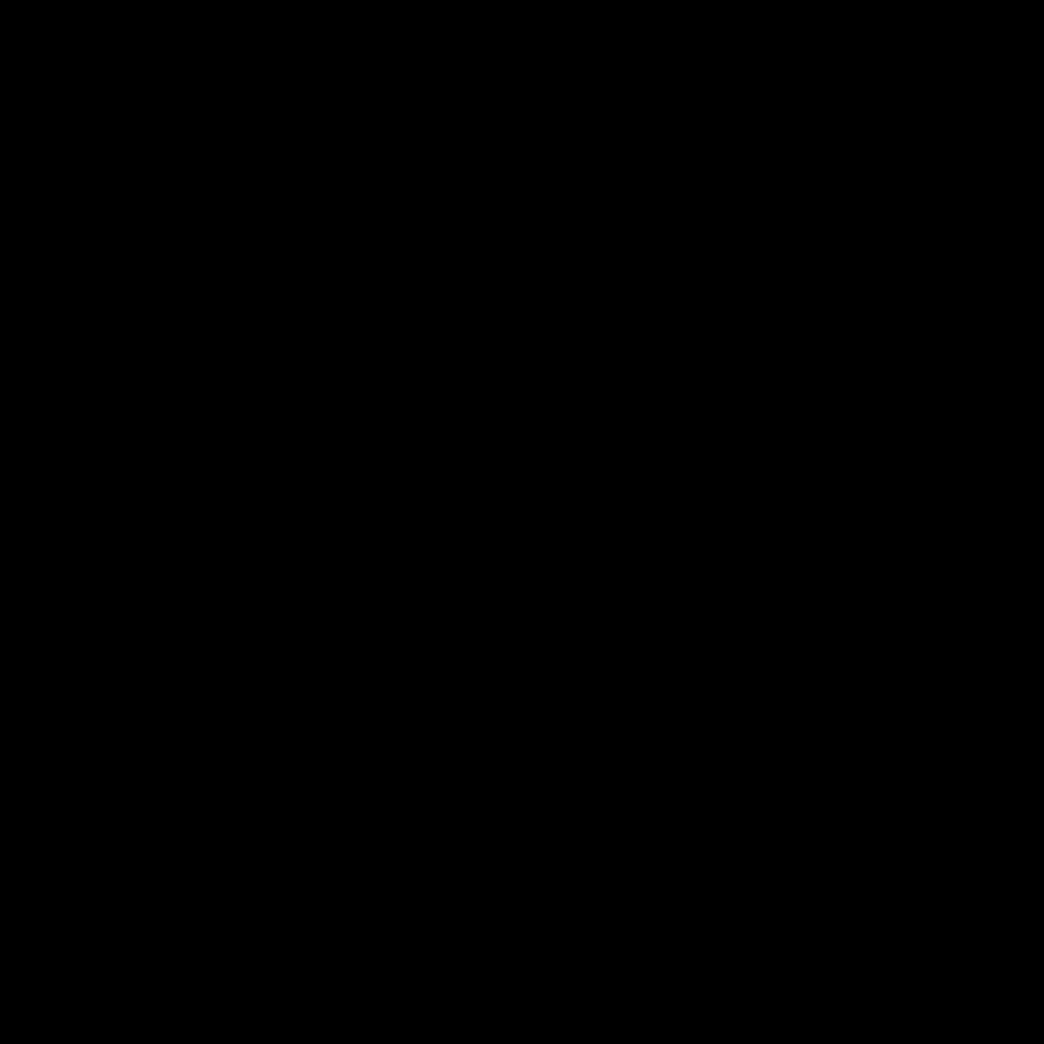}
		\end{minipage}%
	}\hfill
	\subfloat{
		{\footnotesize{~~flat}}
		\begin{minipage}{0.03\linewidth}
			\includegraphics[width=\textwidth]{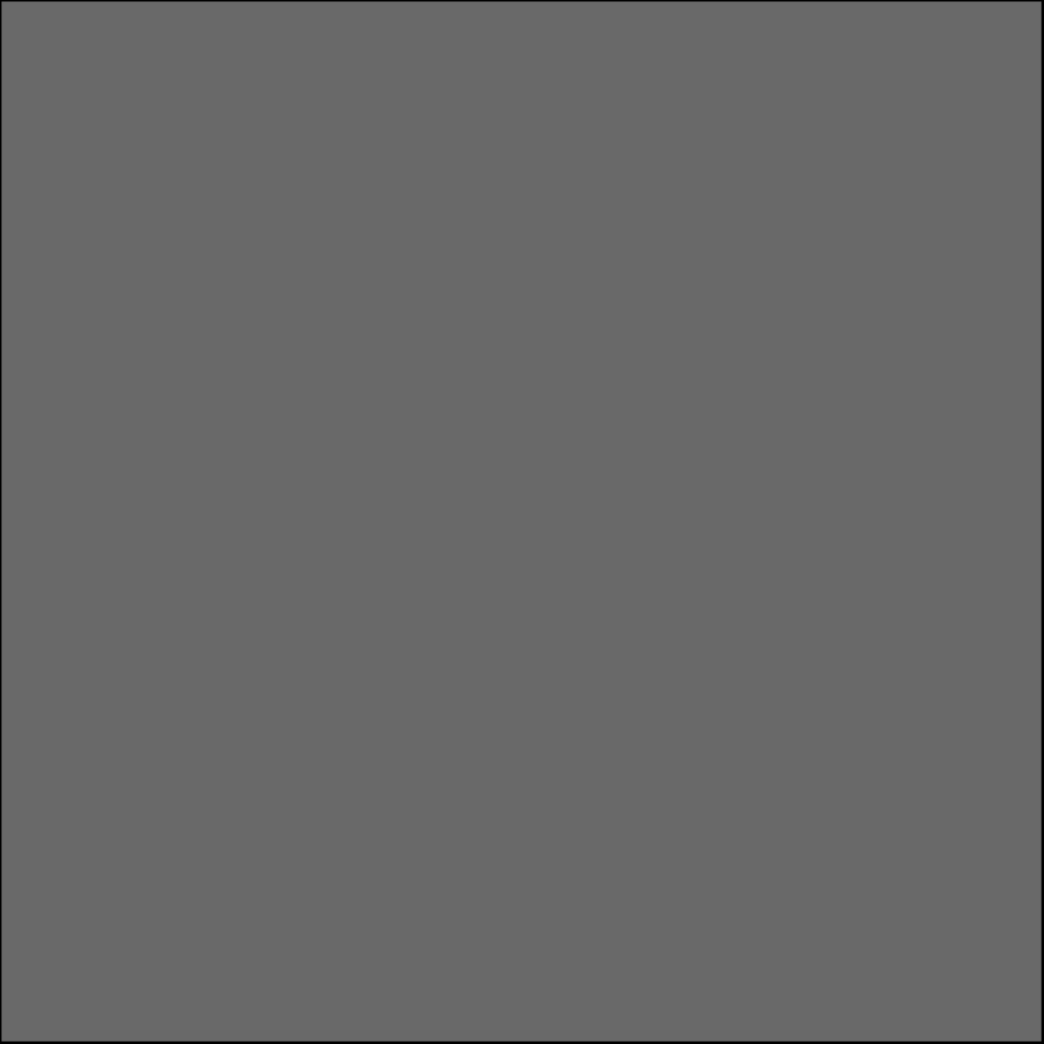}
		\end{minipage}%
	}\hfill
	\subfloat{
		{\footnotesize{~~construction}}
		\begin{minipage}{0.03\linewidth}
			\includegraphics[width=\textwidth]{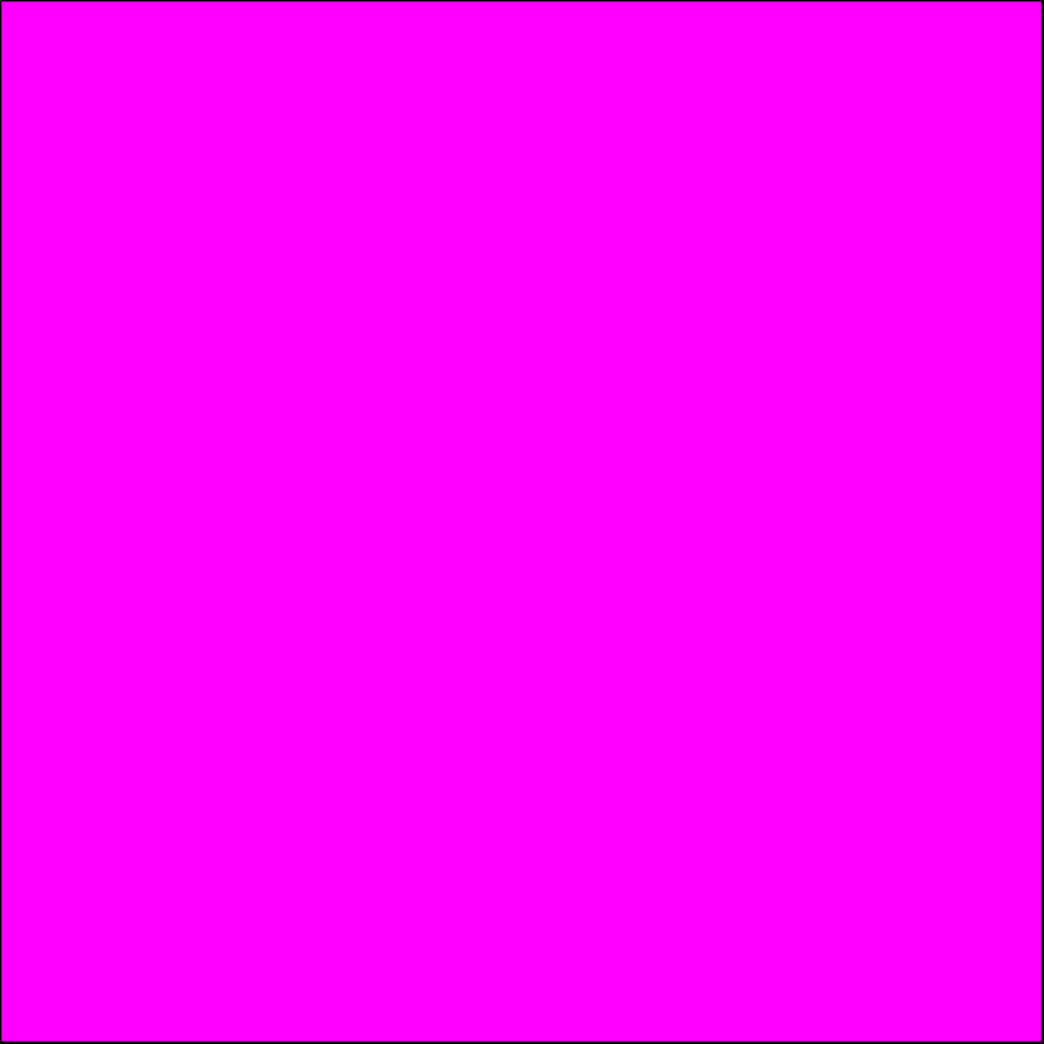}
		\end{minipage}%
	}\hfill
	\subfloat{
		{\footnotesize{~~object}}
		\begin{minipage}{0.03\linewidth}
			\includegraphics[width=\textwidth]{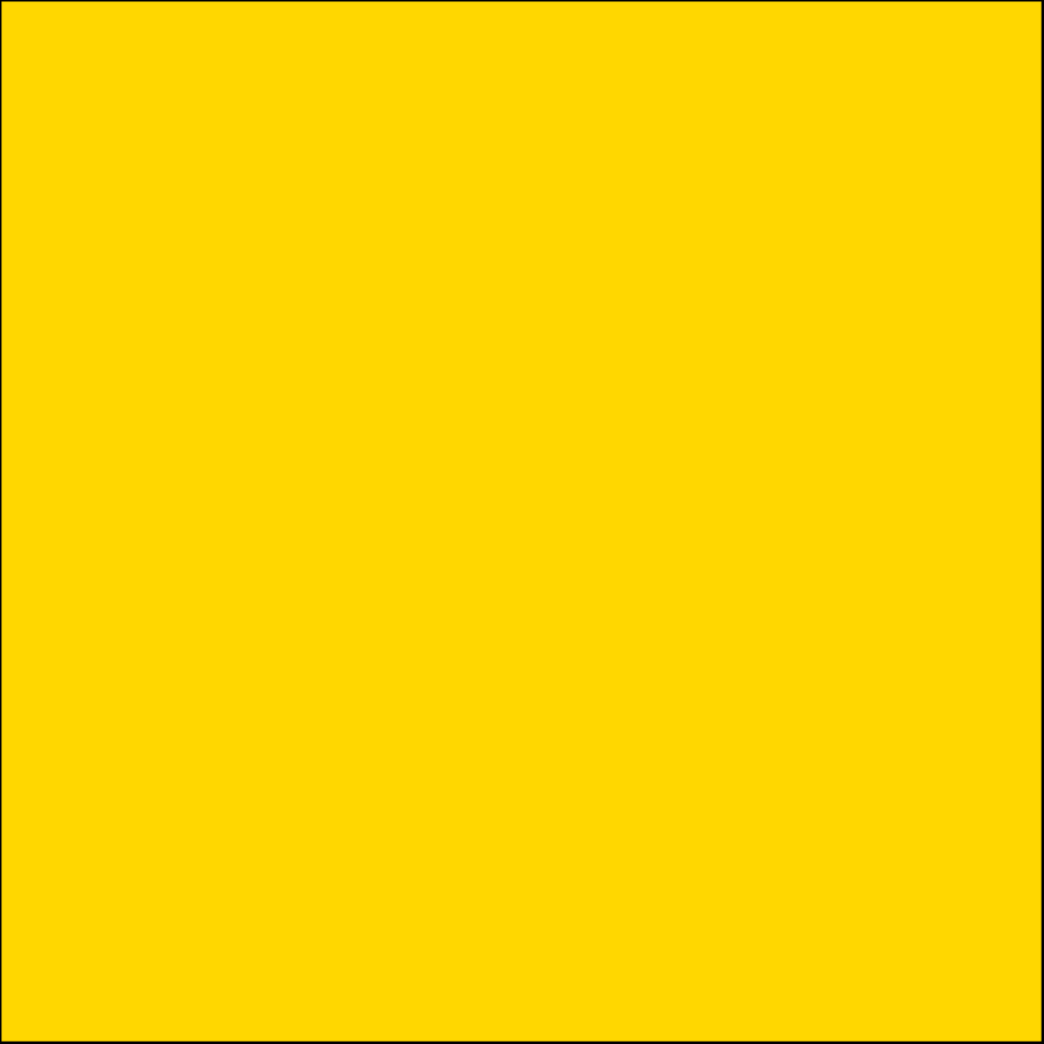}
		\end{minipage}%
	}\hfill
	\subfloat{
		{\footnotesize{~~nature}}
		\begin{minipage}{0.03\linewidth}
			\includegraphics[width=\textwidth]{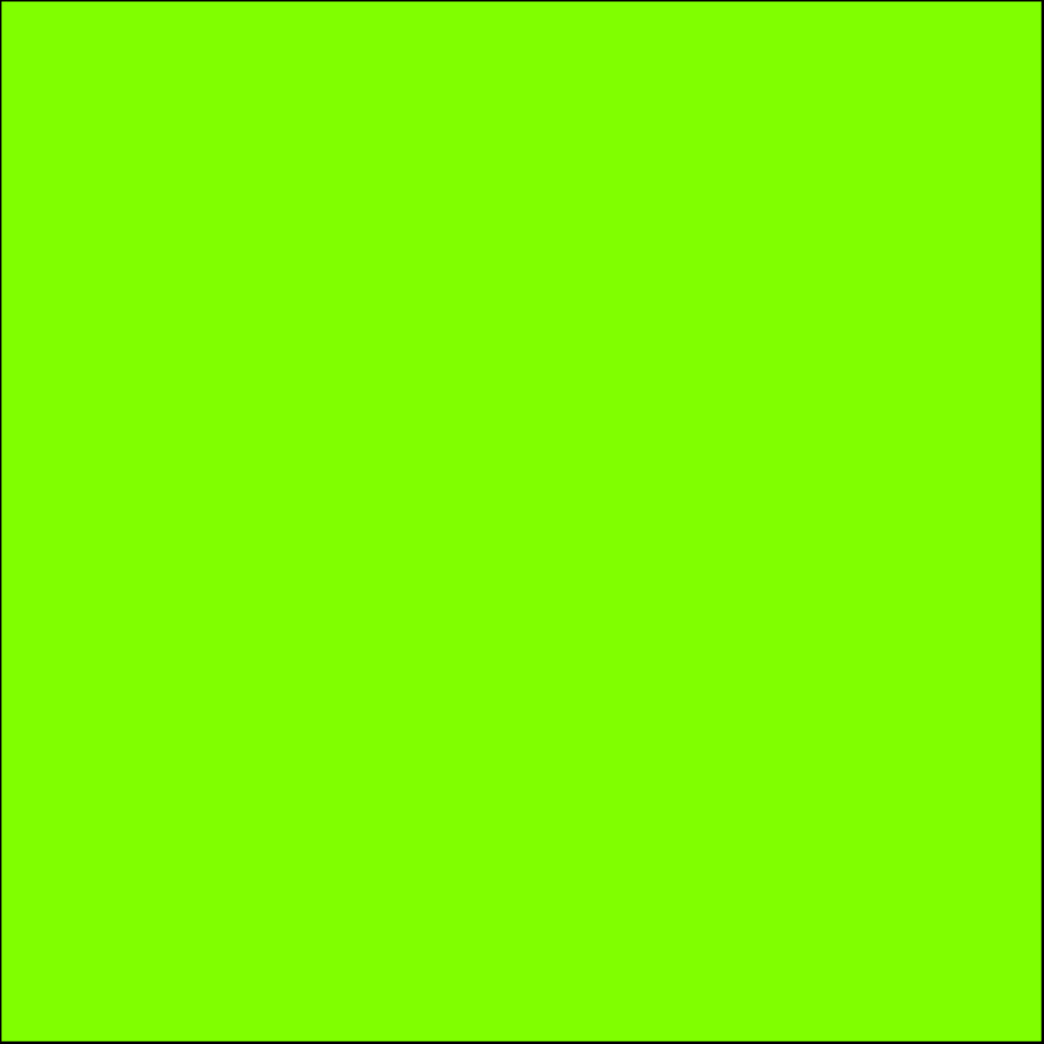}
		\end{minipage}%
	}\hfill
	\subfloat{
		{\footnotesize{~~sky}}
		\begin{minipage}{0.03\linewidth}
			\includegraphics[width=\textwidth]{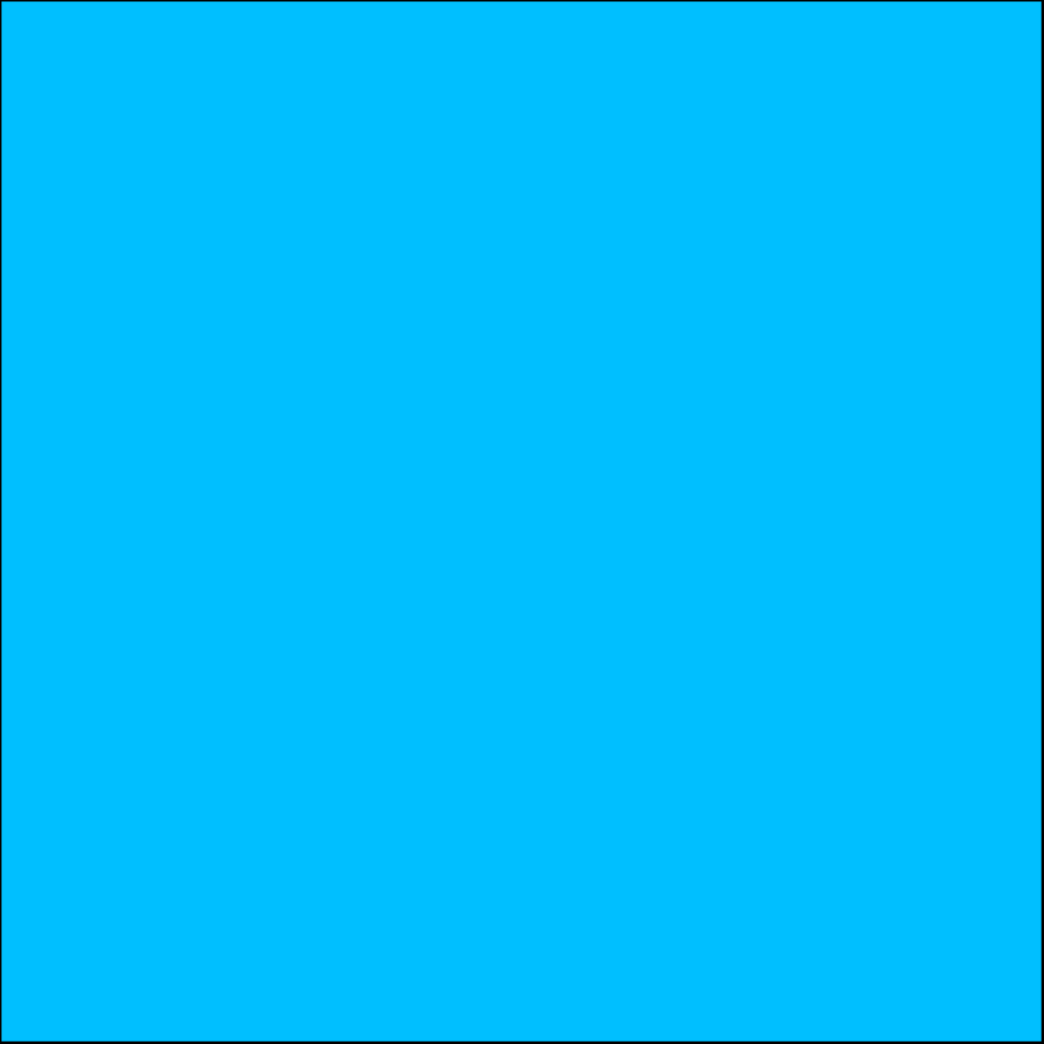}
		\end{minipage}%
	}\hfill
	\subfloat{
		{\footnotesize{~~person}}
		\begin{minipage}{0.03\linewidth}
			\includegraphics[width=\textwidth]{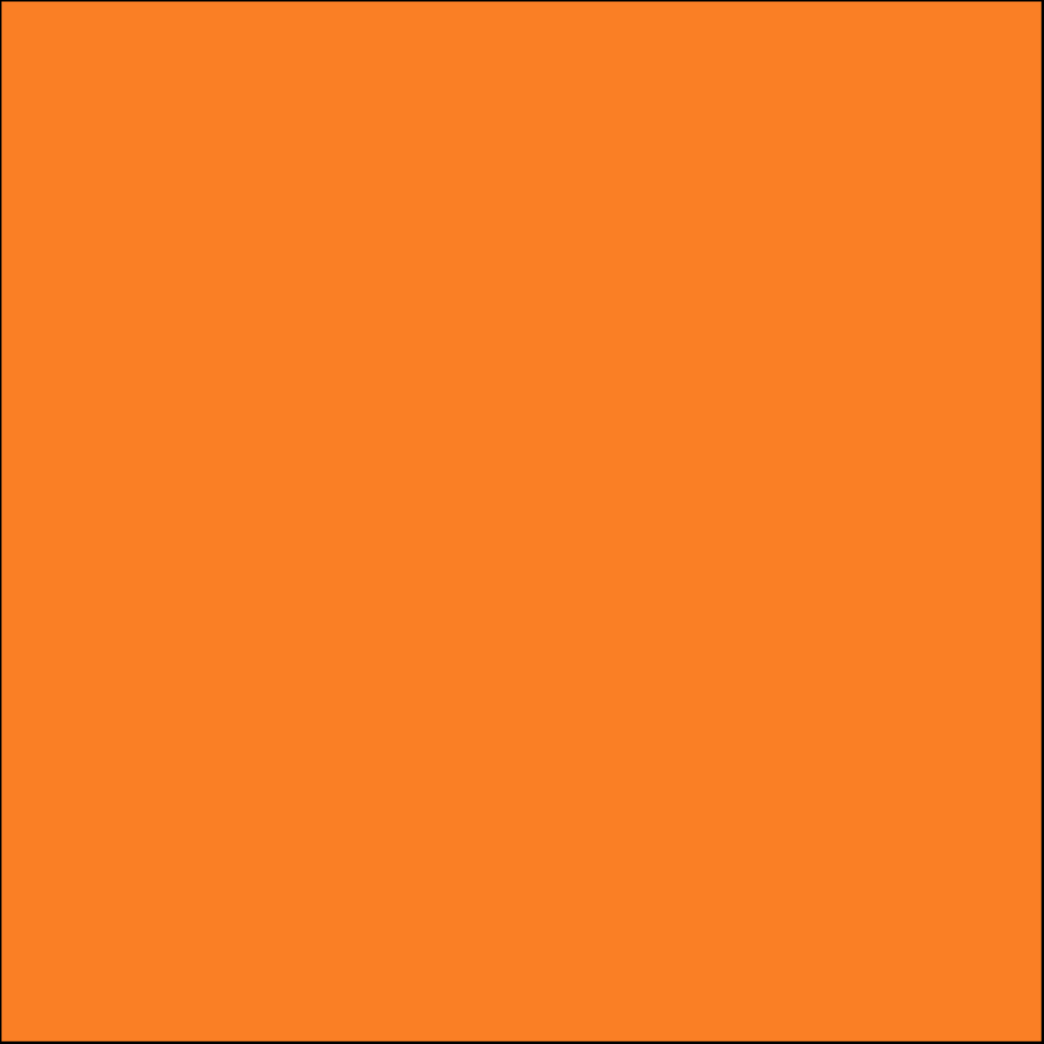}
		\end{minipage}%
	}\hfill
	\subfloat{
		{\footnotesize{~~vehicle}}
		\begin{minipage}{0.03\linewidth}
			\includegraphics[width=\textwidth]{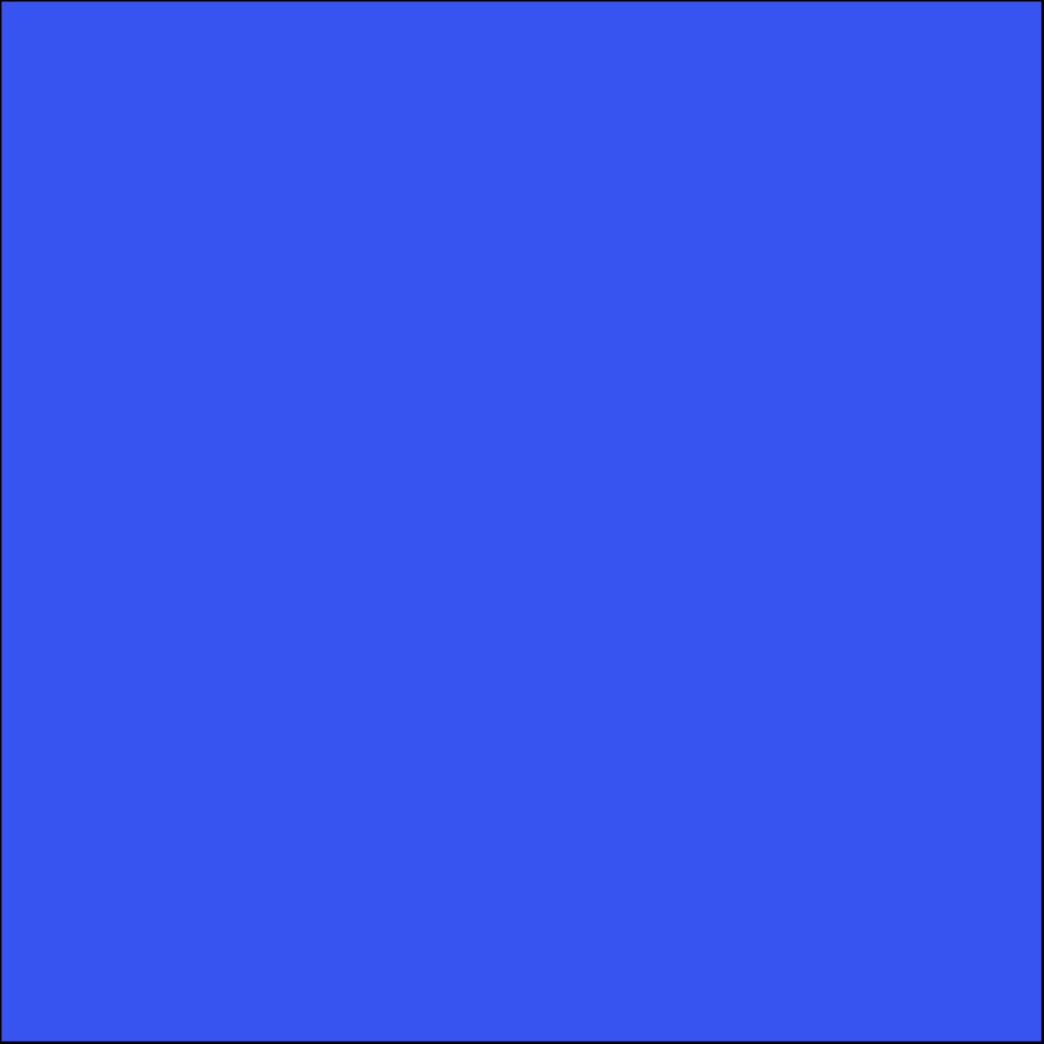}
		\end{minipage}%
	}\hfill
	\caption{Visualized comparisons of the semantic segmentation task. Red circles highlight the examples of errors that do not occur in the proposed SCSC scheme compared with other baselines.}
	\label{visual_seg}
\end{figure*}

\begin{table*} [t]
	\centering
	\caption{Class-wise semantic segmentation accuracy (\%) of different schemes.}
	\begin{tabular}{c|c|c|c|c|c|c|c|c}
		\hline
		Method & mIoU (\%) & ﬂat & construct& object & nature & sky & person & vehicle\\
		\hline
		SCSC (BPG+LDPC1/2+QPSK) & \textbf{56.37} & 69.49 & \textbf{71.03}&\textbf{23.23}&\textbf{68.15}&\textbf{77.85}&\textbf{9.55}&\textbf{63.70}\\
		BPG+LDPC1/2+QPSK  & 30.62&44.59&39.48&10.99&35.00&48.44&5.81&32.13\\
		ProxyNet & 46.83&67.55&60.62&17.71&55.51&68.25&7.84&50.54\\
		ADJSCC & 44.56&{69.87}&57.85&17.11&53.69&67.36&8.47&58.97 \\
		DeepJSCC-MIMO & 54.24&\textbf{70.62}&68.39&19.63&63.52&74.86&8.93&61.68 \\
		\hline
	\end{tabular}
	\label{table0}
\end{table*}

\begin{figure*}[htpb]%
	\centering
	\subfloat[Original Image]{
		\begin{minipage}{0.295\linewidth}
			\centering
			\includegraphics[width=\textwidth]{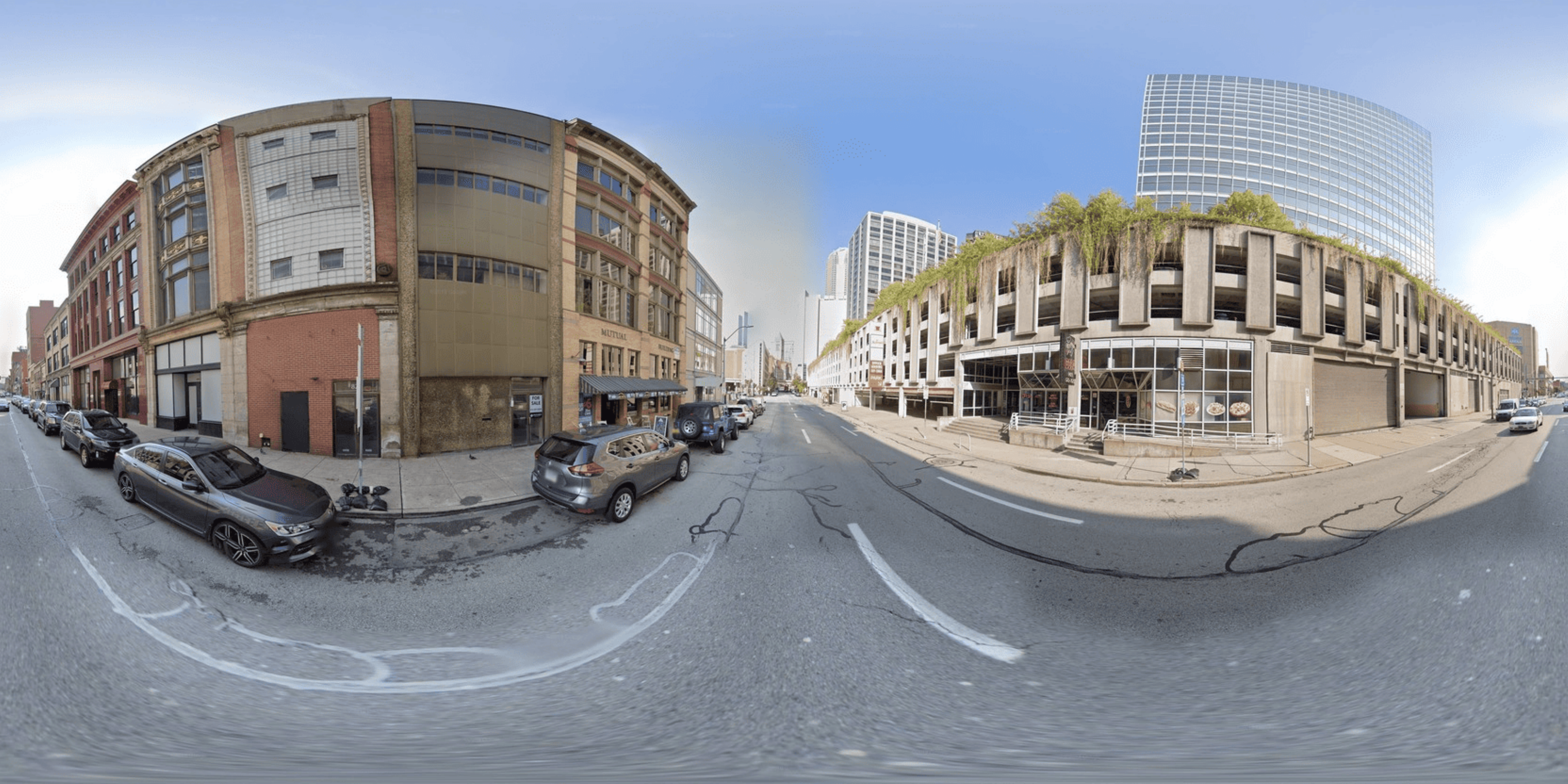}
		\end{minipage}
	}\hfill
	\subfloat[CBR $R=0.053$]{
		\begin{minipage}{0.33\linewidth}
			\centering
			\includegraphics[width=\textwidth]{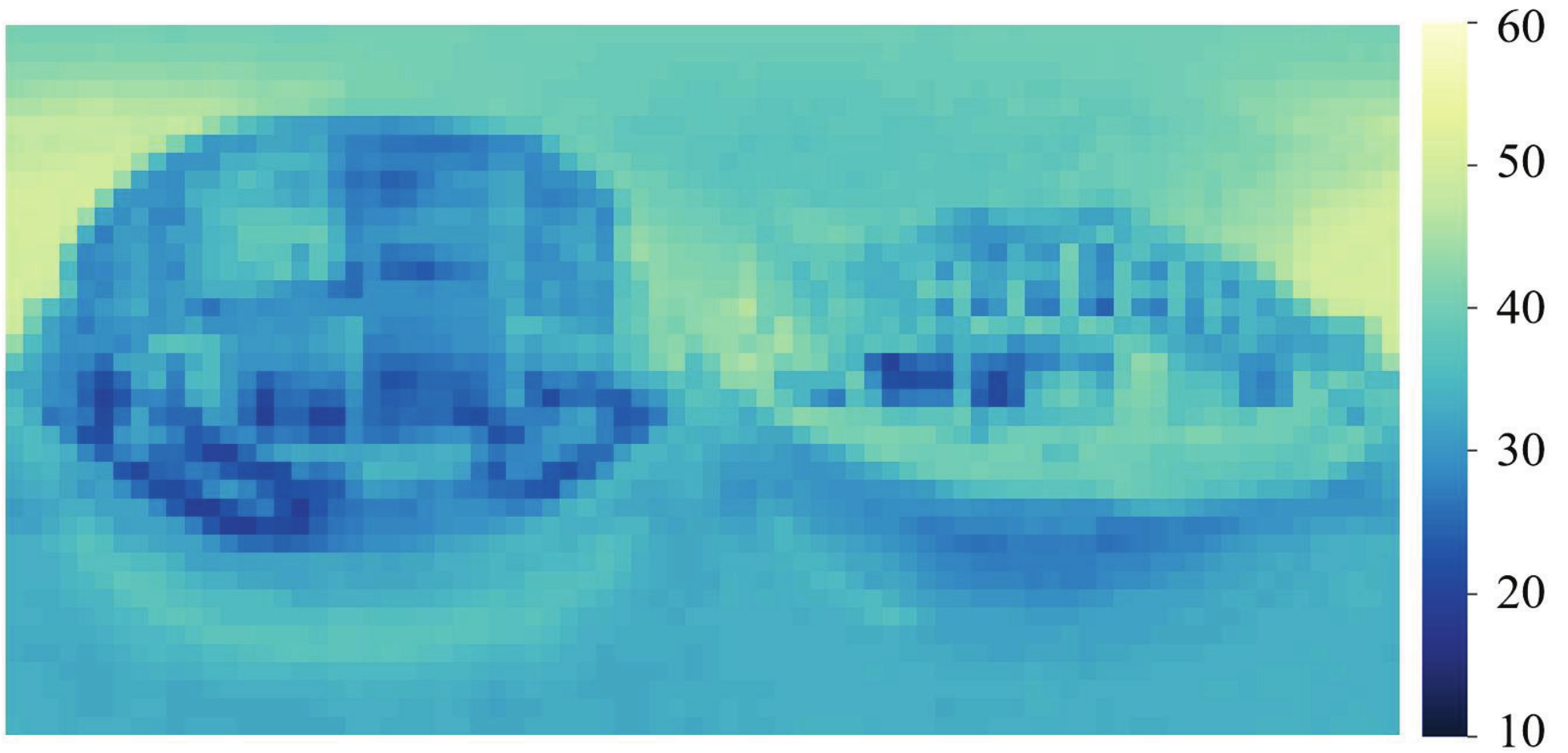}
		\end{minipage}
	}\hfill
	\subfloat[CBR $R=0.126$]{
		\begin{minipage}{0.33\linewidth}
			\centering
			\includegraphics[width=\textwidth]{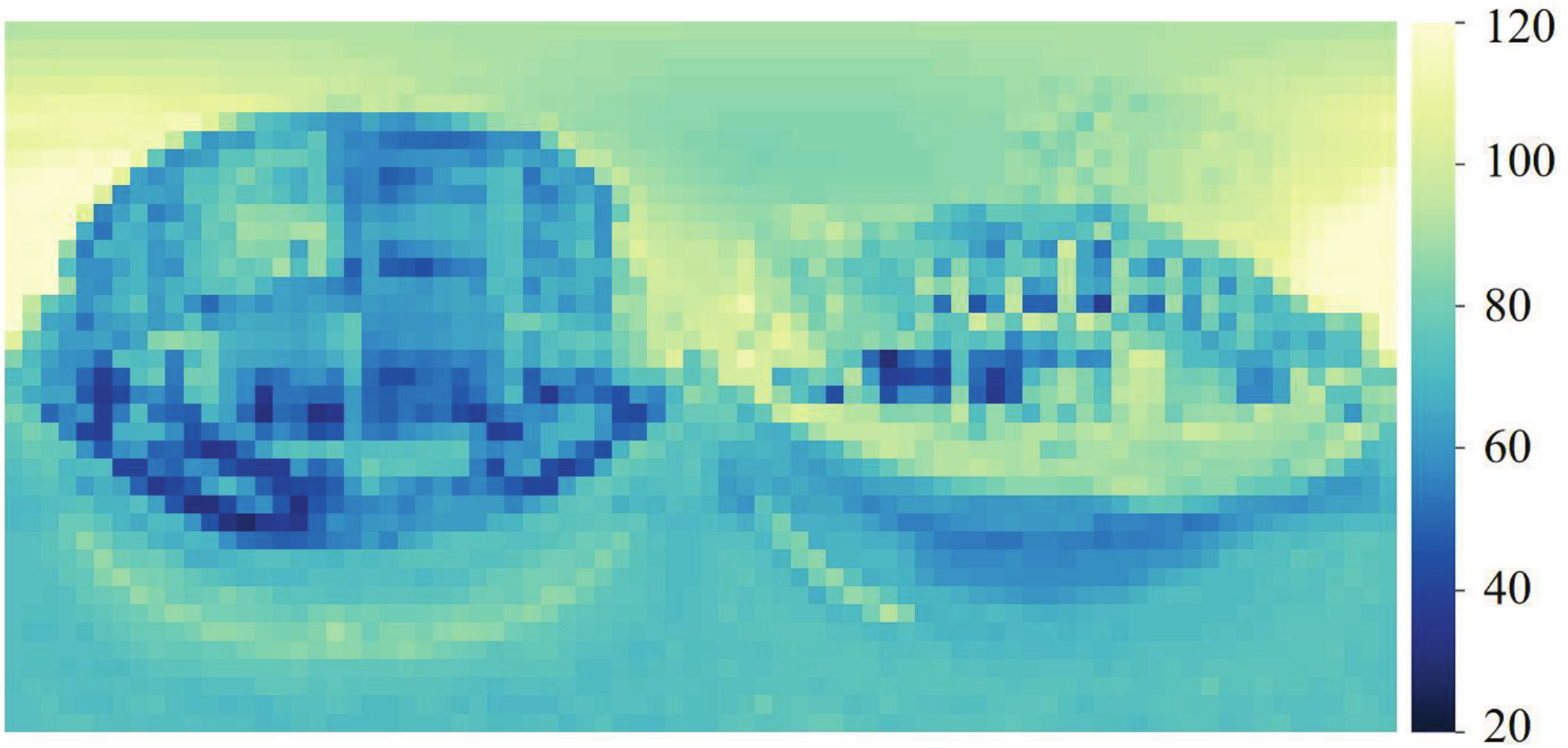} 
		\end{minipage}
	}
	\caption{Visualization of resource allocation map for feature vector over different channel bandwidth ratios. The resolution of the left original image is $1664 \times  832$ while that of both the middle and the right feature images is $78 \times 39$.}
	\label{heatmap}
\end{figure*}

\begin{figure*}[htpb]%
	\centering
	\subfloat[Different compression scheme.]{
		\includegraphics[width=0.35\linewidth]{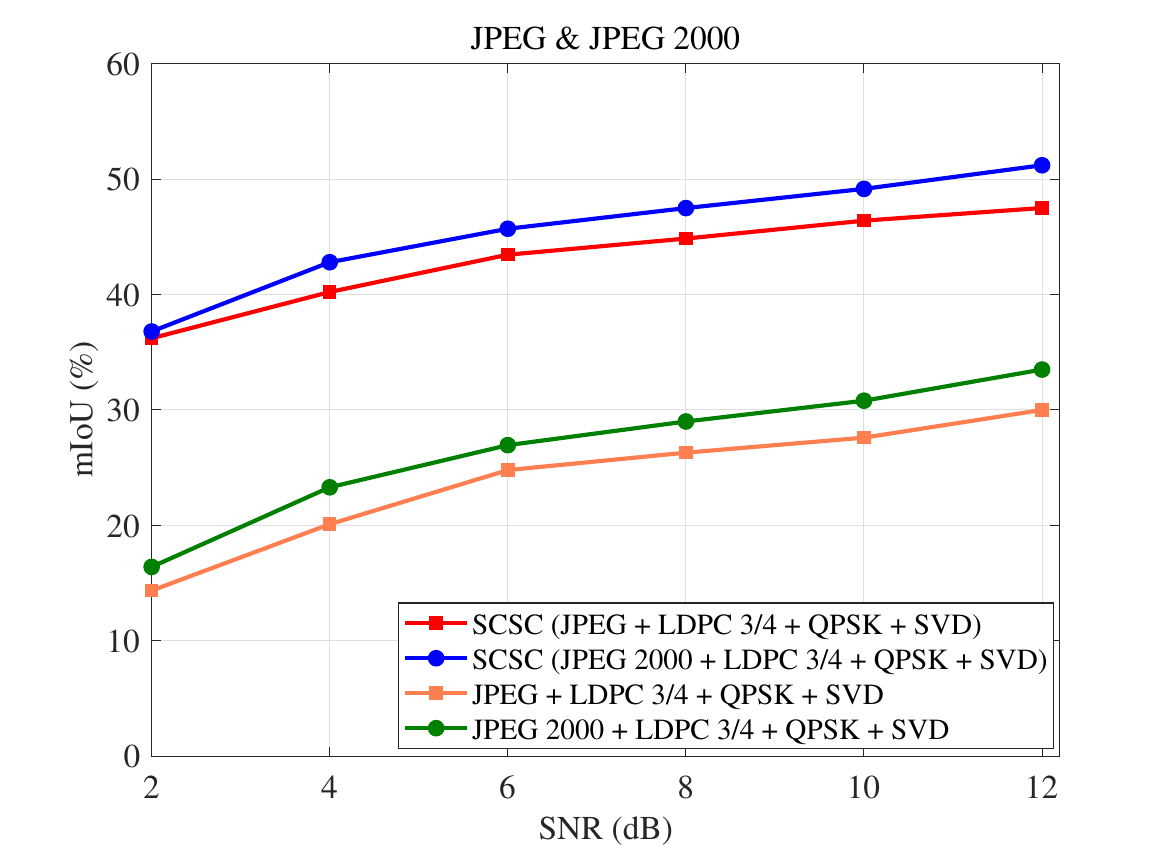}
		\hspace{-1cm}
		\label{miou_JPEG}
	}\hfill
	\subfloat[Coding rate and modulation.]{
		\includegraphics[width=0.35\linewidth]{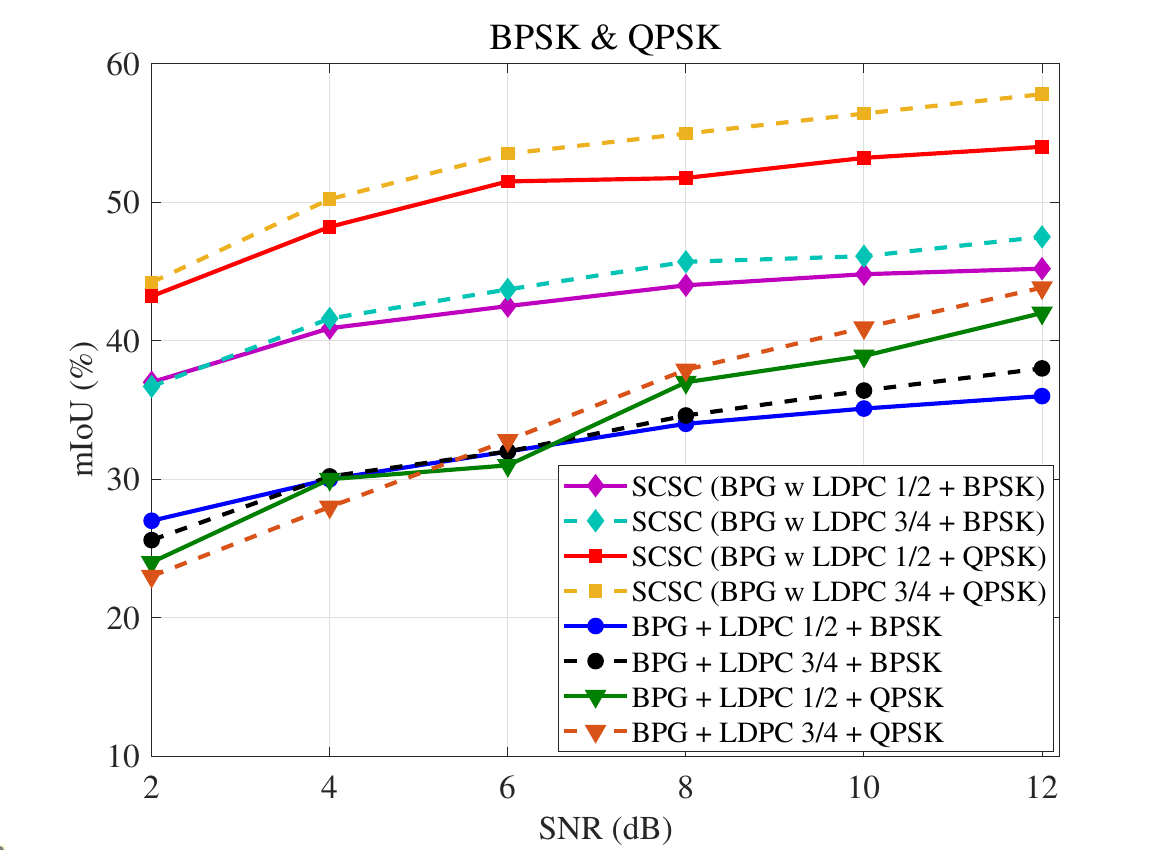}
		\hspace{-1cm}
		\label{miou_BPSK}
	}\hfill
	\subfloat[Different dataset.]{
		\includegraphics[width=0.35\linewidth]{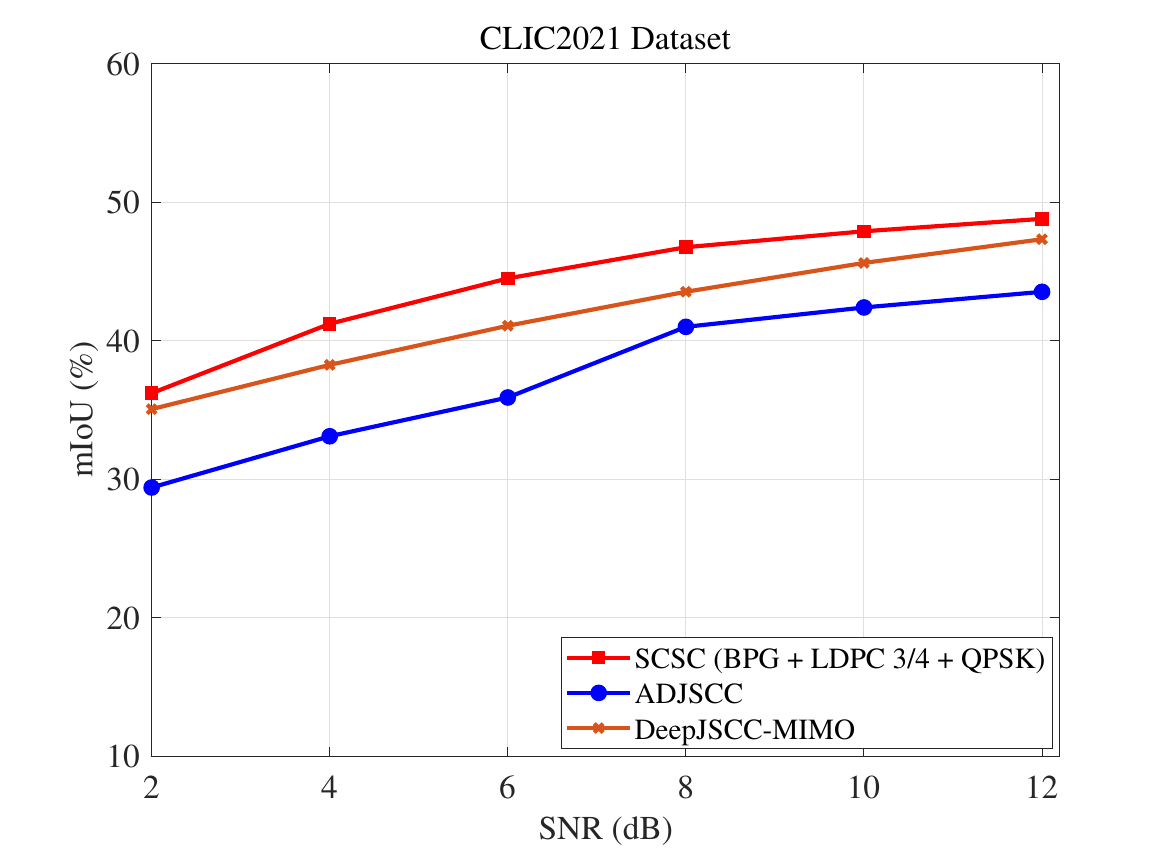} 	
		\label{miou_CLIC}
	}
	\caption{Validation of robustness and generalization. (a) mIoU versus SNR. Comparison of our SCSC (JPEG) and SCSC (JPEG2000) with JPEG and JPEG2000 codecs. (b) Comparison of SCSC trained for different combinations of coding rate and modulation using BPG for source coding, LDPC codes for channel coding, and SVD in the Rayleigh fading channel case. (c) mIoU versus SNR. Comparison of the generalization capability to the CLIC2021 dataset.}
	\label{miou}
\end{figure*}

\subsubsection{mIoU Performance}
Fig. \ref{mIOU} depicts the mIoU performance at different channel SNR and CBR values over the fading channel. Here, we consider the deep learning-based SSCC scheme proposed in \cite{DSSCC} as an additional benchmark, denoted by DSSCC. DSSCC scheme uses the density information of features as side information and employs a variational autoencoder to compress images effectively. The proposed SCSC scheme demonstrates superior performance at most SNRs and CBRs compared to the ADJSCC, ProxyNet, DSSCC, and digital baselines, thanks to its strong adaptability to diverse channel conditions. While our approach slightly underperforms compared to the DeepJSCC-MIMO method, which employs an efficient transformer structure, and freely optimizes the transmitted signal over the channel, our work primarily aims to introduce a framework compatible with existing systems. Notably, the proposed scheme even surpasses DeepJSCC-MIMO in high CBR scenarios. The class-wise segmentation results of different schemes on the CVRG-Pano dataset at $R=1/6$ are presented in Table \ref{table0}. The results indicate that the SCSC scheme not only achieves the best overall mIoU performance, but also outperforms the other schemes in six categories of segmentation. This is because the SCSC approach ensures the restoration of semantic information at the receiver end, greatly avoiding the loss of key information for subsequent tasks upon decompression.

\subsubsection{Visualization Performance}
To intuitively demonstrate the effectiveness of our proposed model, we further provide examples of transmitted and reconstructed images in Fig. \ref{visual1}. It can be observed that SCSC can achieve comparable visual quality with no more channel bandwidth cost than other baselines. 
Moreover, we also show the visualization result of downstream semantic segmentation by SCSC, ADJSCC, and the standard scheme in Fig. \ref{visual_seg}, and it is evident that the SCSC scheme improves the downstream task accuracy. For example, the scene of the reconstructed images produced by our method in the first row can be correctly parsed while the corresponding result from the standard scheme failed, which indicates
the superiority of applying semantics for the efficient data compression and transmission.

Table \ref{table1} presents the BD-CBR and BD-PSNR results, where BD-CBR is used to measure the percentage of saved CBR with the same accuracy, and BD-PSNR indicates the image transmission quality improvement \cite{BDBR}. Here, we set the standard scheme as the benchmark and the BD-X results of the other schemes represent the relative value with respect to the benchmark performance. The results demonstrate that the proposed SCSC method achieves 29.46\% bandwidth savings at channel SNR of 10 dB, or 1.28 dB gain in average PSNR at a CBR of 1/12. As previously analyzed, the DeepJSCC-MIMO scheme shows a 12.19\% increase in the bandwidth cost with 1.08 PSNR gains. ADJSCC exhibits suboptimal performance when applied to high-resolution datasets, leading to a notable 48.65\% increase in the bandwidth cost.

\begin{table} [t]
	\centering
	\caption{A Comparative Analysis of BD-CBR and BD-PSNR Performances.}
	\begin{tabular}{c|c|c}
		\hline
		Method & BD-CBR & BD-PSNR  \\
		\hline
		SCSC(BPG+LDPC1/2+QPSK) & \textbf{-29.46\%} & \textbf{1.28} \\
		BPG+LDPC1/2+QPSK & 0.00\% & 0 \\
		ProxyNet  & 27.56\% & 0.15 \\
		ADJSCC & 48.65\% & 0.72 \\
		DeepJSCC-MIMO & 12.19\% & 1.08\\
		\hline
	\end{tabular}
	\label{table1}
\end{table}

\subsubsection{Resource Allocation Map}
We investigate the resource allocation map to describe the spatial dependencies among the elements of the filtered image. Fig. \ref{heatmap} plots the visualization of the bit allocation for the filtered image $\mathbf{x}$ over a Rayleigh fading channel for SNR = 10 dB. It is observed that the pixels of the same object, e.g., road, sky, and building, have similar bits, whereas the boundaries and contours between different objects possess more distinctive bits. In addition, as the channel bandwidth ratio increases, more bits are allocated to capture intricate object details, leading to a higher fidelity in the reconstructed image. 

\subsection{Robustness and Generalization}
\subsubsection{Trained SCSC with Different Codecs}
To validate the robustness and compatibility of our proposed framework with other digital baselines, we provide more experimental results in Fig. \ref{miou}\subref{miou_JPEG}. Here, we utilize JPEG and JPEG2000 methods to replace BPG within the SCSC framework, without any fine-tuning of the other modules and parameters. Experiment results show that PPEN and PCEN can save more than 10.2\% and 13.5\% channel bandwidth compared with the original JPEG codec and JPEG2000 codec when evaluated on the ERF-PSPNet backbone network, respectively. 

\subsubsection{Evaluation of SCSC with Different Combinations of Coding Rate and Modulation}
Additionally, we consider different combinations of channel coding rate and modulation. Fig. \ref{miou}\subref{miou_BPSK} shows the result of SCSC trained for different coding rates and modulation and tested over a range of channel SNR values. We can observe that the performance gets better at low SNRs by using a lower rate scheme which is consistent with the standard codec results.

\begin{figure}[tpb]%
	\centering
	\includegraphics[width=0.9\linewidth]{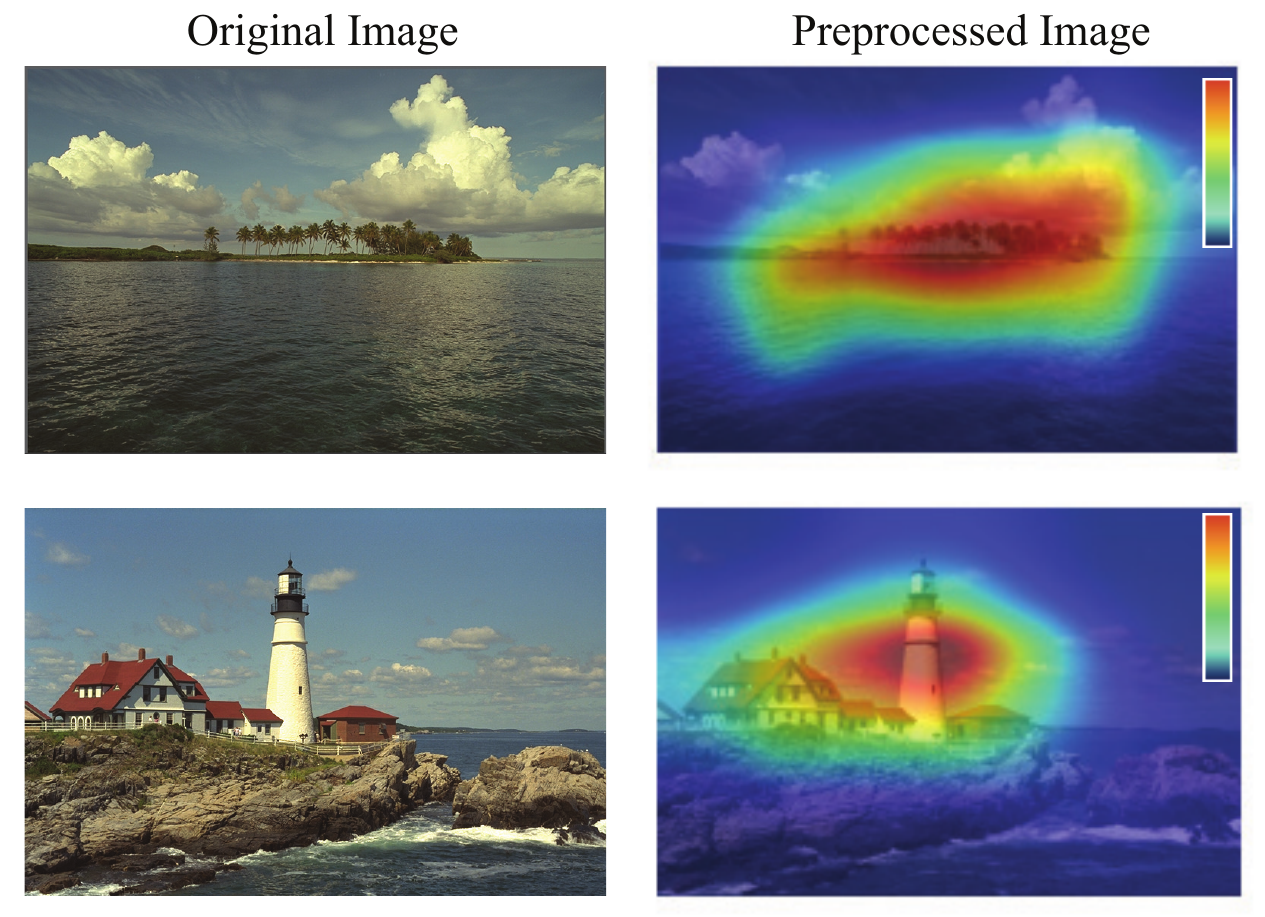}
	\caption{Visualization results of the PPEN. The first column is the original image and the second column is the filtered image after preprocessing. Larger values are denoted by red color.}
	\label{heatmap_kodak}
\end{figure}

\subsubsection{Trained SCSC with Different Datasets}
In order to evaluate the model generalizability and its independence on source data, we further evaluate the proposed SCSC framework on a different dataset (e.g., CLIC2021 dataset \cite{CLIC2021}) over SNR $\in$ [2, 12] dB and $R$ = 1/12. As shown in Fig. \ref{miou}\subref{miou_CLIC}, compared with the ADJSCC and DeepJSCC-MIMO schemes, SCSC maintains better task performance across the whole SNR range. It indicates that when we directly deploy the trained framework without adjustment to different source datasets, SCSC is more compatible and robust. The performance gain mainly comes from the proposed modules and the standard codecs that are not specifically designed for a particular source dataset. Moreover, based on the GradCAM \cite{GradCAM} method, we give an example in Fig. \ref{heatmap_kodak} to show the effectiveness of our preprocessing module on the Kodak dataset \cite{kodak}. We can observe that PPEN preserves the important semantic information with discarded information mainly distributed in the background region where higher values represent more importance for downstream tasks.

\begin{figure}[tbp]%
	\centering
	\subfloat[mAP versus CBR]{
		\label{detection_CBR}
		\includegraphics[width=0.95\linewidth]{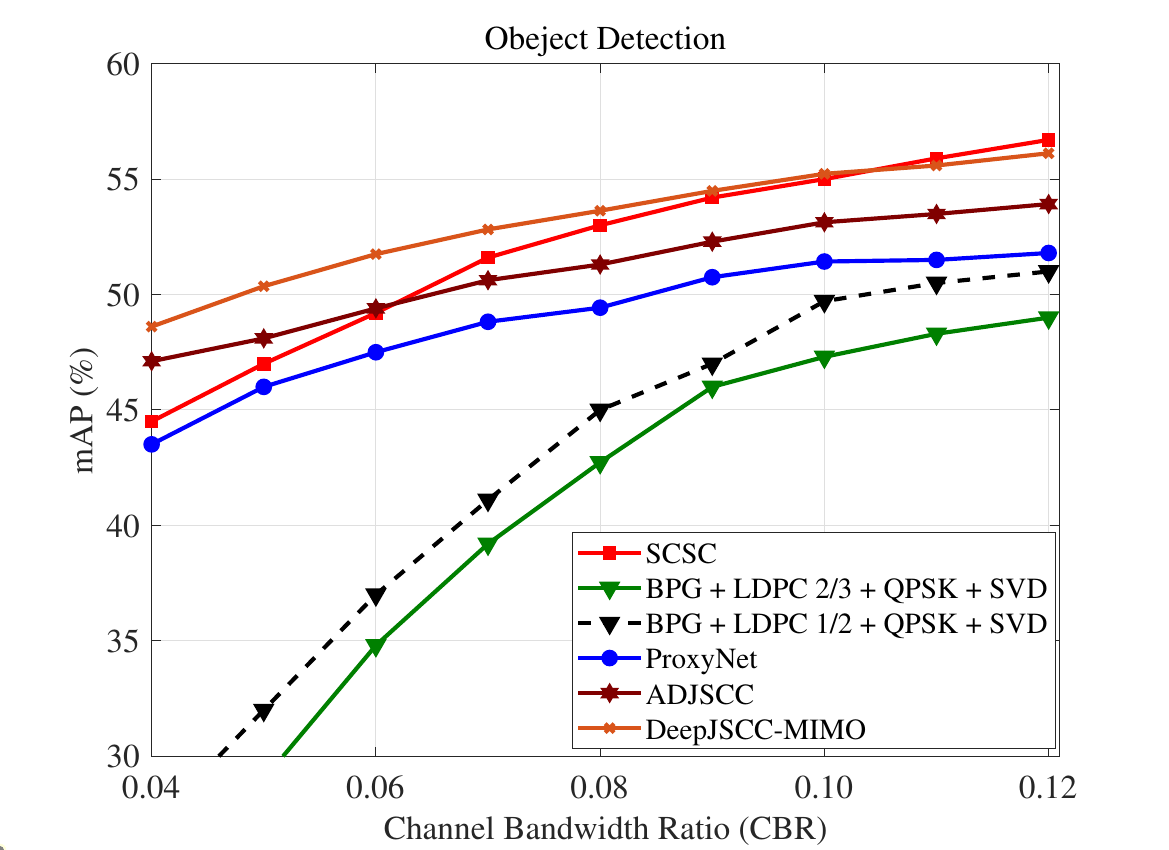}
	}\hfill
	\subfloat[Accuracy versus CBR]{
		\label{classification_CBR}
		\includegraphics[width=0.95\linewidth]{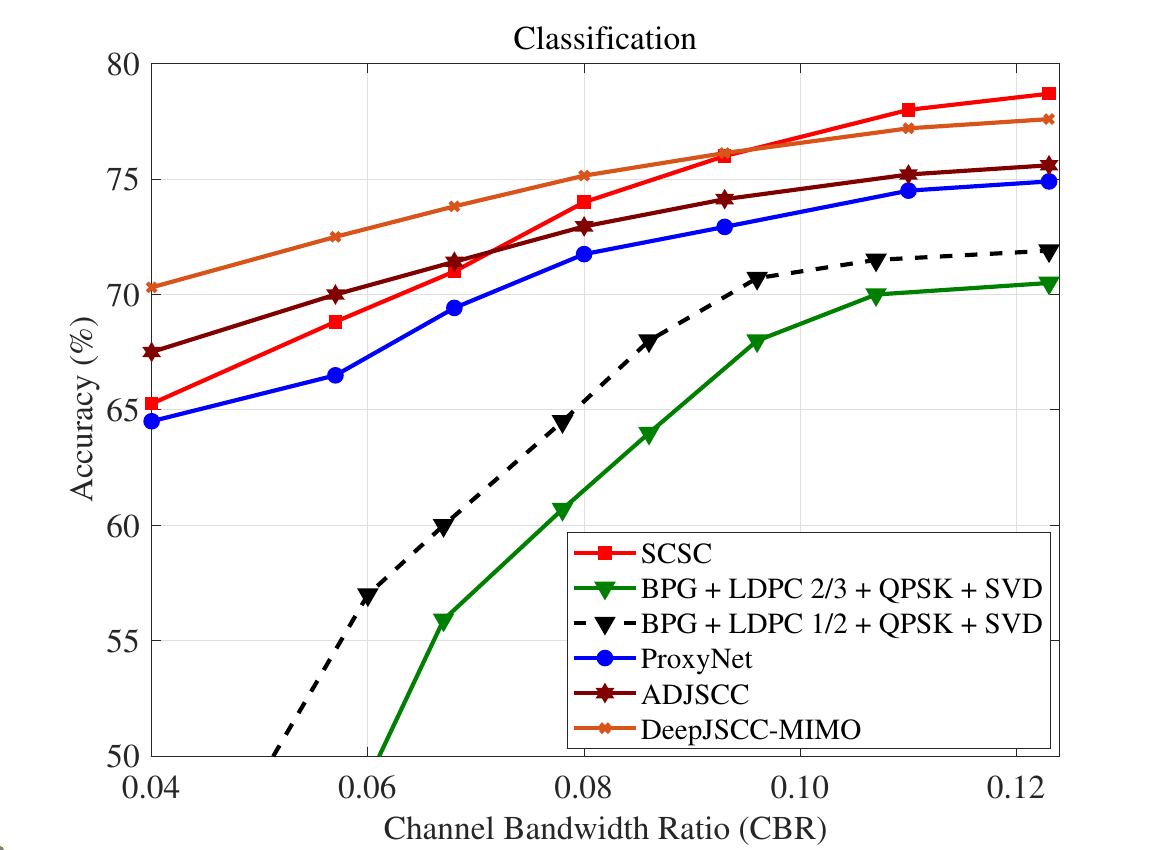}
	}
	\caption{Performance of object detection and image classification over MIMO fading channels; (a) Object detection task performance versus CBR with SNR = 10 dB; (b) Classification task performance versus CBR with SNR = 10 dB.}
	\label{task_CBR}
\end{figure}

\subsubsection{Evaluation of SCSC with Different Downstream Tasks}
To provide a detailed analysis of the effectiveness and generality of SCSC on different downstream tasks, we conduct experiments on object detection and image classification tasks by leveraging EfficientDet \cite{EfficientDet} and ResNet50 \cite{Resnet50} machine analysis networks, respectively. We train and test the corresponding SCSC networks for both classification and detection tasks at SNR = 10 dB, with 2/3 LDPC code and $Q=34$.\\
i) \textit{Object detection.} For the object detection task, we utilize the COCO dataset and the mean average precision (mAP) results as the evaluation metric. As shown in Fig. \ref{task_CBR}\subref{detection_CBR}, the SCSC scheme shows a much better performance compared to ADJSCC, ProxyNet, and digital baselines, while closely aligning with the performance of DeepJSCC-MIMO scheme in high CBR values. Specifically, compared with the ADJSCC scheme, SCSC saves approximately 22.6\% channel bandwidth at the same mAP value. \\
ii) \textit{Image classification.} Fig. \ref{task_CBR}\subref{classification_CBR} shows the top-1 accuracy curves from different schemes on the ImageNet dataset. It is noted that our approach still achieves better task performance and saves more than 23\% channel bandwidth when compared with traditional digital schemes by evaluating it on the ResNet50 model. We can conclude that the non-task-specific and non-source-specific digital communication system can allow our SCSC scheme to perform well on other never-seen tasks for a wide range of channel bandwidth conditions.

\subsubsection{Evaluations of Different MIMO Channel Modes}
\begin{figure}[tbp]%
	\centering
	\includegraphics[width=0.95\linewidth]{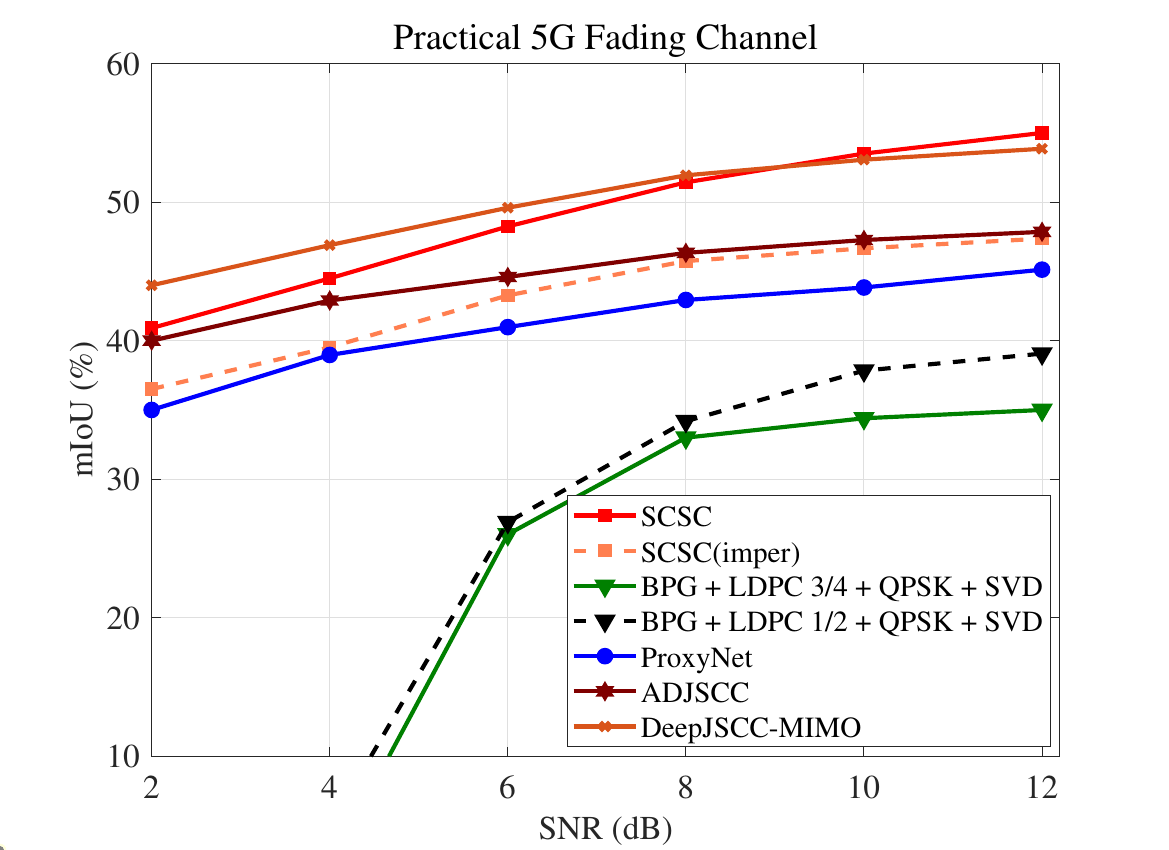}
	\caption{Comparisons of the task performance over a practical fading channel sampled from practical 5G MIMO channel model with $R = 1/12$.}
	\label{channel_SNR}
\end{figure}

To further evaluate the flexibility and robustness of SCSC, we analyze the task performance over the practical 5G MIMO fading channel model \cite{MIMOCSI}. To thoroughly assess the impact of channel estimation error on the system performance, we also conduct simulations under imperfect CSI scenario with channel estimation error following $\mathcal{CN}(0, 0.1)$, labeled as ``SCSC $ (\mathrm{imper})$''. As depicted in Fig. \ref{channel_SNR}, the performance of all the schemes increases with SNR. Notably, when SNR = 6 dB, SCSC achieves about 4.4\% and 23.9\% improvement in mIoU compared to ADJSCC and SSCC schemes, respectively. Moreover, we can observe that, even though the performance of the SCSC system drops with imperfect channel estimation, it still outperforms conventional communication methods. This further demonstrates the effectiveness of our system in handling imperfect channel estimation and maintaining superior performance, which makes it efficient and compatible for practical applications.

\begin{figure}[tpb]%
	\centering
	\includegraphics[width=0.95\linewidth]{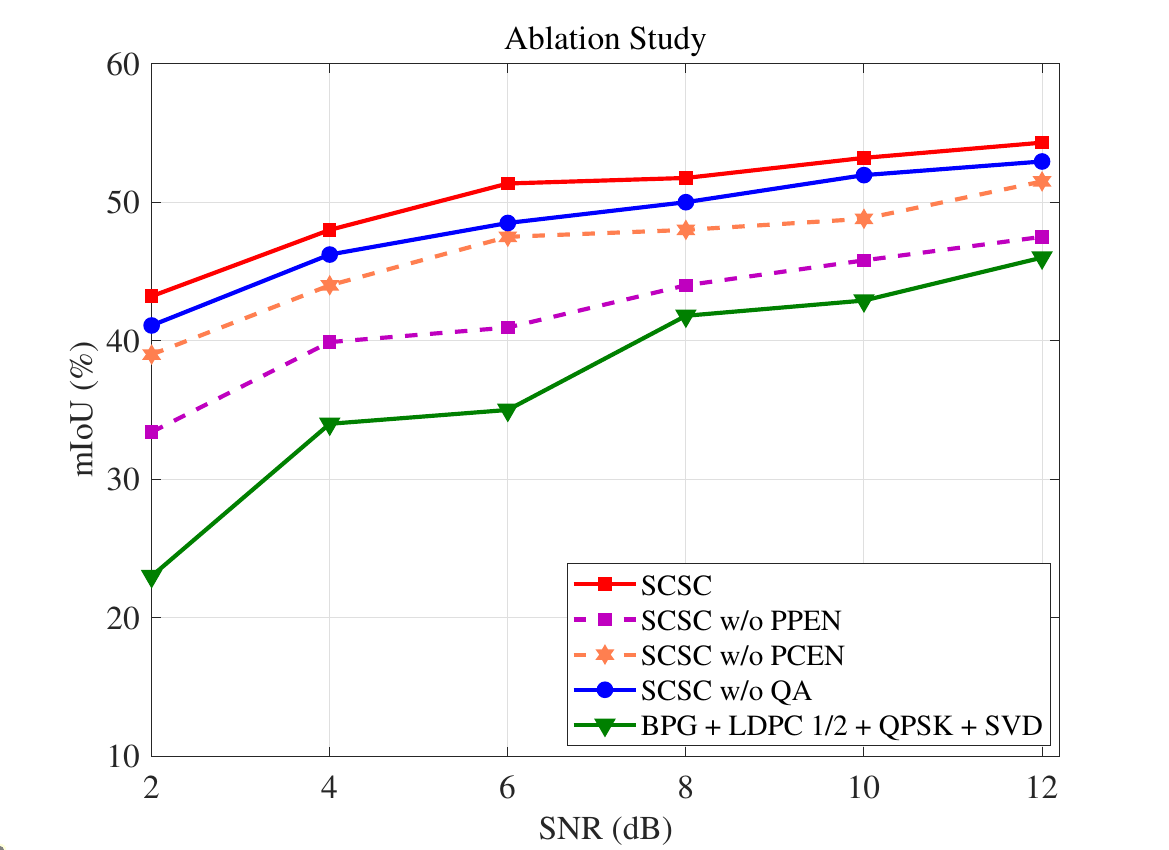}
	\caption{Performance of different ablation schemes to validate the effectiveness of proposed QA layers, PPEN, and PCEN.}
	\label{Ablation_SNR}
\end{figure}

\subsection{Ablation Study}
For SCSC, we employ 1/2 rate LDPC code with $\mathcal Q=37$, $R=1/12$, QPSK modulation, and SVD. The results of SCSC without PPEN (``SCSC w/o PPEN") and SCSC without PCEN (``SCSC w/o PCEN") are presented in Fig. \ref{Ablation_SNR} to verify the benefits of the proposed modules, where the other modules are already optimized without fine-tuning. Compared to the purely digital scheme, the collaborative integration of PPEN and PCEN brings notable performance gains and achieves 20.5\% savings in channel bandwidth. Moreover, the results indicate the effectiveness of PPEN for preserving semantics, and PCEN for efficient transmission, both of which contribute significantly to the improvement of segmentation performance.
Furthermore, our proposed PPEN module is quantization-adaptive and can be used for the standard BPG codec with different quantization values which can be reflected in the CBR value. Here, we conduct an ablation study called ``SCSC w/o QA'' to validate the effectiveness of QA layers, where we remove the quantization adaptive layers and train a set of different PPEN modules for different $\mathcal Q$ values.
We can observe the trend that SCSC exceeds SCSC w/o QA scheme, illustrating the performance gain brought by QA layers. In this way, the proposed modules are verified effective during the wireless MIMO image transmission.

\begin{table} [htbp]
	\centering
	\caption{Evaluation of system throughput.}
	\begin{tabular}{c|c}
		\hline
		Method & Throughput  \\
		\hline
		SCSC (BPG+LDPC2/3+QPSK) & 125.8\\
		BPG+LDPC2/3+QPSK & 6.7  \\
		ProxyNet  & 147.3 \\
		ADJSCC & 225.6 \\
		DeepJSCC-MIMO & 318.8 \\
		\hline
	\end{tabular}
	\label{throughput}
\end{table}

\subsection{Running Time and Complexity} 

The number of parameters of our PPEN module is 12.68M. For any 2048 $\times$ 400 images, the inference time of our PPEN module is only 6.32ms, which means it brings little computational complexity increase to the existing digital transmission framework. On the other hand, the PCEN inherits the superiority of the iterative precoder and DL-based optimization to achieve its improved performance at the expense of a slight increase in computational complexity compared to the standard precoder. The SCSC framework can effectively conserve resources and reduce transmission time while ensuring the quality of panoramic image transmission and downstream task performance. To better evaluate the benefits of SCSC in terms of its processing speed, we evaluate the system throughput to measure the inference speed, which is defined as the number of images that are transmitted and processed at the receiver per second. As shown in Table \ref{throughput}, SCSC scheme significantly outperforms the digital communication scheme due to the algorithm optimization and structural design. Since ProxyNet and ADJSCC are thorough CNN-based frameworks, the inference speed can be much faster than the traditional digital schemes. Due to the fact that the DeepJSCC-MIMO scheme adopts the transformer backbone, it exhibits superior experimental performance while also having a more complicated structure and tolerable extra computation costs. In practical deployments of SCSC, the application of parallel computing and GPU acceleration techniques can significantly enhance the implementation efficiency of the system.

\section{Conclusion}
In this paper, we proposed a standards-compatible semantic communication framework called SCSC for better downstream task performance. We propose the PPEN module to achieve a better trade-off between coding bitrate and the performance of machine vision tasks. Additionally, the PCEN module enables compatibility with finite-alphabet finite-blocklength signals, increasing the throughput over the MIMO system. Our framework considers existing compression and channel coding modules as non-trainable blocks, and treats them as part of the channel. We propose a proxy network to deal with the non-differentiability of standard codecs, and to enable gradient backpropagation for the end-to-end training of PPEN and PCEN modules. Numerical results demonstrate significant improvements in performance in terms of both the reconstruction quality of the image and the segmentation task, compared with the state-of-the-art models across a wide range of SNRs. We have also carried out extensive experiments to show the robustness and flexibility of the proposed framework. The results show that the superior performance of SCSC generalizes to unseen tasks and datasets, and it can be combined with other compression and coding schemes.


\begin{thebibliography}{1}
\bibitem{Yang_2022}
W. Yang, H. Du, Z. Q. Liew, W. Y. B. Lim, Z. Xiong, D. Niyato, X. Chi, X. Shen, and C. Miao, ``Semantic communications for future internet: Fundamentals, applications, and challenges,'' \emph{IEEE Commun. Surv. Tutor.}, vol. 25, no.1, pp. 213--250, Nov. 2022.

\bibitem{ping_sem}
P. Zhang, W. Xu, H. Gao, K. Niu, X. Xu, X. Qin, C. Yuan, Z. Qin, H. Zhao, J. Wei, and F. Zhang, ``Toward wisdom-evolutionary and primitive-concise 6G: A new paradigm of semantic communication networks,'' \emph{Engineering}, vol. 8, pp. 60--73, Jan. 2022.

\bibitem{Gunduz_sem}
D. G{\"u}nd{\"u}z, Z. Qin, I. E. Aguerri, H. S. Dhillon, Z. Yang, A. Yener, and C. B. Chae, ``Beyond transmitting bits: Context, semantics, and task-oriented communications,” \emph{IEEE J. Sel. Areas Commun.}, vol. 41, no. 1, pp. 5--41, Jan. 2022.

\bibitem{DJSCC}
E. Bourtsoulatze, D. B. Kurka, and D. G{\"u}nd{\"u}z, ``Deep joint source-channel coding for wireless image transmission,'' in \emph{Proc. IEEE Int. Conf. Acoust. Speech Signal Process. (ICASSP)}, Brighton, UK, May 2019, pp. 4774--4778.

\bibitem{Jankowski}
M. Jankowski, D. G{\"u}nd{\"u}z and K. Mikolajczyk, ``Wireless image retrieval at the edge,'' \emph{IEEE J. Sel. Areas Commun.}, vol. 39, no. 1, pp. 89--100, Jan. 2021.

\bibitem{Weng}
Z. Weng, Z. Qin, X. Tao, C. Pan, G. Liu, and G. Y. Li, ``Deep learning enabled semantic communications with speech recognition and synthesis,'' \emph{IEEE Trans. Wireless Commun.}, vol. 22, no. 9, pp. 6227--6240, Sept. 2023.

\bibitem{NTSCC}
J. Dai, S. Wang, K. Tan, Z. Si, X. Qin, K. Niu, and P. Zhang, ``Nonlinear transform source-channel coding for semantic communications,'' \emph{IEEE J. Sel. Areas Commun.}, vol. 40, no. 8, pp. 2300--2316, Jun. 2022.

\bibitem{DJSCC-f}
D. Burth Kurka and D. G{\"u}nd{\"u}z, ``DeepJSCC-f: Deep joint source-channel coding of images with feedback,'' \emph{IEEE J. Sel. Areas Info. Theory}, vol. 1, no. 1, pp. 178--193, May 2020.


\bibitem{Wu_2024}
H. Wu, Y. Shao, C. Bian, K. Mikolajczyk, and D. G{\"u}nd{\"u}z, ``Deep joint source-channel coding for adaptive image transmission over MIMO channels,'' \emph{IEEE Trans. Wireless Commun.}, vol. 23, no. 10, pp. 15002--15017, Oct. 2024.

\bibitem{11_DL}
X. Luo, B. Yin, Z. Chen, B. Xia, and J. Wang, ``Autoencoder-based semantic communication systems with relay channels,'' in \emph{Proc. IEEE Int. Conf. Commun. Workshops (ICC Workshops)}, Seoul, South Korea, May 2022, pp. 711--716.

\bibitem{Bian_relay}
C. Bian, Y. Shao, H. Wu, and D. G{\"u}nd{\"u}z, ``Deep joint source-channel coding over cooperative relay networks,'' \emph{arXiv:2211.06705}, Nov. 2022. [Online]. Available: {https://arxiv.org/abs/2211.06705}



\bibitem{15_DL}
Q. Hu, G. Zhang, Z. Qin, Y. Cai, G. Yu, and G. Y. Li, ``Robust semantic communications with masked VQ-VAE enabled codebook,'' \emph{IEEE Trans. Wireless Commun.}, early access, Apr. 2023.

\bibitem{Jiang}
P. Jiang, C.-K. Wen, S. Jin, and G. Y. Li, ``Deep source-channel coding for sentence semantic transmission with HARQ,'' \emph{IEEE Trans. Commun.}, vol. 70, no. 8, pp. 5225--5240, Aug. 2022.

\bibitem{Tung_Jun_2022}
T. -Y. Tung, D. B. Kurka, M. Jankowski, and D. G{\"u}nd{\"u}z, ``DeepJSCC-Q: constellation constrained deep joint source-channel coding,''  \emph{IEEE J. Sel. Areas Inf. Theory}, vol. 3, no. 4, pp. 720--731, Dec. 2022.

\bibitem{Huang_2023}
J. Huang, D. Li, C. Huang, X. Qin, and W. Zhang, ``Joint task and data oriented semantic communications: A deep separate source-channel coding scheme,'' \emph{arxiv:2302.13580}, Jun. 2023. [Online]. Available: {https://arxiv.org/abs/2302.13580}


\bibitem{Yao_2022}
S. Yao, S. Wang, J. Dai, K. Niu, and P. Zhang, ``Versatile semantic coded transmission over MIMO fading channels,'' \emph{arxiv:2210.16741}, Oct. 2022. [Online]. Available: {https://arxiv.org/abs/2210.16741}

\bibitem{Tung_ICC}
T. -Y. Tung and D. G{\"u}nd{\"u}z, ``Deep joint source-channel and encryption coding: Secure semantic communications,'' in \emph{IEEE Int. Conf. Commun. (ICC)}, Rome, Italy, May 2023, pp. 5620--5625.

\bibitem{Schaar_2005}
M. van Der Schaar and Sai Shankar N, ``Cross-layer wireless multimedia transmission: challenges, principles, and new paradigms,'' \emph{IEEE Wireless Commun.}, vol. 12, no. 4, pp. 50--58, Aug. 2005.

\bibitem{Zhai_2007}
F. Zhai and A. K. Katsaggelos, ``Joint source-channel video transmission,'' in \emph{Synthesis Lectures on Image, Video, and Multimedia Processing Series}, Series Editor: Al Bovik, Morgan \& Claypool Publishers, Sept. 2007.


\bibitem{edge_2022}
W. Yang, Z. Q. Liew, W. Y. B. Lim, Z. Xiong, D. Niyato, X. Chi, X. Cao, and K. B. Letaief, ``Semantic communication meets edge intelligence,'' \emph{IEEE Wireless Commun.}, vol. 29, no. 5, pp. 28--35, Oct. 2022.

\bibitem{Luguo}
G. Lu, X. Ge, T. Zhong, J. Geng, and Q. Hu, ``Preprocessing enhanced image compression for machine vision,'' \emph{arxiv:2206.05650}, Jun. 2022. [Online]. Available: {https://arxiv.org/abs/2206.05650}

\bibitem{Yangmingyi}
M. Yang, L. Herranz, F. Yang, L. Murn, M. G. Blanch, and S. Wan, ``Semantic preprocessor for image compression for machines,'' in \emph{Proc. IEEE Int. Conf. Acoust. Speech Signal Process. (ICASSP)}, Rhodes Island, Greece, Jun. 2023, pp. 1--5.

\bibitem{MIMO}
Y. Wang, Z. Gao, D. Zheng, S. Chen, D. G{\"u}nd{\"u}z and H. V. Poor, ``Transformer-empowered 6G intelligent networks: from massive MIMO processing to semantic communication,'' \emph{IEEE Wireless Commun.}, early access, Feb. 2022.

\bibitem{Kang_2022}
J. Kang, H. Du, Z. Li, Z. Xiong, S. Ma, D. Niyato, and Y. Li, ``Personalized saliency in task-oriented semantic communications: Image transmission and performance analysis,'' \emph{IEEE J. Sel. Areas Commun.}, vol. 41, no. 1, pp.186--201, Nov. 2022.

\bibitem{BPG}
F. Bellard, ``BPG image format,'' Apr. 2018. [Online]. Available: https://bellard.org/bpg/

\bibitem{JPEG}
G. K. Wallace, ``The JPEG still picture compression standard,'' \emph{IEEE Trans. Consumer Electron.}, vol. 38, no. 1, pp.18--34, Feb. 1992.


\bibitem{JPEG2000}
D. Taubman and M. Marcellin, ``JPEG2000: Standard for interactive imaging,''  in \emph{Proc. IEEE}, vol. 90, no. 8, pp.1336--1357, Aug. 2002.


\bibitem{LDPC}
T. Richardson and S. Kudekar, ``Design of low-density parity check codes for 5G new radio,'' \emph{IEEE Commun. Mag.}, vol. 56, no. 3, pp. 28--34, Mar. 2018.


\bibitem{Polar}
E. Arıkan, ``Channel polarization: A method for constructing capacity achieving codes for symmetric binary-input memoryless channels,''
\emph{IEEE Trans. Inf. Theory}, vol. 55, no. 7, pp. 3051--3073, Jul. 2009.





\bibitem{Kurka_CRB}
D. B. Kurka and D. G{\"u}nd{\"u}z, ``DeepJSCC-f: Deep joint source-channel coding of images with feedback,'' \emph{IEEE J. Sel. Areas Commun.}, vol. 1, no. 1, pp. 178--193, Apr. 2020.

\bibitem{Marzetta_block_fading}
T. L. Marzetta and B. M. Hochwald, ``Capacity of a mobile multiple antenna communication link in Rayleigh flat fading,'' \emph{IEEE Trans. Inf. Theory}, vol. 45, no. 1, pp. 139--157, Jan. 1999.



\bibitem{Albreem_MIMO_detection}
M. A. Albreem, M. Juntti, and S. Shahabuddin, ``Massive MIMO detection techniques: A survey,'' \emph{IEEE Commun. Surveys Tuts.}, vol. 21, no. 4, pp. 3109--3132, 1st Quart., 2019.



\bibitem{Fangchang_encoder_decoder}
F. Ma and S. Karaman, ``Sparse-to-dense: Depth prediction from sparse depth samples and a single image,'' in {\emph{Proc. Int. Conf. Robot. Automat. (ICRA)}}, Brisbane, QLD, Australia, May 2018, pp. 4796--4803.
     

\bibitem{Jifeng_deformable}
J. Dai, H. Qi, Y. Xiong, Y. Li, G. Zhang, H. Hu, and Y. Wei, ``Deformable convolutional networks,'' in \emph{Proc. IEEE Int. Conf. Comput. Vis. (ICCV)}, Venice, Italy, Oct. 2017, pp. 764--773.

\bibitem{Qibin_SPM}
Q. Hou, L. Zhang, M.-M. Cheng, and J. Feng, ``Strip pooling: Rethinking spatial pooling for scene parsing,'' in \emph{Proc. IEEE/CVF Conf. Comput. Vis. Pattern Recognit. (CVPR)}, Seattle, WA, USA, Jun. 2020, pp. 4003--4012.

\bibitem{water-filling}
G. Scutari, D. P. Palomar, and S. Barbarossa, ``The MIMO iterative waterfilling algorithm,'' \emph{IEEE Trans. Signal Process.}, vol. 57, no. 5, pp. 1917--1935, May 2009.


\bibitem{chengshan_precoder}
C. Xiao, Y. R. Zheng, and Z. Ding, ``Globally optimal linear precoders for finite alphabet signals over complex vector Gaussian channels,'' \emph{IEEE Trans. Signal Process.}, vol. 59, no. 7, pp. 3301--3314, Jul. 2011.

\bibitem{3GPP}
X. Lin, ``An overview of 5G advanced evolution in 3GPP release 18,'' in \emph{ IEEE Commun. Stand.},  vol. 6, no. 3, pp. 77--83, Sept. 2022.

\bibitem{unfolding}
Q. Hu, Y. Cai, G. Zhang, G. Yu, and G. Y. Li, ``Deep-unfolding for next-generation transceivers,'' \emph{arxiv:2305.08303}, May 2023. [Online]. Available: {https://arxiv.org/abs/2305.08303}


\bibitem{damping}
C. -J. Wang, C. -K. Wen, S. Jin and S. -H. Tsai, ``Finite-alphabet precoding for massive MU-MIMO with low-resolution DACs,'' \emph{IEEE Trans. Wireless Commun.}, vol. 17, no. 7, pp. 4706--4720, Jul. 2018.


\bibitem{finite}
H. He, C. -K. Wen, S. Jin and G. Y. Li, ``Model-Driven Deep Learning for MIMO Detection,'' {IEEE Trans. Signal Process.}, vol. 68, pp. 1702--1715, Feb. 2020.

\bibitem{Cityscapes}
M. Cordts, M. Omran, S. Ramos, T. Rehfeld, M. Enzweiler, R. Benenson, U. Franke, S. Roth, and B. Schiele, ``The cityscapes dataset for semantic urban scene understanding,'' in \emph{Proc. IEEE Conf. Comput. Vis. Pattern Recognit. (CVPR)}, Las Vegas, NV, USA, Jun. 2016, pp. 3213--3223.

\bibitem{CVRG}
S. Orhan and Y. Bastanlar, ``Semantic segmentation of outdoor panoramic images,'' \emph{Signal Image Video Process.}, vol. 16, pp. 643--650, Aug. 2021.


\bibitem{ERF-PSPNet}
K. Yang, L. M. Bergasa, E. Romera, R. Cheng, T. Chen, and K. Wang, ``Unifying terrain awareness through real-time semantic segmentation,'' in \emph{Proc. IEEE Intell. Vehicles Symp. (IV)}, Changshu, China, Jun. 2018, pp. 1033--1038.

\bibitem{Adam}
D. P. Kingma and J. L. Ba, ``Adam: A method for stochastic optimization,'' in \emph{Proc. Int. Conf. Learn. Rep. (ICLR)}, San Diego, CA, USA, 2015, pp. 1--15.

\bibitem{MIMOCSI}
S. Wu, C. Wang, E. Aggoune, M. Alwakeel, and X.You, ``A general 3-D non-stationary 5G wireless channel model'', \emph{IEEE Trans. Commun.}, vol. 66, no. 7, pp. 3065--3078, Jul. 2018.

\bibitem{ADJSCC}
J. Xu, B. Ai, W. Chen, A. Yang, P. Sun, and M. Rodrigues, ``Wireless image transmission using deep source channel coding with attention modules,'' \emph{arxiv:2012.00533}, Apr. 2022. [Online]. Available: {https://arxiv.org/abs/2012.00533}

\bibitem{DSSCC}
J. Huang, D. Li, C. Huang, X. Qin, and W. Zhang, ``Joint task and data oriented semantic communications: A deep separate source-channel coding scheme,'' \emph{IEEE Internet Things J.}, vol. 11, no. 2, pp. 2255--2272, Jan. 2024.

\bibitem{BDBR}
G. Bjontegaard, ``Calculation of average PSNR differences between RD-curves,'' VCEG-M33, Austin, TX, USA, Apr. 2001, pp. 2--4.


\bibitem{CLIC2021}
``CLIC 2021: Challenge on learned image compression,'' [Online]. Available: http://compression.cc, 2021.

\bibitem{GradCAM}
R. R. Selvaraju, M. Cogswell, A. Das, R. Vedantam, D. Parikh, and D. Batra, “Grad-cam: Visual explanations from deep networks via gradient-based localization,” in \emph{Proc. IEEE Int. Conf. Comput. Vis. (ICCV)}, Venice, Italy, Oct. 2017, pp. 618--626.

\bibitem{kodak}
G. Toderici, D. Vincent, N. Johnston, S. J. Hwang, D. Minnen, J. Shor, and M. Covell, ``Full resolution image compression with recurrent neural networks,''  in \emph{Proc. IEEE/CVF Conf. Comput. Vis. Pattern Recognit. (CVPR)}, Honolulu, Hawaii, USA, Jul. 2017, pp. 5435--5443.

\bibitem{EfficientDet}
M. Tan, R. Pang, and Q. V. Le, ``EfficientDet: Scalable and efficient object detection,'' in \emph{Proc. IEEE/CVF Conf. Comput. Vis. Pattern Recognit. (CVPR)}, Seattle, WA, USA, Jun. 2020, pp. 10781--10790.

\bibitem{Resnet50}
K. He, X. Zhang, S. Ren, and J. Sun, ``Deep residual learning for image recognition,'' in \emph{Proc. IEEE/CVF Conf. Comput. Vis. Pattern Recognit. (CVPR)}, Las Vegas, NV, USA, Jun. 2016, pp. 770--778.

\end{thebibliography}
\end{document}